\documentclass[twocolumn]{aastex631}

\received{February 24, 2022}
\submitjournal{AAS Journals}

\shorttitle{The Composition of HVSs}
\shortauthors{Reggiani et al.}

\begin{document}

\title{The Chemical Composition of Extreme-Velocity Stars\footnote{This paper
includes data gathered with the 6.5-meter Magellan Telescopes located
at Las Campanas Observatory, Chile.}\footnote{This paper
includes data gathered with the 3.5-meter Telescope at Apache Point Observatory.}}

\correspondingauthor{Henrique Reggiani}
\email{hreggiani@carnegiescience.edu}

\author[0000-0001-6533-6179]{Henrique Reggiani}
\affiliation{The Observatories of the Carnegie Institution for Science,
813 Santa Barbara St., Pasadena, CA 91101, USA}

\author[0000-0002-4863-8842]{Alexander P.\ Ji}
\affiliation{Department of Astronomy and Astrophysics, University of Chicago, 
Chicago IL 60637, USA}

\author[0000-0001-5761-6779]{Kevin C. Schlaufman}
\affiliation{William H.\ Miller III Department of Physics and Astronomy, Johns Hopkins
University, 3400 N Charles St, Baltimore, MD 21218, USA}

\author[0000-0002-2139-7145]{Anna Frebel}
\affiliation{Massachusetts Institute of Technology, Cambridge,
MA, USA}

\author[0000-0003-2806-1414]{Lina Necib}
\affiliation{Massachusetts Institute of Technology, Cambridge,
MA, USA}

\author[0000-0003-3707-5746]{Tyler Nelson}
\affiliation{Department of Astronomy, The University of Texas at Austin, 2515 Speedway Boulevard, Austin, TX 78712, USA}

\author[0000-0002-1423-2174]{Keith Hawkins}
\affiliation{Department of Astronomy, The University of Texas at Austin, 2515 Speedway Boulevard, Austin, TX 78712, USA}

\author[0000-0001-9261-8366]{Jhon Yana Galarza}
\affiliation{CRAAM, Mackenzie Presbyterian University, Rua da Consolação, 896, São Paulo, Brazil}

%ana frebel, lina ecip

\begin{abstract}

\noindent
Little is known about the origin of the fastest stars in the Galaxy. Our understanding 
of the Milky Way and surrounding dwarf galaxies chemical evolution history allows us to 
use the chemical composition of a star to investigate its origin, and say 
whether a star was formed in-situ or was accreted. However, the fastest stars, 
the hypervelocity stars, are young and massive and their chemical 
composition has not yet been analyzed. Though it is difficult to analyze the chemical composition 
of a massive young star, we are well versed in the analysis of late-type stars. 
We have used high-resolution ARCES/3.5m Apache 
Point Observatory, MIKE/Magellan spectra to study the chemical details of 15 late-type 
hypervelocity stars candidates. With Gaia EDR3 astrometry and
spectroscopically determined radial velocities we found total velocities 
with a range of $274$ - $520$ km s$^{-1}$ and mean value of $381$ km s$^{-1}$. Therefore, 
our sample stars are not fast enough to be classified as Hypervelocity stars, and 
are what is known as extreme-velocity stars. Our sample has a wide iron abundance range of 
$-2.5 \le \mathrm{[Fe/H]} \le -0.9$. Their chemistry indicate that at least 
50\% of them are accreted extragalactic stars, with iron-peak elements consistent with prior  
sub-Chandrasekhar mass type Ia supernova enrichment. Without indication of binary companions,
their chemical abundances and orbital parameters are indicative that they 
are the accelerated tidal debris of disrupted dwarf galaxies.

\end{abstract}


\section{Introduction}\label{intro}

The existence of Hypervelocity stars (HVSs) was first proposed by \cite{hills1988}. 
\cite{hills1988} coined the term Hypervelocity stars to describe stars ejected with total 
velocities exceeding $1000$ km s$^{-1}$ after a stellar encounter with a massive black hole. 
Hills theorized that an interaction of a pair of binary stars with the supermassive 
black hole Sagitarius A could result in the ejection of one of the stars at such velocities. 
The proposed acceleration mechanism is known as the Hills mechanism. 
Finding a HVS coming from the Galactic center, argued Hills, would be definitive evidence of the 
presence of a massive black hole 
in the Galactic center. Hills continued his theoretical study of these objects in \cite{hills1991,hills1992}, 
but it took almost 20 years for a HVS to be found. 
\cite{brown2005} reported the discovery of SDSS J090745.0+024507, a 
$853$ km s$^{-1}$ star found to be unbound from the Galactic potential. 
To explain such a speed, a dynamical interaction with a compact object is needed \citep{brown2015}.

Since the discovery of SDSS J090745.0+024507 the interest in HVSs has considerably 
increased, both from theoretical and observational perspectives. 
Theorists have put forward different ways to explain how stars can be 
accelerated to high velocities: a variation of the Hills mechanism via three-body interactions 
\citep{yu2003}; ejection of the companion when one star in a binary system explodes as a supernova  \citep{blaauw1961}; 
tidal debris from a dwarf galaxy recently disrupted near the center of the Milky Way \citep{abadi2009}; dynamical ejection 
from dense stellar clusters \citep{poveda1967,leonard1991,bromley2009}; ejection from star-forming dwarf 
galaxies, in particular it has been proposed that stars can be ejected from the Large Magellanic Cloud (LMC) 
with extremely large velocities \citep{boubert2016}; among others.

It is important to clarify that the different mechanisms eject stars with different velocities. While the 
Hills mechanism was initially proposed to explain stars with velocities exceeding $1000$ km s$^{-1}$, the different 
interactions and mechanisms proposed above can produce stars with velocities as ``low'' 
as $\sim400$ km s$^{-1}$. Though the term HVS has initially been used to refer to unbound 
stars accelerated 
by Hills mechanism, it was also used to describe unbound stars and even stars with extreme velocities ($\sim300-600$ km s$^{-1}$) that are still bound to the Galacitc Potential \citep{hattori2018b}. 
In this paper we will use the term HVSs to describe extreme-velocity stars, but 
not necessarily unbound, nor stars that directly interacted with the central 
Sagittarius A*. via the Hills mechanism.

From an observational perspective, only about 20 stars with extreme velocities were found prior to 2018 
\citep{hattori2018b,brown2015}. At that point, most of the HVSs were young, massive stars, such as 
B-type main sequence stars \citep[e.g.,][and references therein]{brown2015}. These stars are believed to 
have been recently ejected from the Galactic Center, stellar disk, or star-forming dwarf satellites. 

Opportunities to discover and characterize HVSs have considerably increased since 
the initial discoveries, specially owing to the Gaia data releases. For example,  
\cite{shen2018} reported the discovery of three white dwarfs with velocities exceeding $1000$ km s$^{-1}$, or the discovery of S5-HVS 1, an unbound A type star located at 9 kpc from the Sun 
travelling at a staggering $1700$ km s$^{-1}$ \citep{koposov2020,irrgang2021}. \cite{bromley2018}, \cite{marchetti2018a,marchetti2018b}, \cite{boubert2018}, and \cite{hattori2018a} have all 
searched for HVSs candidates, with a focus on late-type stars, and found over 30 stars.

The discovery of fast-moving late-type stars provides a powerful tool to study both the origin 
of HVSs and the mechanism 
that ejects stars to high velocities. While a chemical analysis of massive, young, A, O, and B-type stars is 
limited due to the high temperatures in the stellar photospheres, the analysis of late-type stars chemistry is far 
less complicated. The first work to study the chemistry of late-type HVSs was \cite{hawkins2018}. They 
did a comprehensive study of four extreme-velocity stars and found that all of them have chemical 
patterns consistent with Milky Way stars, and are part of the halo velocity distribution tail.

In this paper we continue the effort of carrying out a comprehensive spectroscopic analysis of 
HVSs. We observed 15 HVs candidates, updated their total velocities using 
Gaia EDR3 parallax and proper motions, and radial velocities from our high-resolution spectra, 
and performed a comprehensive chemical analysis of the targets.
The paper is divided as follows: In Section \ref{sample_obs} we outline our sample selection and 
observation, in Section \ref{atm_param_orbits} we describe how we estimated stellar parameters 
and orbits. We describe the chemical patterns of the stars in 
Section \ref{chem_abundances_sec}, discuss our results in Section \ref{discussion} 
and provide our final remarks in Section \ref{conclusion}.

\section{Sample Selection and Observations}\label{sample_obs}

We analyzed a total of 15 stars. Our sample was selected from 
extreme-velocity stars from \cite{hattori2018a} 
and \cite{herzog2018} (Gaia DR1 data). 
We selected stars that had quoted total 
velocities $\gtrapprox 300$ km s$^{-1}$, without previous spectroscopic analysis. 
A lack of a former spectroscopic analysis also mean their radial velocity 
estimates could be improved upon. 
We selected the targets, from the two studies above, based on their 
magnitudes (prioritizing brighter stars) and observability at the time of 
our observing runs. 
We also observed the well-known rapid neutron-capture process (r-process) 
enhanced star HE 1523$-$0901, which is also an extreme-velocity star. The extreme 
r-process enhancement of this star raises the interesting question of either 
or not other extreme-velocity stars are also r-process enhanced 
(Section \ref{neutrn_capt}). 

We observed the targets with two instruments: the Magellan Inamori Kyocera
Echelle (MIKE) spectrograph on the Magellan Clay Telescope at Las Campanas
Observatory \citep{bernstein2003,shectman2003} and the ARC echelle spectrograph 
(ARCES) on the 3.5m Apache Point Observatory telescope. For MIKE observations we 
used either the $0\farcs35$ slit, the $0\farcs5$ slit or the $0\farcs7$ slit with standard blue 
and red grating azimuths, yielding spectra between 335 nm and 950 nm with resolving power  
varying from $R \approx83,\!000$ to $R \approx40,\!000$ in the blue arm and 
from $R \approx65,\!000$ to $R \approx 31,\!000$ in the red arm. We collected all
calibration data (e.g., bias, quartz \& ``milky" flat field, and ThAr
lamp frames) in the afternoon before each night of observations.
We reduced the raw spectra and calibration frames using the
\texttt{CarPy}\footnote{\url{http://code.obs.carnegiescience.edu/mike}}
software package \citep{kelson2000,kelson2003,kelson2014}. On ARCES, we used 
the standard 1\farcs6 slit yielding effective spectra with continuous coverage 
between $390$ nm and $900$ nm with resolving power 
of $R \approx 31,\!000$. All calibration data were 
collected prior to the science observations. We reduced ARCES data using the 
CERES pipeline \citep{brahm2017}. 
We present a log of these observations in Table~\ref{obs_log}.
We used \texttt{iSpec}\footnote{\url{https://www.blancocuaresma.com/s/iSpec}}
\citep{blanco-cuaresma2014,blanco-cuaresma2019} to calculate radial
velocities and barycentric corrections and to normalize the spectra. Our 
estimated radial velocities are compatible with Gaia DR2 radial velocities, 
therefore we did not identify any signs of variability based on our data plus Gaia DR2.

\begin{deluxetable*}{lccccccccc}
\tablecaption{Log of Observations\label{obs_log}}
\tablewidth{0pt}
\tablehead{
\colhead{Source ID} & \colhead{R.A.} & \colhead{Decl.} & 
\colhead{UT Date} & \colhead{Start} & \colhead{Slit} & \colhead{Exposure} &
\colhead{S/N} & \colhead{S/N} & \colhead{Instrument}\\
\colhead{EDR3} & \colhead{(h:m:s)} & \colhead{(d:m:s)} &
\colhead{} & \colhead{} & \colhead{Width} & \colhead{Time (s)} & 
\colhead{$4500 \ \rm{\AA}$} & \colhead{$6500 \ \rm{\AA}$} & \colhead{}
}
\startdata
2853089398265954432 & 00 00 21.88 & $+$25 19 21.81 & 2018 Oct 26 & 6:00:13 & 1.6 & 870 & 35 & 55 & ARCES\\ 
330414789019026944 & 01 56 52.65 & $+$36 39 57.89 & 2018 Oct 28 & 5:48:53 & 1.6 & 1200 & 78 & 123 & ARCES\\ 
2260163008363761664 & 18 07 53.09 & $+$70 12 47.92 & 2018 Oct 28 & 2:05:55 & 1.6 & 1300 & 66 & 104 & ARCES\\ 
2233912206910720000 & 19 57 21.65 & $+$55 30 23.00 & 2018 Oct 28 & 2:36:30 & 1.6 & 4110 & 41 & 65 & ARCES\\ 
1765600930139450752 & 21 49 48.66 & $+$10 48 43.08 & 2018 Oct 26 & 5:26:18 & 1.6 & 1350 & 35 & 55 & ARCES\\ 
3252546886080448384 & 04 09 02.19 & $-$02 12 27.07 & 2018 Nov 16 & 3:53:10 & 0.5 & 1000 & 86 & 118 & MIKE\\ 
5212110596595560192 & 06 16 51.67 & $-$78 26 20.78 & 2018 Nov 16 & 8:30:20 & 0.5 & 300 & 56 & 115 & MIKE\\ 
5212817273334550016 & 07 08 47.74 & $-$76 13 10.16 & 2018 Nov 16 & 8:36:53 & 0.5 & 400 & 63 & 126 & MIKE\\ 
3599974177996428032 & 11 44 40.87 & $-$04 09 51.27 & 2018 Jan 25 & 7:28:56 & 0.7 & 600 & 110 & 198 & MIKE\\ 
6317828550897175936 & 15 26 01.07 & $-$09 11 38.88 & 2017 Aug 25 & 1:42:34 & 0.7 & 600 & 91 & 210 & MIKE\\ 
6479574961975897856 & 21 05 58.65 & $-$49 19 33.51 & 2017 Aug 17 & 4:12:25 & 0.35 & 3030 & 296 & 342 & MIKE\\ 
6558932694746826240 & 21 55 50.81 & $-$51 05 37.21 & 2017 Aug 17 & 2:49:14 & 0.7 & 600 & 104 & 178 & MIKE\\ 
2629296824480015744 & 22 23 19.86 & $-$02 31 10.61 & 2018 Nov 16 & 1:12:31 & 0.5 & 450 & 76 & 122 & MIKE\\ 
6505889848642319872 & 22 47 40.24 & $-$55 37 31.43 & 2018 Nov 16 & 1:22:18 & 0.5 & 300 & 85 & 145 & MIKE\\ 
6556192329517108352 & 23 08 00.95 & $-$33 38 06.94 & 2017 Aug 17 & 3:01:00 & 0.7 & 600 & 97 & 180 & MIKE
\enddata
\end{deluxetable*}

\section{Atmospheric Parameters}
\label{atm_param_orbits}

The stellar parameters were obtained using the same method described in 
\cite{reggiani2020,reggiani2021} and \cite{reggiani2022}, a hybrid 
isochrone/spectroscopy approach. Isochrones are especially useful
for effective temperature $T_{\text{eff}}$ inferences in this case,
as high-quality multiwavelength photometry is available for all stars.  
Similarly, the known distances, from \cite{bailer-jones2021} using 
Gaia EDR3 parallaxes (with uncertainties $\le0.05$ \textit{mas}), 
make the calculation of
surface gravity $\log{g}$ via isochrones straightforward. We estimate 
the stellar microturbulent velocity using the empirical relation 
from \cite{kirby2009}. 

To measure the stellar metallicity we use atomic data for 
\ion{Fe}{1} and \ion{Fe}{2} atomic absorption 
lines from \cite{ji2020} and \cite{reggiani2018}. The atomic and molecular data 
for the selected lines are from the \texttt{linemake}
code\footnote{\url{https://github.com/vmplacco/linemake}}
\citep{sneden2009,sneden2016,placco2021}, maintained by Vinicius Placco and 
Ian Roederer. 
We first measure the equivalent widths of \ion{Fe}{1} and
\ion{Fe}{2} atomic absorption lines by fitting Gaussian profiles with the
\texttt{splot} task in \texttt{IRAF} to our continuum-normalized spectra.
Whenever necessary, we use the \texttt{deblend} task to disentangle
absorption lines from adjacent spectral features. We report our input
atomic data, measured equivalent widths, and line-abundances in
Table \ref{measured_ews}.

We use 1D plane-parallel $\alpha$-enhanced ATLAS9 model atmospheres
\citep{castelli2004}, the 2019 version of the \texttt{MOOG}
radiative transfer code \citep{sneden1973}, and the \texttt{q$^2$}
\texttt{MOOG} wrapper\footnote{\url{https://github.com/astroChasqui/q2}}
\citep{ramirez2014} to derive stellar metallicities.

We use the \texttt{isochrones} 
package\footnote{\url{https://github.com/timothydmorton/isochrones}}
\citep{morton2015} in an iterative process to self-consistently infer
atmospheric and fundamental stellar parameters for each star using
as input:
\begin{enumerate}
\item
Gaia DR2 $G$ magnitudes \citep{gaia2016,gaia2018,evans2018};
\item
Distance prior from \cite{bailer-jones2021} and 
Gaia EDR3 parallaxes \citep{gaia2016,gaia2020,lindegren2020};
\item
$u$, $v$, $g$, $r$, $i$, and $z$ magnitudes and associated uncertainties from Data
Release (DR) 2 of the SkyMapper Southern Sky Survey \citep{wolf2018} when available;
\item
$J$, $H$, and $K_{s}$ magnitudes and associated uncertainties from the
2MASS PSC \citep{skrutskie2006};
\item
$W1$, and $W2$ magnitudes and associated uncertainties from
the Wide-field Infrared Survey Explorer (WISE) AllWISE Source Catalog
\citep{wright2010,mainzer2011};
\item $\mathrm{A_V}$ extinction from the Bayestar2019 3D extinction maps 
\citep{green2019};

\end{enumerate}

We use \texttt{isochrones} to fit the MESA Isochrones and Stellar Tracks
\cite[MIST;][]{dotter2016,choi2016,paxton2011,paxton2013,paxton2015}
library to these data using
\texttt{MultiNest}\footnote{\url{https://ccpforge.cse.rl.ac.uk/gf/project/multinest/}}
\citep{feroz2008,feroz2009,feroz2019} through \textit{PyMultinest} \citep{buchner2014}.  We restricted the  MIST 
library based on the photo-geometric distances from \cite{bailer-jones2021}, 
and extinction $A_{V}$ in the range $0 \leq \rm{A_{V} \ (mag)} \leq 0.3$, based on 
the A$_{\rm{V}}$ values and errors proposed by 
the 3D extinction maps of \cite{green2019}. 

We first use \texttt{isochrones}  
to find self-consistent and physically motivated effective temperature and 
surface gravity results. Using the inferred log$g$ we estimate the 
microturbulent velocity $\xi$ using the empirical relation from 
\cite{kirby2009}. 
We impose the $T_{\text{eff}}$, $\log{g}$, and $\xi$ inferred in this way
to derive $[\text{Fe/H}]$. The metallicities are inferred exclusively 
from \ion{Fe}{2} lines, as they are less affected by non-LTE effects. Though at times 
there are non negligible differences between \ion{Fe}{1} and \ion{Fe}{2} 
inferred abundances we adopt the [\ion{Fe}{2}/H] ratios as our model metallicity. 
We then execute another \texttt{isochrones} calculation, this time using this
updated set of atmospheric stellar parameters in the likelihood.
We iterate this process until the metallicities inferred from both the
\texttt{isochrones} analysis and the equivalent widths analysis are 
consistent (i.e., the isochrone metallicity is the same as the spectroscopic 
within the uncertainties found by the isochrone analysis, which is typically 
on the order of 0.1 dex). The adopted stellar parameters are presented in 
Table~\ref{stellar_params}. 

We derive the uncertainties in our adopted $[\text{Fe/H}]$ and $\xi$
values due to the uncertainties in our adopted $T_{\text{eff}}$
and $\log{g}$ values using a Monte Carlo simulation.  We randomly
sample  200 self-consistent pairs of $T_{\text{eff}}$ and $\log{g}$ from our
\texttt{isochrones} posteriors and calculate the values of $[\text{Fe/H}]$
and $\xi$. 
We save the result of each iteration and find that the contributions of
$T_{\text{eff}}$ and $\log{g}$ uncertainties to our final $[\text{Fe/H}]$
and $\xi$ uncertainties are not relevant. 
%We derive our adopted metallicities and uncertainties imposing the adopted
%$T_{\text{eff}}$ and $\log{g}$ values described above. The result of this
%calculation agrees well with the result of the Monte Carlo simulation.

The random uncertainties derived via isochrones are under the unlikely assumption 
that the MIST isochrone grid perfectly reproduces all stellar properties. 
There are almost certainly 
larger systematic uncertainties that we have not investigated. To account for 
them we adopted the conservative photospheric stellar parameters uncertainties 
of $\Delta\mathrm{T_{eff}}=100$ K and 
$\Delta \mathrm{log}g=0.15$ for the isochrone-based parameters. 
As a change of $0.15$ dex in surface gravity only changes 
our estimated microturbulences by $0.03$ km s$^{-1}$, we 
conservatively adopted an uncertainty of $\Delta\xi=0.1$ km s$^{-1}$. 
The reported metallicity uncertainties 
were derived by adding in 
quadrature the uncertainties from our line-by-line abundance
dispersions and the $T_{\text{eff}}$, $\log{g}$, and $\xi$ adopted 
uncertainties. We report our derived stellar parameters, along with the stellar photometric and 
astrometric data, in Table \ref{stellar_params}.

One of our targeted stars, identified here as Gaia EDR3 6317828550897175936, is 
the well-known r-process enhanced metal-poor star HE 1523$-$0901 \citep{frebel07}. 
We use this star to roughly quantify the systematics generated by our 
methodology. Based on our isochrone analysis combined with the 
\ion{Fe}{2} lines, we derived the following stellar parameters for HE 1523$-$0901: 
T$_{\rm{eff}}=4742\pm100$ K, log$g=1.29\pm0.15$, 
[Fe/H]$=-2.65\pm0.22$, and $\xi=1.83\pm0.10$ km s$^{-1}$.

\cite{frebel07} derived T$_{\rm{eff}}=4630\pm40$ K, log$g=1.0\pm0.3$, 
[Fe/H]$=-2.95\pm0.20$, and $\xi=2.63\pm0.3$ km s$^{-1}$. Their effective temperature 
was derived applying the \cite{alonso99} calibration from  BVRI photometry.  
Surface gravity, metallicity, and microturbulent velocity were derived via the classic spectroscopic
method. Despite the differences in our analysis methods the derived stellar parameters are 
fully compatible within the proposed analyzes uncertainties. 

A more recent analysis of this star was performed by \cite{casey17} using a 
data-driven analysis of RAVE \citep{kunder17} data using \textit{The Cannon} 
\citep{ness15,ness16} pipeline. They reported  
T$_{\rm{eff}}=4789\pm108$ K, log$g=1.54\pm0.23$, and 
[Fe/H]$=-2.56\pm0.09$. These recent results are very similar to our derived photospheric  parameters. 
Our method to determine the  stellar parameters are in agreement with the most recent 
data-driven methods. 

Finally we compare our results for this star to 
\cite{sakari2018}, who derived T$_{\rm{eff}}=4530$ K, log$g=0.8$, 
[Fe/H]$=-2.72$, and $\xi=2.26$ km s$^{-1}$. Their stellar 
parameters were estimated using a differential spectroscopic analysis, taking into account 
$<3D>$ non-LTE corrected iron abundances. In their paper \cite{sakari2018} 
estimates uncertainties to be on the order of 20-200 K for effective temperature, 
0.05-0.3 dex in surface gravity, and 0.10-0.35  in microturbulent velocity. Therefore 
their results are also compatible with our derived stellar parameters.

In summary, our method is compatible with a recent, well tested, data-driven method and 
also consistent with the more classical approach from \cite{frebel07}. There is a 
larger difference between our work and that of \cite{sakari2018} but it can be accounted 
by the uncertainties. 
All four studies briefly discussed here 
were performed with fundamentally different techniques but they obtain similar results within their 
uncertainties. We can conclude that our methodology can be directly compared with 
the above-mentioned results and that our adopted uncertainties are reasonable.
\subsection{Stellar Orbits}
\label{stellar_orbits_sec}

We calculated the Galactic orbits of our sample stars using
\texttt{galpy}\footnote{\url{https://github.com/jobovy/galpy}}. 
\texttt{Galpy} uses a left-handed coordinate system, with the Sun at positive X 
and positive L$_{\mathrm{Z}}$ indicates a prograde orbit.
We sampled 1,000 Monte Carlo realizations from the Gaia EDR3 astrometric
solutions for each star using the distance posterior that results from
our isochrone analysis while taking full account of the covariances
between position, parallax, and proper motion.  We used the radial
velocities derived from our spectra and assumed
no covariance between our measured radial velocity and the Gaia EDR3
astrometric solution.  We used each Monte Carlo realization as an
initial condition for an orbit and integrated it forward 10 Gyr in
a Milky Way-like potential.  We adopted the \texttt{MWPotential2014}
described by \citet{bovy2015}.  In that model, the bulge is parameterized
as a power-law density profile that is exponentially cut-off at 1.9
kpc with a power-law exponent of $-1.8$.  The disk is represented by
a Miyamoto--Nagai potential with a radial scale length of 3 kpc and
a vertical scale height of 280 pc \citep{miyamoto1975}.  The halo is
modeled as a Navarro--Frenk--White halo with a scale length of 16 kpc
\citep{navarro1996}.  We set the solar distance to the Galactic center
to $R_{0} = 8.122$ kpc, the circular velocity at the Sun to $V_{0} =
238$ km s$^{-1}$, the height of the Sun above the plane to $z_{0} =
25$ pc, and the solar motion with the respect to the local standard
of rest to ($U_{\odot}$, $V_{\odot}$, $W_{\odot}$) = (10.0, 11.0,
7.0) km s$^{-1}$ \citep{juric2008,blandhawthorn2016,gravity2018}.
We report the resulting orbital parameters (radial velocities, 
total galactic velocities $v$, pericenter R$_{\mathrm{peri}}$, 
apocenter R$_{\mathrm{apo}}$, eccentricity $e$, maximum distance 
from the galactic plane Z$_{\mathrm{max}}$, total orbital energy E$_{\mathrm{tot}}$, 
and angular momentum L$_{\mathrm{Z}}$) in Table~\ref{stellar_params}. 

\section{Chemical Abundances}
\label{chem_abundances_sec}

We measured the equivalent widths of atomic absorption lines for
\ion{Na}{1}, \ion{Mg}{1}, \ion{Al}{1}, \ion{Si}{1}, \ion{K}{1}, \ion{Ca}{1},
\ion{Sc}{2}, \ion{Ti}{1}, \ion{Ti}{2}, \ion{Cr}{1}, \ion{Cr}{2}, 
\ion{Mn}{1}, \ion{Fe}{1}, \ion{Fe}{2}, \ion{Co}{1}, \ion{Cu}{1}, 
\ion{Ni}{1}, \ion{Zn}{1}, \ion{Sr}{1}, \ion{Sr}{2}, \ion{Y}{2}, \ion{Ba}{2}, 
and \ion{La}{2}, in our
continuum-normalized spectra by fitting Gaussian profiles with the
\texttt{splot} task in \texttt{IRAF}. We used the \texttt{deblend} task
to disentangle absorption lines from adjacent spectral features whenever
necessary. We employed the 1D plane-parallel $\alpha$-enhanced ATLAS9
model atmospheres and the 2019 version of \texttt{MOOG} to calculate
abundances for each equivalent width. We report our input
atomic data from \cite{ji2020} and \cite{reggiani2018}. The atomic and 
molecular data are from the \textit{linemake} 
code \citep{sneden2009,sneden2016, placco2021}. We measured 
equivalent widths, and individual
inferred abundances in Table~\ref{measured_ews}. We measured Eu abundances 
from \ion{Eu}{2} lines through spectrum synthesis, from up to five lines: 
4129, 4205, 4435, 4522, and 6645 \AA. 
We present our adopted
mean chemical abundances, and associated errors, in Table~\ref{chem_abundances_tbl}. 
The $\sigma_{\mathrm{[X/Fe]}}$ 
are a measure of the line dispersion divided by the number of lines, 
added in quadrature to the 
uncertainties due to the adopted stellar parameters uncertainties 
($\Delta\mathrm{T_{eff}}=100$ K, $\Delta\mathrm{log}g=0.15$ cm.s$^{-2}$, 
and $\xi=0.1$ km s$^{-1}$). We show our abundances, 
with data from additional bibliographical sources 
in Figures~\ref{alphas_fig} to \ref{neutron_capture_fig}. In Figures 
\ref{alphas_fig} through \ref{neutron_capture_fig} we 
use the more commonly employed mean value of all \ion{Fe}{1} and \ion{Fe}{2} lines 
to estimate the [X/Fe] ratios and for the [Fe/H] ratio. Though it differs from our 
model metallicities, these [Fe/H] ratios and [X/Fe] ratios are directly comparable 
to the literature data.

To quantify possible systematics we compared our derived abundances for star HE 1523$-$0901 with those 
from \cite{sakari2018}. We chose \cite{sakari2018} because neither \cite{frebel07} nor 
\cite{casey17} provided full sets of abundances for this star, 
and because \cite{sakari2018} also calculated most abundances from equivalent widths 
and used the same radiative transfer code (MOOG). 

We found a mean absolute abundance difference of  
$\Delta\rm{[X/Fe]}=-0.14$ dex. This value is well within the errors of the analyzes. 
HE 1523$-$0901 is known to be highly r-process enhanced, and 
our inferred neutron-capture process elemental abundances are, as expected, enhanced. 
There is a difference when we compare our results to those inferred by 
\cite{sakari2018}: $\Delta\rm{[Y/Fe]}=-0.16$, $\Delta\rm{[Ba/Fe]}=-0.3$, 
$\Delta\rm{[Eu/Fe]}=0.07$, and $\Delta\rm{[La/Fe]}=-0.24$. These differences are mostly due to the fact that our mean metalillicity (derived from \ion{Fe}{1} and \ion{Fe}{2} lines) 
is higher than adopted in their work. Instead, if our \ion{Fe}{2} 
[Fe II/H] is used to estimate [X/Fe] the differences are considerably reduced: $\Delta\rm{[Y/Fe]} = 0.04$, $\Delta\rm{[Ba/Fe]} = 0.1$, $\Delta\rm{[Eu/Fe]} = -0.01$, and 
$\Delta\rm{[La/Fe]} = 0.04$.

Although there are non-negligible differences in a few elements, they are on the order of the inferred 
uncertainties, as evidenced by our mean abundance difference \citep[from][]{sakari2018}.
We conclude that any systematics in our analysis are lower than about $\sim0.2$ dex and 
are accounted for in our uncertainties. 

\subsection{$\alpha$-Elements}
\label{alpha_abnd}

\begin{figure*}
\includegraphics[scale=0.4]{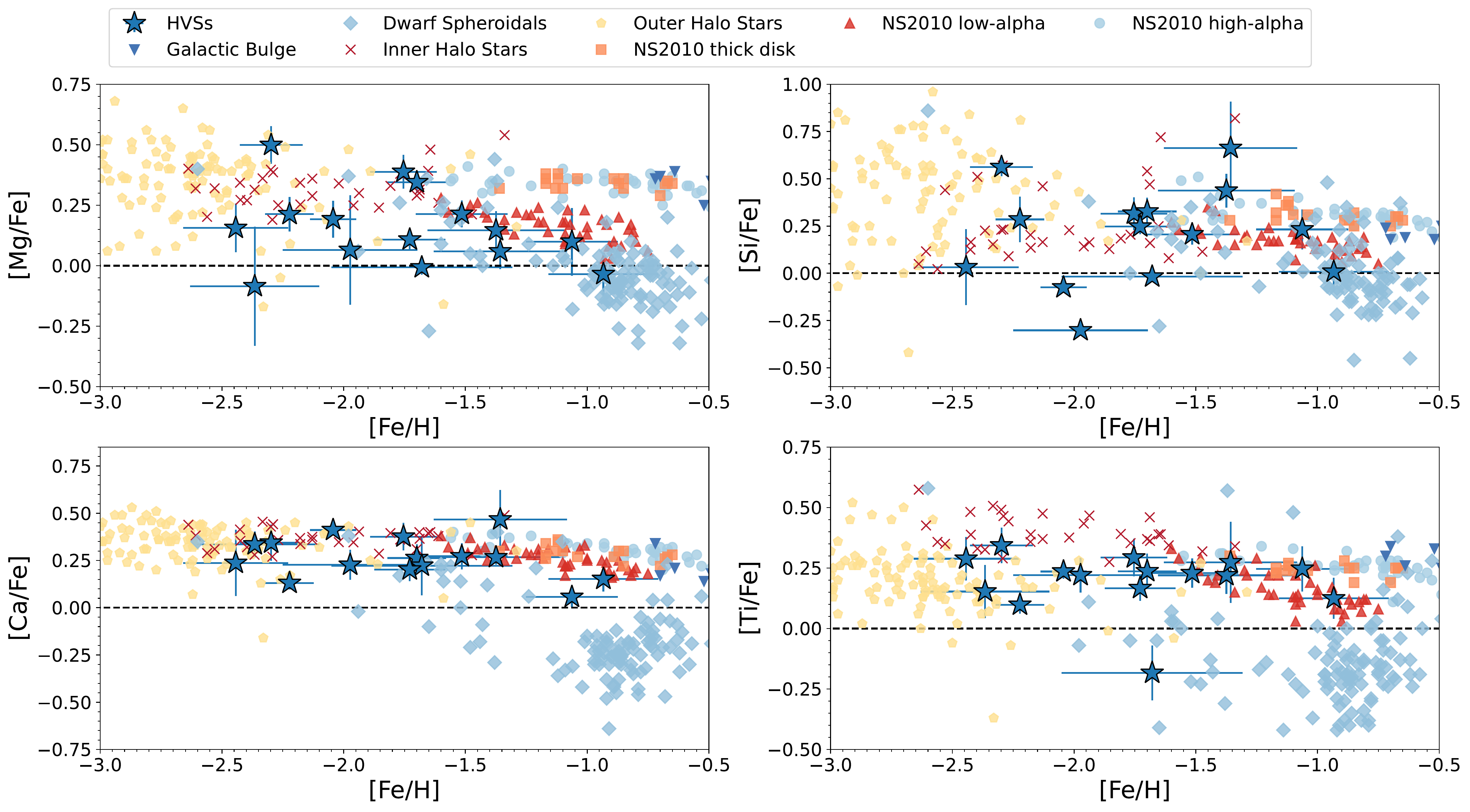}
\caption{Abundances of the $\alpha$-elements magnesium, 
silicon, and calcium, and titanium.  We plot our HVSs as dark-blue stars. 
We plot in yellow the abundances of Milky Way metal-poor stars from
\cite{cayrel2004} and \cite{jacobson15} outer halo stars; in red we show abundances 
of inner halo stars from Reggiani et al. (in prep), \cite{reggiani2017} and \cite{nissen2010}; 
we display the \cite{bensby2010} bulge stars in orange; in light-blue 
we include stars from the dwarf spheroidal galaxies 
Carina, Sculptur, Fornax, and Sagittarius from \cite{shetrone2003}, 
\cite{geisler2005}, \cite{monaco2005}, and \cite{letarte2010}. \label{alphas_fig}}
\end{figure*}

In Figure \ref{alphas_fig} we show the chemical abundances of the $\alpha$-elements 
(Mg, Si, Ca, and Ti - elements synthesized through $\alpha$ reaction, and Ti, which is not 
formed through the same nucleosynthetic channel but is often considered to be an $\alpha$-element 
due to its chemical abundances similarities). 
These elements indicative of how fast chemical evolution 
took place in any given environment. This can be done via the ``knee'', an inflection in [$\alpha$/Fe] 
indicating ``when'' iron production start to be dominated by type Ia supernovae instead of 
type II supernovae. The position of the knee ultimately depends on the complex details of the 
star formation history in a given environment, but as a general rule it can be associated 
with the mass of the host galaxy \citep[e.g.,][]{tinsley1979,matteucci1990,suda2017}. 

In the Milky Way the position of the ``knee'' is at [Fe/H] $\sim -1.0$ and the evolution of the 
[$\alpha$/Fe] abundances is well understood. Below the position of the knee there is an 
abundance plateau in the $\alpha$-abundances that spans all the metallicity range of our study.  
The abundance plateau, for metallicities lower than [Fe/H]$\le-1.0$, 
is observed to be at $\sim0.3$ dex for [Mg/Fe], [Si/Fe], and [Ca/Fe]. However, 
models diverge from the observations and predict a plateau at [Mg/Fe]$\sim0.45$ dex, 
[Si/Fe]$\sim0.5$ dex, and [Ca/Fe]$\sim0.3$ dex 
\citep[e.g.][ to site a few]{tinsley1979,matteucci1990,reggiani2017,hayes2018,amarsi2020,kobayashi2020}. 

The HVSs analyzed by \cite{hawkins2018} all had Galactic chemistry. Their $\alpha$-abundances 
were typical high-$\alpha$ halo stars, with metallicities between $-2 \le $ [Fe/H] $\le -1$. 
In Figure \ref{alphas_fig} we show our $\alpha$-abundance ratios, along with 
different bibliographic samples of the Milky Way halo \citep{cayrel2004,nissen2010,reggiani2017}, 
Milky Way thick disk \citep{nissen2010}, 
Milky Way bulge \citep{bensby2010}, and the dwarf galaxies 
Carina \citep{shetrone2003}, Sculptor \citep{shetrone2003,geisler2005}, Sagittarius \citep{monaco2005}, and Fornax \citep{letarte2010}.

At metallicities higher than 
[Fe/H]$\gtrsim -1.5$, our mean abundances are [Mg/Fe]$=0.1$ dex, [Ca/Fe]$=0.2$ dex, 
[Si/Fe]$=0.3$ dex, and [Ti/Fe]$=0.2$ dex. Apart from silicon abundances, for which 
non-LTE effects can be important at the measured lines, all our $\alpha$-elements have 
abundances consistently lower than the observed Milky Way plateau, and 
considerably lower than the chemical evolution models predict \citep{kobayashi2020}. 
These abundances are consistent with abundances of stars formed after the knee, which 
is expected to be at lower metallicities for lower-mass galaxies \citep{suda2017}. 
In Figure \ref{alphas_fig} we can 
also visibly infer that the $\alpha$-abundances for these 5 stars are on top of 
either the dwarf galaxies abundances or the \cite{nissen2010} low-$\alpha$ sequence, which are 
also interpreted as accreted stars. this is particularly true for the two highest metallicity 
stars. The other three stars have somewhat higher calcium abundances, but as can be seen in 
Figure \ref{alphas_fig} the calcium dispersion is not as high and the low- and high-$\alpha$ 
sequences from \cite{nissen2010} are not as distinguishable as they are in the magnesium ratios. 
Therefore, their $\alpha$-abundance ratios are more consistent with those found in 
dwarf galaxies. 

At metallicities lower than [Fe/H]$\lesssim-1.5$ our mean abundances are different: 
[Mg/Fe]$=0.2$ dex, [Si/Fe]$=0.2$ dex, [Ca/Fe]$=0.3$ dex, and [Ti/Fe]$=0.2$ dex. Again, as 
silicon might be affected by non-LTE effects, our main indicators are the higher magnesium and 
calcium abundances, consistent with Milky Way studies. It is also important to point that the 
stars with the lowest $\alpha$-abundances below [Fe/H]$\le-1.5$ also have the largest 
uncertainties. Although we have stars at this metallicity range that are compatible with 
the low-$\alpha$ sequence \citep{nissen2010} and can be accreted stars, the overall behavior 
at lower metallicities is more similar to Milky Way stars. It is also important to remark that 
the scatter at lower metallicities is higher (see the \cite{cayrel2004} abundance spread 
compared to the \cite{reggiani2017} and \cite{nissen2010} spread at Figure \ref{alphas_fig}, 
for example).

To summarize, our $\alpha$-abundances indicate that the high metallicity end of our sample 
is likely composed of accreted stars, while the low-metallicity sample is more likely 
to be Milky Way in situ stars, apart from a couple of stars with higher uncertainties that could 
also be accreted stars. The chemical abundances of additional elements will provide further 
information regarding the stellar origin.

\subsection{Light odd-Z}
\label{light_odd_z_abnd}

\begin{figure*}
\includegraphics[scale=0.4]{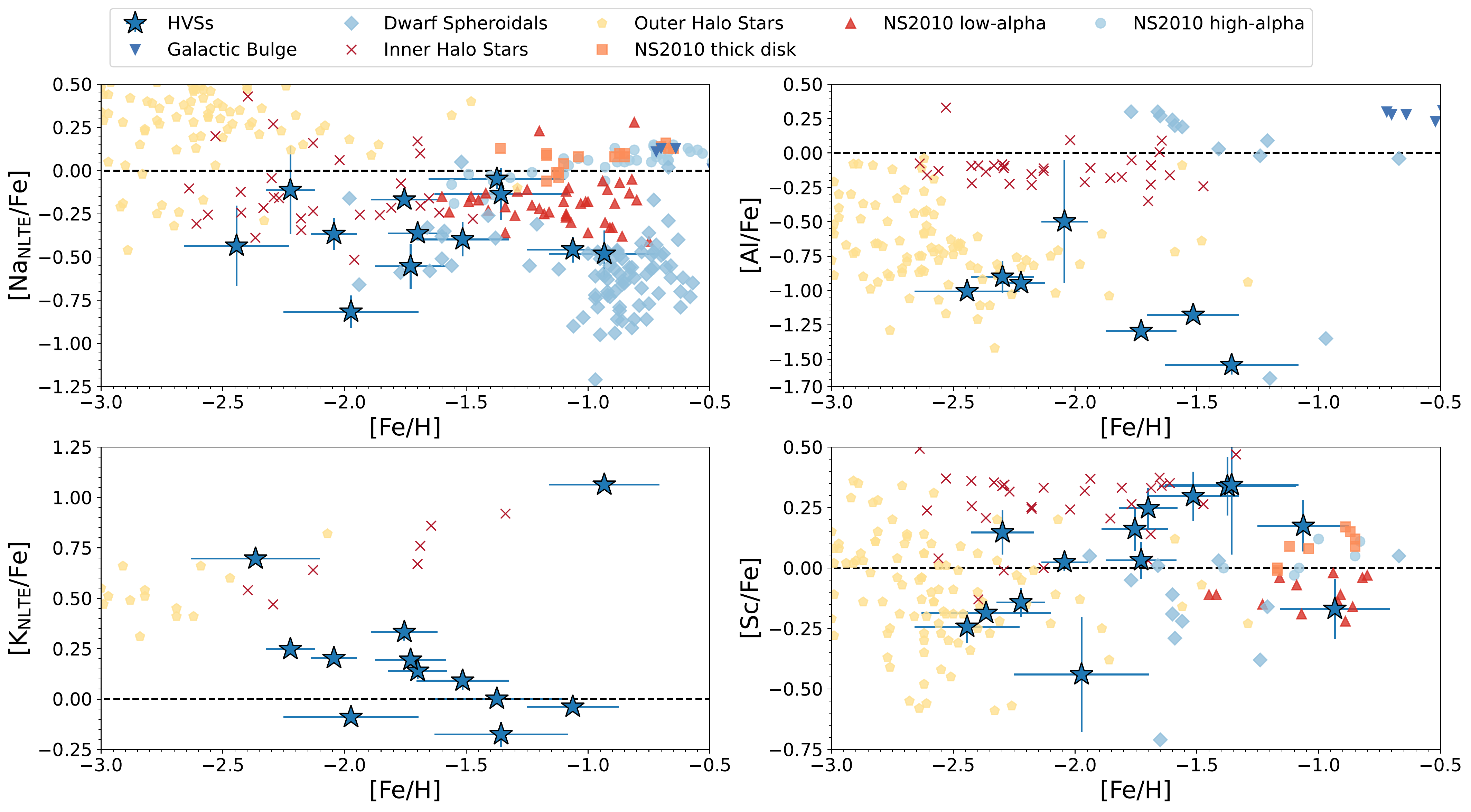}
\caption{Abundances of the light odd-Z elements sodium, aluminum, potassium 
and scandium. We plot our HVSs as dark-blue stars. 
We plot in yellow the abundances of Milky Way metal-poor stars from
\cite{cayrel2004} and \cite{jacobson15} outer halo stars; in red we show abundances 
of inner halo stars from Reggiani et al. (in prep), \cite{reggiani2017}, \cite{nissen2010}, 
and \cite{fishlock2017}; 
we display the \cite{bensby2010} bulge stars in orange; in light-blue 
we include stars from the dwarf spheroidal galaxies 
Carina, Sculptur, Fornax, and Sagittarius from \cite{shetrone2003}, 
\cite{geisler2005}, \cite{monaco2005}, and \cite{letarte2010}. \label{light_odd_fig}}
\end{figure*}

Figure \ref{light_odd_fig} shows the abundances of the light odd-Z elements 
(Na, Al, K, and Sc). 

While sodium and aliminum are mostly produced in core-collapse supernovae, their 
yields are dependent on metallicity and consequently their chemical
evolution is not as easily interpreted as the chemical evolution of the
$\alpha$-elements. Both the exact nucleosynthetic origin and chemical
evolution of scandium and potassium are likewise hard to identify and
interpret \citep[e.g.,][]{clayton2007,zhao2016,prantzos2018,reggiani2019,kobayashi2020}.  
We plot in the top panels of
Figure \ref{light_odd_fig} our measured $[\text{Na/Fe}]$
and $[\text{Al/Fe}]$, and $[\text{K/Fe}]$ and $[\text{Sc/Fe}]$ 
abundances as a function of $[\text{Fe/H}]$ on the lower panels.

Sodium, aluminum and potassium abundance inferences can be strongly affected
by departures from LTE, but scandium abundances from \ion{Sc}{2} lines are not
strongly affected \citep{zhao2016}. We corrected sodium abundances using the \citet{lind2011}
grid through the INSPECT project\footnote{http://inspect-stars.com/}. 
The aluminum non-LE abundance corrections are from \cite{nordlander2017}. 
We obtained potassium abundance corrections via a linear interpolation
of the \citet{reggiani2019} grid of corrections for abundances inferred
from the equivalent width of the \ion{K}{1} line at 7698 \AA.

As already found for the $\alpha$-elements the abundance ratios of 
sodium, aluminum and potassium are lower for the higher metallicity stars. 
Particularly for sodium we can see that the abundance of our high metallicity stars 
are closer to the abundances in dSphs than the abundances in the 
Milky Way. Scandium abundances are near [Sc/Fe]$\sim0.0$ with a scatter, similar 
to the stars both in the Milky Way and dwarf galaxies. 

\subsection{Iron-peak elements}
\label{iron_peak_abnd}

\begin{figure*}
\includegraphics[scale=0.4]{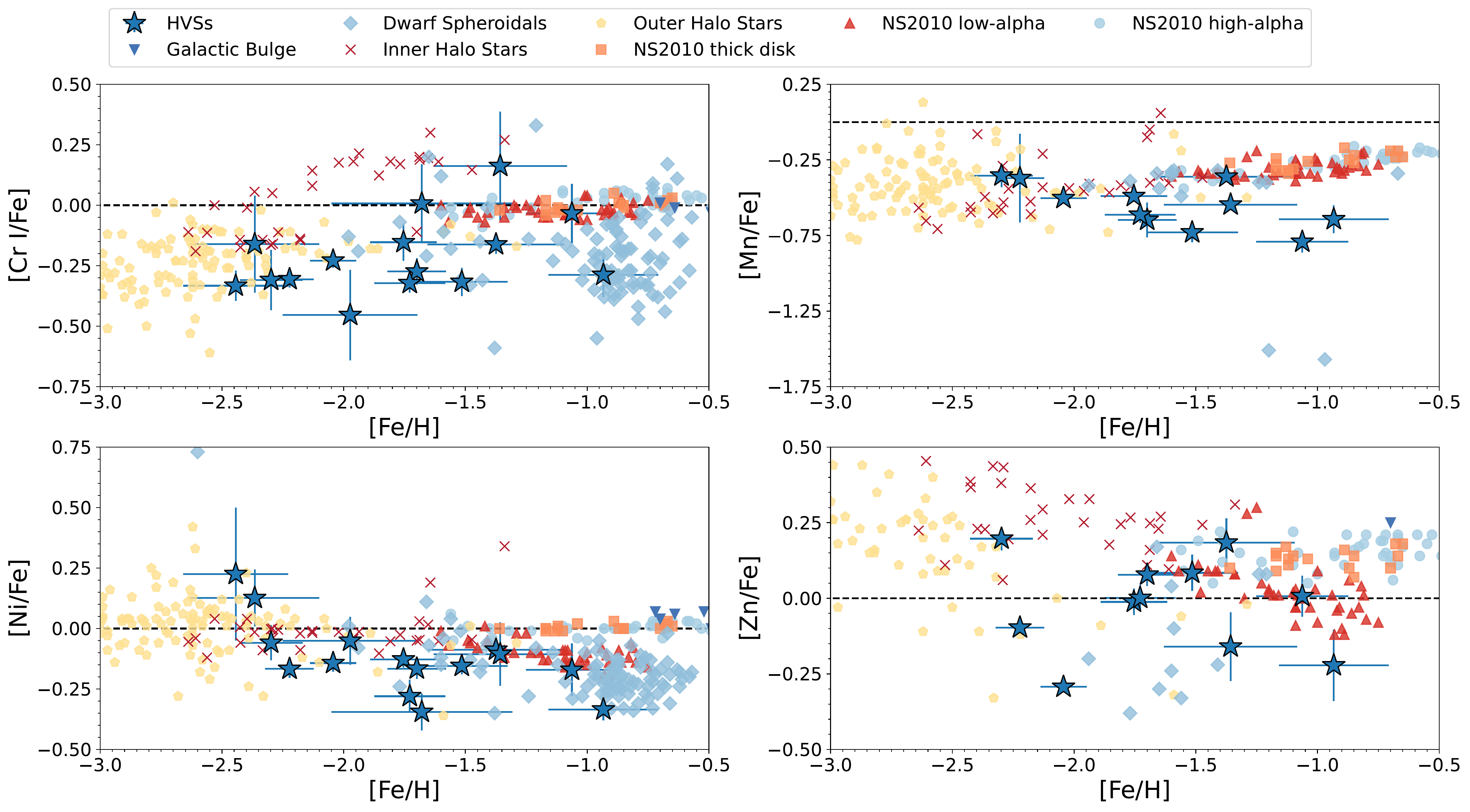}
\caption{Abundances of the iron-peak elements cromium, manganese, 
nickel, and zinc. We plot our HVSs as dark-blue stars. 
We plot in yellow the abundances of Milky Way metal-poor stars from
\cite{cayrel2004} and \cite{jacobson15} outer halo stars; in red we show abundances 
of inner halo stars from Reggiani et al. (in prep), \cite{reggiani2017} and \cite{nissen2010}; 
we display the \cite{bensby2010} bulge stars in orange; in light-blue 
we include stars from the dwarf spheroidal galaxies 
Carina, Sculptur, Fornax, and Sagittarius from \cite{shetrone2003}, 
\cite{geisler2005}, \cite{monaco2005}, and \cite{letarte2010}. \label{iron_peak_fig}}
\end{figure*}

Iron-peak elements are mostly synthesized in type Ia supernovae explosions, 
and their abundance pattern typically follow that of iron. 
In Figure \ref{iron_peak_fig} we show our iron-peak [X/Fe] ratios (Cr, Mn, Ni, 
and Zn). 

Chromium abundances were calculated both from \ion{Cr}{1} and 
\ion{Cr}{2} lines and both abundances are presented in Table \ref{chem_abundances_tbl}. In 
Figure \ref{iron_peak_fig}, we show the results for \ion{Cr}{1}. 
The chromium abundances 
are scattered, but follow the same abundance path with increasing 
metallicity that is observed in other samples based on \ion{Cr}{1} lines. At higher 
metallicities the stars also appear to have abundances that overlap with those observed 
in dwarf galaxies.

The other iron-peak elements diverge more from the Milky Way. 
At lower metallicities the inferred manganese abundances 
are compatible with the abundances of the Milky Way, but the abundances of the 
more metal-rich stars are lower than in the Milky Way. %These differences are evidence 
%of an enrichment history incompatible with that of the Milky Way. 
Other than our target stars, in Figure \ref{iron_peak_fig}, there are 
few stars with very low manganese abundances, and all of them 
are from dwarf galaxies. The same happens for nickel and 
zinc. It is particularly interesting that we find low [Ni/Fe] ratios, because in Milky Way stars 
nickel abundances are [Ni/Fe]$\sim0.0$ regardless of the stellar metallicity 
because its enrichment history is extremely well tied to that of iron. This is exemplified 
in \cite{reggiani2017}, who showed the extremely tight correlation of nickel abundances 
in the solar neighborhood. On the other hand there is considerable scatter in the nickel 
abundances of dwarf galaxies. This is tied to the different nucleosynthetic sources expected for 
the Milky Way and dwarf galaxies. While a large enrichment from sub-Chandrasekhar mass 
type Ia explosions (lower nickel production) 
is observed for dwarf galaxies, in the Milky Way the enrichment is dominated 
by Chandrasekhar-mass type Ia explosions \citep[larger nickel production, e.g., ][]{kobayashi2020b}. 
Zinc in the Milky Way is also 
defined by an increase in the abundance ratio as metallicity decreases. This is associated 
with higher explosion energies of type II supernovae at earlier times \citep{kobayashi2020}. 
On the other hand, 
high explosion energies in dwarf galaxies, where the gravitational potential is shallow, might 
cause the gas to leave the galaxy and the observed abundances of the next generation of stars 
do not include this high zinc signature as it does in the Milky Way. Therefore, the  
abundance ratios of [Ni/Fe] and [Zn/Fe] are strong evidence of the 
extragalactic origin of these stars.

\subsection{Neutron capture elements}
\label{neutrn_capt}
\begin{figure*}
\includegraphics[scale=0.4]{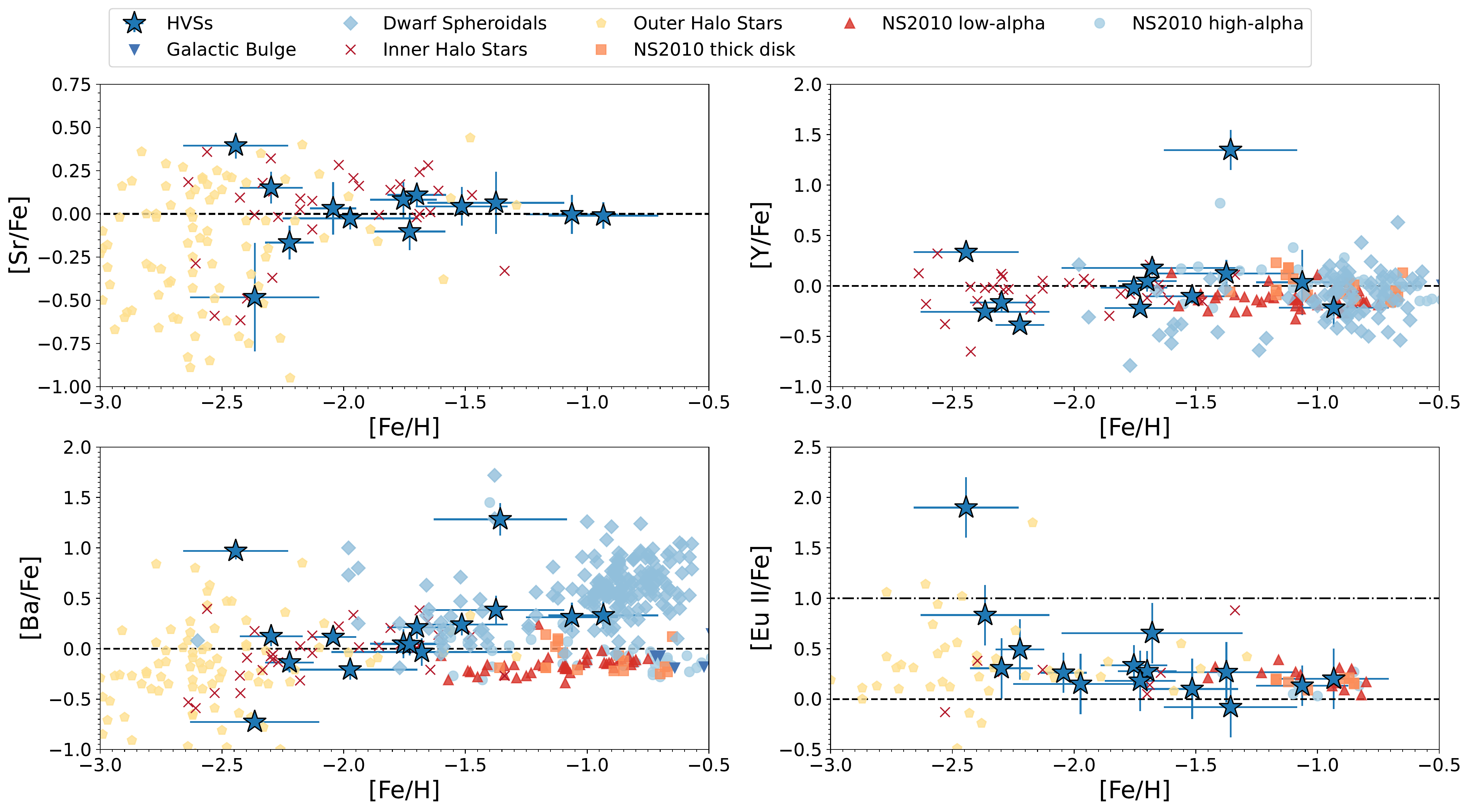}
\caption{Abundances of the neutron capture elements strontium, yttrium, 
and barium. We plot our HVSs as dark-blue stars. 
We plot in yellow the abundances of Milky Way metal-poor stars from
\cite{cayrel2004} and \cite{jacobson15} outer halo stars; in red we show abundances 
of inner halo stars from Reggiani et al. (in prep), \cite{reggiani2017} and \cite{nissen2010}; 
we display the \cite{bensby2010} bulge stars in orange; in light-blue 
we include stars from the dwarf spheroidal galaxies 
Carina, Sculptur, Fornax, and Sagittarius from \cite{shetrone2003}, 
\cite{geisler2005}, \cite{monaco2005}, and \cite{letarte2010}. \label{neutron_capture_fig}}
\end{figure*}

We also measured neutron-capture elements (Sr, Y, Ba, La, and Eu), and we show our 
abundances in Figure \ref{neutron_capture_fig}. We show the \ion{Sr}{2} abundances 
because they are not strongly affected by non-LTE effects \citep{hansen2013}. 
Yttrium abundances were estimated 
using hyperfine structure data from the 
Kurucz\footnote{http://kurucz.harvard.edu/linelists.html} linelists. Barium abundances 
were calculated using the isotopic splitting from \cite{mcwilliam1998} and \cite{klose2002}. 
Europium abundances were estimated using spectral synthesis from 
up to 5 lines ($4129$, $4205$, $4435$, $4522$ and $6645$ \AA).
%Once again, our results are within what is expected for thick disk/inner halo stars. 

Contrary to what we saw so far for the $\alpha$, light-odd, and iron-peak elements, the 
abundances of the neutron-capture elements Y and Ba are supersolar for the high 
metallicity stars in our sample. This is compatible with the observed patterns of extragalactic 
environments, but not with extreme r-process enhancement found, for example, in the 
ultra-faint dwarf galaxy Reticulum II \cite{ji2016}. The europium 
in our sample is not highly enhanced. The r-process distribution of our sample is 
similar to the Milky Way r-process abundance distribution, with the exception of 
HE 1523$-$0901 which was already known to have [Eu/Fe]$>1.7$ dex. Again, it indicates 
our stars are not from an extreme environment such as Reticulum II.

\section{Discussion}
\label{discussion}

\subsection{Stellar Origin}
We observed 15 extreme-velocity stars with either 
MIKE/Magellan or ARCES at the 3.5m Apache Point observatory. 
The chemical abundances ([$\alpha$, iron-peak, and neutron-capture/Fe] ratios) 
of our sample of extreme-velocity stars indicate that 
at least 50\% of our sample are of extragalactic origin (these 
stars are identified by a $c$ at the External ID field of Table \ref{stellar_params}). 

The stars that do have more similarities with Milky Way chemical abundances are at the 
low metallicity end of our sample. In particular the three most metal-poor stars we 
analyzed have chemical abundances more consistent with Milky Way halo stars, 
(high-$\alpha$ and slow neutron-capture 
scattered around $\sim0.0$). Their light odd-Z abundances, and in particular the iron-peak 
[Mn/Fe] and [Ni/Fe] abundances are Milky Way-like. There are stars at the low-metallicity end 
that have low [Ni/Fe] abundances but Milky Way-like abundances of other species, and the 
nickel abundance alone is not enough evidence of an 
extragalactic origin. As was concluded by \cite{hawkins2018} for their sample, 
these stars are also consistent with the tail of the velocity distribution of the Milky Way.

The chemical abundances of the high metallicity end of the sample 
([Fe/H]$\gtrapprox -1.5$) resemble the 
chemical patterns of stars formed in other galaxies, in particular the 
low [$\alpha$/Fe] ratios, and high [s-process/Fe] ratios. The knee of the Milky Way, 
the position in [Fe/H] space where there is an inflection in [$\alpha$/Fe] 
towards [$\alpha$/Fe]$\le0.0$, is known to be at [Fe/H]$=-1.0$ 
\citep[e.g.,][]{matteucci1990,suda2017}. Therefore, 
stars with near-solar or sub-solar $\alpha$ abundances below or close to that [Fe/H]$=-1.0$ 
are likely to have formed in another galaxy. 

These conclusions can be made exclusively based on the $\alpha$-elements because 
their abundances in our targets stars are extremely informative. Nevertheless, the 
abundances of Na, Al, Mn, Ni, Zn, Y, and 
Ba in our higher metallicity stars also show signatures that diverge from 
the behavior found in Milky Way stars.

In particular, their [Mn/Fe], [Ni/Fe], and [Zn/Fe] are very diagnostic. Both 
\cite{delosreyes20} and \cite{sanders21} analyzed extragalactic stars and studied 
the enrichment of iron-peak elements via sub-Chandrasekhar Type Ia supernovae. 
While \cite{delosreyes20} used 
medium resolution spectra to study stars in dwarf galaxies, \cite{sanders21} used 
high-resolution results from the literature and APOGEE to study the chemical history 
of the Gaia-Enceladus-Sausage. Both studies concluded that the dominant mechanism 
of iron-peak enrichment in dwarf galaxies is the sub-Chandrasekhar type Ia supernova. 
\cite{sanders21} also looked at [Ni/Fe] and [Zn/Fe] and found both are sub-solar and 
indicative of substansive 
sub-Chandrasekhar mass enrichment. We see in our stars the same qualitative behavior, 
meaning sub-solar abundances of these three key elements. However, our abundances, manganese 
in particular, are lower than those reported by the studies above and our sample 
is not monotonically selected to have been formed in the same dwarf galaxy/Galactic structure. 
Therefore it is difficult to make further conclusions regarding their type Ia supernova 
enrichment history. 
We can say though that near-Chandrasekhar or Chandrasekhar mass thermonuclear 
explosions would yield larger manganese, nickel, and zinc abundances 
\citep[see, e.g., ][]{reggiani2020}, and we are therefore 
seeing contributions from sub-Chandrasekhar mass type Ia supernovae.

\begin{figure}
\includegraphics[scale=0.55]{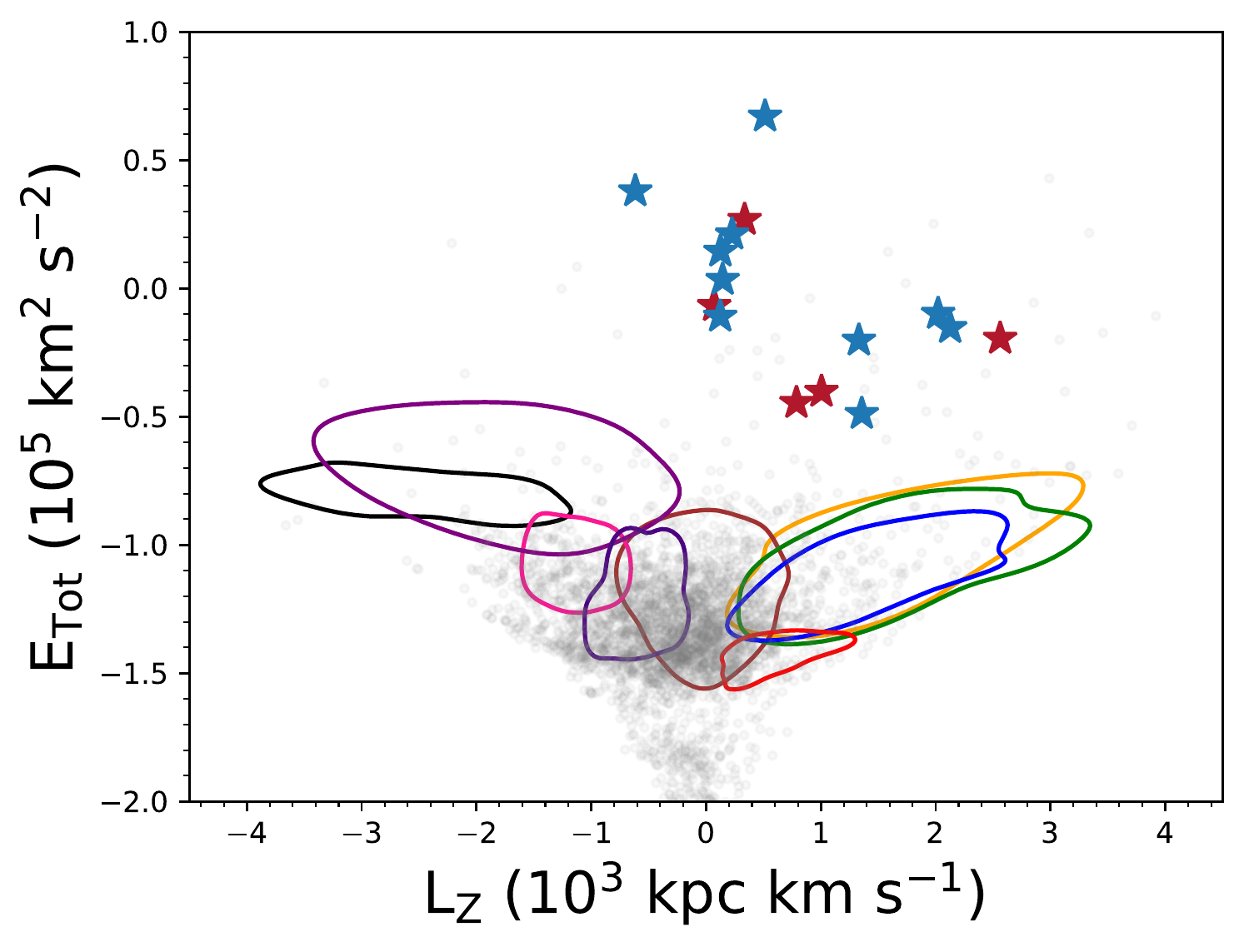}
\caption{Total energy as a function of angular momentum. We separate our stars in two groups. 
In blue are our stars with [Mg/Fe]<0.2 (more likely to be extragalactic) and in red 
stars with [Mg/Fe]>0.2. The background gray data are galactic halo stars 
from \cite[][private com]{naidu2021}. 
The colored ellipses are the substructures identified 
in \cite{naidu2020}. We include Arjuna, Cetus, GSE, Helmi, Iitoi, Sequoia, Sgr., Thamnos, and 
Wukong (orange, black, brown, deep pink, green, blue, purple, red and indigo, respectively). 
\label{kinematics_fig}}
\end{figure}

Further evidence of extragalactic origin for our HVSs sample come from their 
kinematics. 
We used the kinematic and astrometric data (Gaia EDR3 parallaxes and proper motions), distances 
estimated from our isochrones posteriors (compatible with the \cite{bailer-jones2021} 
prior-informed distances) and radial velocities from our spectra to calculate their orbital 
parameters and substructure membership probabilities. In Figure \ref{kinematics_fig} 
we show their total energy as a function of angular momentum. We also show in that figure 
the Galactic substructures as identified in \cite{naidu2020}. All our targets have quite different 
kinematic properties from the known substructures and the estimated membership probabilities are likely unreliable. 

This is not unexpected as they were selected to have higher velocities and 
therefore higher energies. It is surprising, though, to see at least two separate clusters of 
stars in the angular momentum plane. Assuming these stars did not originate in the 
Milky Way we can assume that the stars clustered around L$_{\rm{z}}<0.55 \times 10^3$ 
kpc km s$^{-1}$ 
were accreted to the galaxy during a radial accretion event. 
The clustering at higher L$_{\rm{z}}$ is 
more difficult to explain. This could hint to a previously unidentified substructure, a hypothesis 
that would be compatible with our chemical results, but we do not have enough data (stars) 
to conclude it, and our data is biased because we only targeted high velocity stars. 
We also do not see any sign of one clear substructure in our velocities distribution. 
We see stars that are both rotation in the same direction and in the opposite direction as 
the galactic rotation. This is not unexpected and can be attributed to the selection of our 
targets. We just selected for total velocity, and did not search for substructures in the data.

\begin{figure}
\includegraphics[scale=0.6]{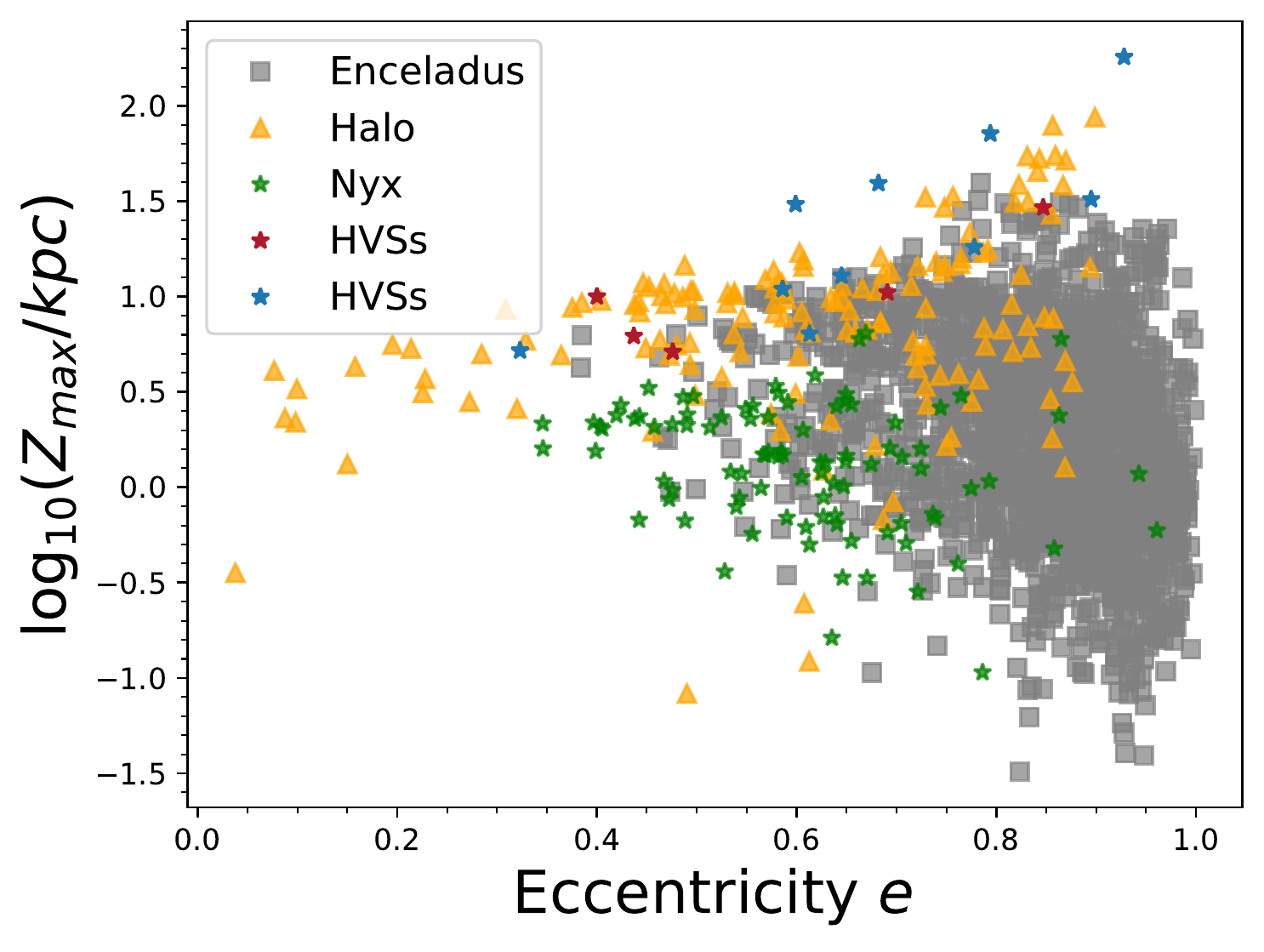}
\caption{Distance from the Galactic Plane as a function of eccentricity. Orange triangles 
are stars from the Galactic Halo, gray squares are Enceladus stars, and green stars are from 
the Nyx stream \citep{necib2020}. We separate our stars in two groups. In blue are our stars 
with [Mg/Fe]<0.2 (more likely to be extragalactic) and in red stars with [Mg/Fe]>0.2. Notice how 
all the stars with deviating eccentricity have low [Mg/Fe] abundance.
\label{evszmax}}
\end{figure}

Additionally we show in Figure \ref{evszmax} the maximum distance from the 
Galactic plane as a function of eccentricity for the stars in our sample. We compare our results 
to those of \cite{necib2020}, which include Galactic Halo stars, Enceladus Stars and 
stars from the Nyx stream. We can see that a few stars from our sample clearly deviate 
from our comparison data. The high eccentricities, along with the large total distances 
from the Galactic plane are yet another evidence of accretion. While a few stars are 
compatible with the halo sample, in particular those with smaller eccentricities, most 
stars in our sample could have been accreted (like Enceladus stars). There is a caveat 
that the eccentricities and large distances from the Galactic plane could have originated 
if a star is ``kicked'' through some mechanism, being accelerated by, for example, 
a supernova explosion. This is yet another evidence pointing these stars 
are likely accreted stars.

We also remark that our calculations did not include the Magellanic Clouds potential and 
we are unable to infer a probability of one of the stars having originated from the 
Large Magellanic Cloud (LMC). 
According to \cite{evans2021} the LMC can be an important source of hypervelocity 
stars, particularly for ``lower" velocity stars, and could contribute more than the 
Galactic center to the number of observed HVSs, but so far only one HVS star has 
been confirmed to have originated in the LMC \citep{erkal2019}. 
\cite{evans2021} also points that the high-velocity tail, above 500 
km s$^{-1}$, should be mostly composed of stars accelerated by the Milky Way Galactic center. 
As most of our stars have total velocities below $500$ km s$^{-1}$ it is possible 
that orbital calculations including the Magellanic Clouds potential could  
provide valuable additional information. 

At least eight, out 
of our 15 targets, have chemical abundances and/or orbital parameters 
that diverge from the typical 
Milky Way. We identify those stars in 
Table \ref{stellar_params} (identified by a $c$ at the External ID field). 
At this point it is not possible to say with certainty that all our targets 
are accreted stars. However, based on the fact that at least 50\% of our targets show evidence 
of being accreted we conclude that it is likely that the majority of the 
extreme-velocity stars are accreted.

\subsection{The Acceleration Mechanism}

The Hills mechanism \citep{hills1988,hills1991,hills1992} accelerates stars 
via three body interactions between two binary stars and a massive black hole. It is 
understood that this mechanism would be prevalent in Milky Way stars that were 
formed in the bulge and then accelerated after interacting with Sagittarius A* (Sag. A*). 
In this scenario the stars can be accelerated up to $\sim4000$ km s$^{-1}$. An example 
of such an ocurrence is S5-HVS 1, the $1700$ km s$^{-1}$ unbound star. Its orbit is indicative 
of interaction with Sag. A* \citep{koposov2020,irrgang2021}. 
Though the Hills mechanism does accelerate stars at low velocities as well, 
the stars in our sample are not compatible with the Hills mechanism. They are neither 
travelling at the unique high velocities only produced by this mechanism, 
nor have chemical abundances consistent with 
the bulge (Figures \ref{alphas_fig} to \ref{neutron_capture_fig}). Besides, the pericenters of 
the stars in our sample are too large to be consistent with ejection by the Hills mechanism 
(the pericenters of our targets vary from $\mathrm{R_{peri}}=4.98$ kpc to 
$\mathrm{R_{peri}}=8.55$ kpc). 

We also do not see chemical evidence that these stars were stripped from globular clusters. 
The stars do not show the anti-correlations observed for globular cluster stars (e.g. Mg-Al). 
This leaves us with two main possibilities: ejection from the LMC or tidal debris from 
disrupted dwarf galaxies. 
Even though we did not include the Magellanic Clouds potential in our orbital 
calculation, we see more than one possible cluster in orbital parameters among our stars, 
which is indicative of more than one origin. 
They do have high total energies, expected given their velocities, but their 
other orbital (and chemical) parameters vary considerably, thus indicating that 
they do not necessarily share the same origin.

Their chemistry does not point to the LMC either. As recently showed by \cite{reggiani2021} 
Magellanic Clouds stars are r-process enhanced, and our targets 
do not show any clear level of r-process enhancement. Though our most metal-rich 
star have $\alpha$-abundances that are consistent with the Magellanic Cloud, its 
apocenter is not large enough (31 kpc). It is also important to remark that most 
of our targets have metallicities lower than [Fe/H]$\le-1.5$, a metallicity region 
that has not been thoroughly studied for the Magellanic Clouds, making any chemical 
comparison difficult \citep[e.g.][]{nidever2020,reggiani2021}. 
Additional chemical abundances of Magellanic Clouds stars with [Fe/H]$\lesssim-1.5$ 
could provide further evidence, but with the data at hand it is not possible to confirm 
these stars are LMC runaways. 

An alternative mechanism for accelerating stars into extreme velocities has been proposed 
by \cite{abadi2009}. They argued that the known population of HVSs and their anisotropic 
distribution in the sky, and a preferred travel time, was inconsistent, at a population 
level, with the Hills mechanism scenario. They proposed that tidal disruptions of dwarf galaxies 
in the Galactic Potential could have accelerated at least a portion (the lower velocity portion) 
of the known HVSs. Using a set of cosmological simulations they showed that a few million 
years after the disruption of a dwarf galaxy a tidal tail is formed, one in which stars are 
accelerated into extreme velocities, and can even become unbound from the main galaxy. 
\cite{abadi2009} predicted that if this is a viable mechanism we would observe high-velocity stars 
preferably at low metallicities, including evolved stars. They also predicted we could 
find a radial velocity 
angular gradient among the high-velocity stars. Another prediction is that this 
mechanism does not accelerate stars into velocities as high as those expected by the 
Hills mechanism. Therefore, though the young A and O stars, and other stars with velocities 
$\sim1000$ km s$^{-1}$, are more likely to have been accelerated either by the Hills mechanism, or 
have been ejected from the LMC, the \cite{abadi2009} provides an explanation for extreme-velocity 
stars such as those analyzed in this work. 

As predicted by \cite{abadi2009} our extreme-velocity stars are evolved metal-poor stars 
that, although fast, do not exceed the Galactic Potential escape velocity. The scenario 
outlined in \cite{abadi2009} provide the most natural explanation for the conclusions 
about the origins of our targeted stars based on their chemical abundances, and different 
than the LMC scenario, it does not require a shared origin for most of our targets. 
It is also tempting to interpret the orbital clustering seen in Figure \ref{kinematics_fig} as  
another prediction from their paper, though it is not, at this time, a statistically 
significant result.

\section{Conclusion}\label{conclusion}
We obtained high-resolution spectra of 
15 low-mass hypervelocity stars candidates using either the MIKE 
spectrograph at the 6.5 meter Magellan telescope, or the ARCES spectrograph at the 
3.5 meter Apache point Observatory telescope. We confirmed the 
radial velocities from Gaia DR2 and did not find signs of RV variability in 
these stars. The total velocities calculated using Gaia EDR3 astrometry and 
the inferred radial velocities confirm previous Gaia DR2 findings that they are 
bound to the Galactic potential and are not as fast as expected for classical 
hypervelocity stars.

The chemical patterns of the metal-rich end of our 
sample stars are consistent with that of 
extragalactic stars, with sub-solar $\alpha$-abundances and 
supersolar neutron-capture abundances, 
and the iron-peak elements indicate 
sub-Chandrasekhar mass type Ia enrichment. 
Our orbital analysis is consistent with the chemical 
findings and our targets are not part of any Galactic substructure, 
although that is a biased result, as the very goal 
of the study is to study high velocity stars.

Our analysis of these extreme-velocity stars indicate that they are 
likely the tidal debris from disrupted dwarf 
galaxies accreted by the Milky Way, accelerated by the disruption of their 
host galaxy by the Galactic Potential \citep{abadi2009}.

With the increasing identification of HVSs candidates, follow-up studies with larger samples 
will help us continue to determine the acceleration mechanism(s) mainly responsible 
for the fast tail of the Milky Way stellar velocity distribution. In particular, the upcoming 
Gaia DR3 updated radial velocities, along with the precise astrometric solutions, for 
millions of stars will allow us to 
identify both these extreme-velocity stars, and bona-fide hypervelocity stars. With a 
large enough sample we will be able to confirm/falsefy the \cite{abadi2009} predictions based 
on orbital parameters and clustering information. A chemical analysis of additional 
candidates will provide further information about the stellar origins. 
Therefore, we strongly encourage a continued chemical characterization of these stars, in order 
to provide us with the information to characterize their acceleration mechanism(s) in a 
population level.

\section*{Acknowledgments}
%\begin{acknowledgments}
\noindent
We thank the referee for the careful analysis of our manuscript and the suggestions 
that helped us improve our study. We thank Ana Bonaca and Rohan Naidu for the valuable discussion on the paper and for kindly 
providing us with the orbital comparison data used in this study.
HR acknowledges support from a Carnegie Fellowship. KH \& TN acknowledge support 
from the National Science Foundation grant AST-1907417 and AST-2108736. 
KH is partially sup-ported through the Wootton Center for Astrophysical 
Plasma Properties funded under the United States Department of Energy 
collaborative agreement. JYG acknowledges support from CNPq (166042/2020-0). 
This work was partially performed at the Aspen Center for Physics, which is supported by National Science Foundation grant PHY-1607611.
This paper includes data gathered with the 6.5-meter Magellan Telescopes located
at Las Campanas Observatory, Chile. This paper is partially 
based on observations obtained with the Apache Point Observatory 
3.5-meter telescope, which is owned and operated by the Astrophysical Research Consortium. 
This work has made use of data from the European Space Agency (ESA) mission
{\it Gaia} (\url{https://www.cosmos.esa.int/gaia}), processed by the {\it Gaia}
Data Processing and Analysis Consortium (DPAC,
\url{https://www.cosmos.esa.int/web/gaia/dpac/consortium}). Funding for the DPAC
has been provided by national institutions, in particular the institutions
participating in the {\it Gaia} Multilateral Agreement. 
This publication makes use of data products from the Wide-field Infrared
Survey Explorer, which is a joint project of the University of California,
Los Angeles, and the Jet Propulsion Laboratory/California Institute
of Technology, and NEOWISE, which is a project of the Jet Propulsion
Laboratory/California Institute of Technology.  WISE and NEOWISE
are funded by the National Aeronautics and Space Administration.
This publication makes use of data products from the Two Micron All
Sky Survey, which is a joint project of the University of Massachusetts
and the Infrared Processing and Analysis Center/California Institute of
Technology, funded by the National Aeronautics and Space Administration
and the National Science Foundation. This research has made use of the SkyMapper survey data release 2. 
The national facility capability for SkyMapper has been funded through ARC 
LIEF grant LE130100104 from the Australian Research Council, awarded to the 
University of Sydney, the Australian National University, 
Swinburne University of Technology, the University of Queensland, the University 
of Western Australia, the University of Melbourne, Curtin University of 
Technology, Monash University and the Australian Astronomical Observatory. 
SkyMapper is owned and operated by The Australian National University's 
Research School of Astronomy and Astrophysics. The survey data were processed and provided by the SkyMapper Team at ANU. The SkyMapper node of the All-Sky Virtual 
Observatory (ASVO) is hosted at the National Computational 
Infrastructure (NCI). Development and support of the SkyMapper node of 
the ASVO has been funded in part by Astronomy Australia Limited (AAL) and the 
Australian Government through the Commonwealth's Education Investment Fund 
(EIF) and National Collaborative Research Infrastructure Strategy (NCRIS), 
particularly the National eResearch Collaboration Tools and Resources 
(NeCTAR) and the Australian National Data Service Projects (ANDS).
This research has made use of  NASA/IPAC Infrared Science Archive, 
which is funded by the National
Aeronautics and Space Administration and operated by the California
Institute of Technology \citep{ipac_2mass,ipac_allwise}.

\software{\texttt{astropy} \citep{astropy2013,astropy2018},
          \texttt{CarPy} \citep{kelson2000,kelson2003},
          \texttt{galpy} \citep{bovy2015},
          \texttt{q2} \citep{ramirez2014},
          \texttt{isochrones} \citep{morton2015},
          \texttt{numpy} \citep{harris2020},
          \texttt{MultiNest} \citep{feroz2008,feroz2009,feroz2019},
          \texttt{pandas} \citep{mckinney2010,reback2020},
          \texttt{scipy} \citep{virtanen2020}, 
          \texttt{IRAF} \citep{iraf1986,iraf1993}, 
          \texttt{matplotlib} \citep{hunter2007}
          }

\facilities{ARC (ARCES), Magellan:Clay (MIKE),	CDS, Gaia, IRSA, Skymapper, CTIO:2MASS, WISE}

%\end{acknowledgments}

\bibliographystyle{aasjournal}
\bibliography{hvs}{}

\begin{thebibliography}{}
\expandafter\ifx\csname natexlab\endcsname\relax\def\natexlab#1{#1}\fi
\providecommand{\url}[1]{\href{#1}{#1}}
\providecommand{\dodoi}[1]{doi:~\href{http://doi.org/#1}{\nolinkurl{#1}}}
\providecommand{\doeprint}[1]{\href{http://ascl.net/#1}{\nolinkurl{http://ascl.net/#1}}}
\providecommand{\doarXiv}[1]{\href{https://arxiv.org/abs/#1}{\nolinkurl{https://arxiv.org/abs/#1}}}

\bibitem[{{Abadi} {et~al.}(2009){Abadi}, {Navarro}, \& {Steinmetz}}]{abadi2009}
{Abadi}, M.~G., {Navarro}, J.~F., \& {Steinmetz}, M. 2009, \apjl, 691, L63,
  \dodoi{10.1088/0004-637X/691/2/L63}

\bibitem[{{Alonso} {et~al.}(1999){Alonso}, {Arribas}, \&
  {Mart{\'\i}nez-Roger}}]{alonso99}
{Alonso}, A., {Arribas}, S., \& {Mart{\'\i}nez-Roger}, C. 1999, \aaps, 140,
  261, \dodoi{10.1051/aas:1999521}

\bibitem[{{Amarsi} {et~al.}(2020){Amarsi}, {Lind}, {Osorio}, {Nordlander},
  {Bergemann}, {Reggiani}, {Wang}, {Buder}, {Asplund}, {Barklem}, {Wehrhahn},
  {Sk{\'u}lad{\'o}ttir}, {Kobayashi}, {Karakas}, {Gao}, {Bland-Hawthorn}, {de
  Silva}, {Kos}, {Lewis}, {Martell}, {Sharma}, {Simpson}, {Zucker},
  {{\v{C}}otar}, {Horner}, \& {Galah Collaboration}}]{amarsi2020}
{Amarsi}, A.~M., {Lind}, K., {Osorio}, Y., {et~al.} 2020, \aap, 642, A62,
  \dodoi{10.1051/0004-6361/202038650}

\bibitem[{{Astropy Collaboration} {et~al.}(2013){Astropy Collaboration},
  {Robitaille}, {Tollerud}, {Greenfield}, {Droettboom}, {Bray}, {Aldcroft},
  {Davis}, {Ginsburg}, {Price-Whelan}, {Kerzendorf}, {Conley}, {Crighton},
  {Barbary}, {Muna}, {Ferguson}, {Grollier}, {Parikh}, {Nair}, {Unther},
  {Deil}, {Woillez}, {Conseil}, {Kramer}, {Turner}, {Singer}, {Fox}, {Weaver},
  {Zabalza}, {Edwards}, {Azalee Bostroem}, {Burke}, {Casey}, {Crawford},
  {Dencheva}, {Ely}, {Jenness}, {Labrie}, {Lim}, {Pierfederici}, {Pontzen},
  {Ptak}, {Refsdal}, {Servillat}, \& {Streicher}}]{astropy2013}
{Astropy Collaboration}, {Robitaille}, T.~P., {Tollerud}, E.~J., {et~al.} 2013,
  \aap, 558, A33, \dodoi{10.1051/0004-6361/201322068}

\bibitem[{{Astropy Collaboration} {et~al.}(2018){Astropy Collaboration},
  {Price-Whelan}, {Sip{\H{o}}cz}, {G{\"u}nther}, {Lim}, {Crawford}, {Conseil},
  {Shupe}, {Craig}, {Dencheva}, {Ginsburg}, {Vand erPlas}, {Bradley},
  {P{\'e}rez-Su{\'a}rez}, {de Val-Borro}, {Aldcroft}, {Cruz}, {Robitaille},
  {Tollerud}, {Ardelean}, {Babej}, {Bach}, {Bachetti}, {Bakanov}, {Bamford},
  {Barentsen}, {Barmby}, {Baumbach}, {Berry}, {Biscani}, {Boquien}, {Bostroem},
  {Bouma}, {Brammer}, {Bray}, {Breytenbach}, {Buddelmeijer}, {Burke},
  {Calderone}, {Cano Rodr{\'\i}guez}, {Cara}, {Cardoso}, {Cheedella}, {Copin},
  {Corrales}, {Crichton}, {D'Avella}, {Deil}, {Depagne}, {Dietrich}, {Donath},
  {Droettboom}, {Earl}, {Erben}, {Fabbro}, {Ferreira}, {Finethy}, {Fox},
  {Garrison}, {Gibbons}, {Goldstein}, {Gommers}, {Greco}, {Greenfield},
  {Groener}, {Grollier}, {Hagen}, {Hirst}, {Homeier}, {Horton}, {Hosseinzadeh},
  {Hu}, {Hunkeler}, {Ivezi{\'c}}, {Jain}, {Jenness}, {Kanarek}, {Kendrew},
  {Kern}, {Kerzendorf}, {Khvalko}, {King}, {Kirkby}, {Kulkarni}, {Kumar},
  {Lee}, {Lenz}, {Littlefair}, {Ma}, {Macleod}, {Mastropietro}, {McCully},
  {Montagnac}, {Morris}, {Mueller}, {Mumford}, {Muna}, {Murphy}, {Nelson},
  {Nguyen}, {Ninan}, {N{\"o}the}, {Ogaz}, {Oh}, {Parejko}, {Parley}, {Pascual},
  {Patil}, {Patil}, {Plunkett}, {Prochaska}, {Rastogi}, {Reddy Janga},
  {Sabater}, {Sakurikar}, {Seifert}, {Sherbert}, {Sherwood-Taylor}, {Shih},
  {Sick}, {Silbiger}, {Singanamalla}, {Singer}, {Sladen}, {Sooley},
  {Sornarajah}, {Streicher}, {Teuben}, {Thomas}, {Tremblay}, {Turner},
  {Terr{\'o}n}, {van Kerkwijk}, {de la Vega}, {Watkins}, {Weaver}, {Whitmore},
  {Woillez}, {Zabalza}, \& {Astropy Contributors}}]{astropy2018}
{Astropy Collaboration}, {Price-Whelan}, A.~M., {Sip{\H{o}}cz}, B.~M., {et~al.}
  2018, \aj, 156, 123, \dodoi{10.3847/1538-3881/aabc4f}

\bibitem[{{Bailer-Jones} {et~al.}(2021){Bailer-Jones}, {Rybizki}, {Fouesneau},
  {Demleitner}, \& {Andrae}}]{bailer-jones2021}
{Bailer-Jones}, C.~A.~L., {Rybizki}, J., {Fouesneau}, M., {Demleitner}, M., \&
  {Andrae}, R. 2021, \aj, 161, 147, \dodoi{10.3847/1538-3881/abd806}

\bibitem[{{Bensby} {et~al.}(2010){Bensby}, {Feltzing}, {Johnson}, {Gould},
  {Ad{\'e}n}, {Asplund}, {Mel{\'e}ndez}, {Gal-Yam}, {Lucatello}, {Sana},
  {Sumi}, {Miyake}, {Suzuki}, {Han}, {Bond}, \& {Udalski}}]{bensby2010}
{Bensby}, T., {Feltzing}, S., {Johnson}, J.~A., {et~al.} 2010, \aap, 512, A41,
  \dodoi{10.1051/0004-6361/200913744}

\bibitem[{{Bernstein} {et~al.}(2003){Bernstein}, {Shectman}, {Gunnels},
  {Mochnacki}, \& {Athey}}]{bernstein2003}
{Bernstein}, R., {Shectman}, S.~A., {Gunnels}, S.~M., {Mochnacki}, S., \&
  {Athey}, A.~E. 2003, in Society of Photo-Optical Instrumentation Engineers
  (SPIE) Conference Series, Vol. 4841, \procspie, ed. M.~{Iye} \& A.~F.~M.
  {Moorwood}, 1694--1704, \dodoi{10.1117/12.461502}

\bibitem[{{Blaauw}(1961)}]{blaauw1961}
{Blaauw}, A. 1961, \bain, 15, 265

\bibitem[{{Blanco-Cuaresma}(2019)}]{blanco-cuaresma2019}
{Blanco-Cuaresma}, S. 2019, \mnras, 486, 2075, \dodoi{10.1093/mnras/stz549}

\bibitem[{{Blanco-Cuaresma} {et~al.}(2014){Blanco-Cuaresma}, {Soubiran},
  {Heiter}, \& {Jofr{\'e}}}]{blanco-cuaresma2014}
{Blanco-Cuaresma}, S., {Soubiran}, C., {Heiter}, U., \& {Jofr{\'e}}, P. 2014,
  \aap, 569, A111, \dodoi{10.1051/0004-6361/201423945}

\bibitem[{{Bland-Hawthorn} \& {Gerhard}(2016)}]{blandhawthorn2016}
{Bland-Hawthorn}, J., \& {Gerhard}, O. 2016, \araa, 54, 529,
  \dodoi{10.1146/annurev-astro-081915-023441}

\bibitem[{{Boubert} \& {Evans}(2016)}]{boubert2016}
{Boubert}, D., \& {Evans}, N.~W. 2016, \apjl, 825, L6,
  \dodoi{10.3847/2041-8205/825/1/L6}

\bibitem[{{Boubert} {et~al.}(2018){Boubert}, {Guillochon}, {Hawkins},
  {Ginsburg}, {Evans}, \& {Strader}}]{boubert2018}
{Boubert}, D., {Guillochon}, J., {Hawkins}, K., {et~al.} 2018, \mnras, 479,
  2789, \dodoi{10.1093/mnras/sty1601}

\bibitem[{{Bovy}(2015)}]{bovy2015}
{Bovy}, J. 2015, \apjs, 216, 29, \dodoi{10.1088/0067-0049/216/2/29}

\bibitem[{{Brahm} {et~al.}(2017){Brahm}, {Jord{\'a}n}, \&
  {Espinoza}}]{brahm2017}
{Brahm}, R., {Jord{\'a}n}, A., \& {Espinoza}, N. 2017, \pasp, 129, 034002,
  \dodoi{10.1088/1538-3873/aa5455}

\bibitem[{{Bromley} {et~al.}(2009){Bromley}, {Kenyon}, {Brown}, \&
  {Geller}}]{bromley2009}
{Bromley}, B.~C., {Kenyon}, S.~J., {Brown}, W.~R., \& {Geller}, M.~J. 2009,
  \apj, 706, 925, \dodoi{10.1088/0004-637X/706/2/925}

\bibitem[{{Bromley} {et~al.}(2018){Bromley}, {Kenyon}, {Brown}, \&
  {Geller}}]{bromley2018}
---. 2018, \apj, 868, 25, \dodoi{10.3847/1538-4357/aae83e}

\bibitem[{{Brown}(2015)}]{brown2015}
{Brown}, W.~R. 2015, \araa, 53, 15, \dodoi{10.1146/annurev-astro-082214-122230}

\bibitem[{{Brown} {et~al.}(2005){Brown}, {Geller}, {Kenyon}, \&
  {Kurtz}}]{brown2005}
{Brown}, W.~R., {Geller}, M.~J., {Kenyon}, S.~J., \& {Kurtz}, M.~J. 2005,
  \apjl, 622, L33, \dodoi{10.1086/429378}

\bibitem[{{Buchner} {et~al.}(2014){Buchner}, {Georgakakis}, {Nandra}, {Hsu},
  {Rangel}, {Brightman}, {Merloni}, {Salvato}, {Donley}, \&
  {Kocevski}}]{buchner2014}
{Buchner}, J., {Georgakakis}, A., {Nandra}, K., {et~al.} 2014, \aap, 564, A125,
  \dodoi{10.1051/0004-6361/201322971}

\bibitem[{{Casey} {et~al.}(2017){Casey}, {Hawkins}, {Hogg}, {Ness}, {Rix},
  {Kordopatis}, {Kunder}, {Steinmetz}, {Koposov}, {Enke}, {Sanders}, {Gilmore},
  {Zwitter}, {Freeman}, {Casagrande}, {Matijevi{\v{c}}}, {Seabroke},
  {Bienaym{\'e}}, {Bland-Hawthorn}, {Gibson}, {Grebel}, {Helmi}, {Munari},
  {Navarro}, {Reid}, {Siebert}, \& {Wyse}}]{casey17}
{Casey}, A.~R., {Hawkins}, K., {Hogg}, D.~W., {et~al.} 2017, \apj, 840, 59,
  \dodoi{10.3847/1538-4357/aa69c2}

\bibitem[{{Castelli} \& {Kurucz}(2004)}]{castelli2004}
{Castelli}, F., \& {Kurucz}, R.~L. 2004, arXiv Astrophysics e-prints

\bibitem[{{Cayrel} {et~al.}(2004){Cayrel}, {Depagne}, {Spite}, {Hill}, {Spite},
  {Fran{\c{c}}ois}, {Plez}, {Beers}, {Primas}, {Andersen}, {Barbuy},
  {Bonifacio}, {Molaro}, \& {Nordstr{\"o}m}}]{cayrel2004}
{Cayrel}, R., {Depagne}, E., {Spite}, M., {et~al.} 2004, \aap, 416, 1117,
  \dodoi{10.1051/0004-6361:20034074}

\bibitem[{{Choi} {et~al.}(2016){Choi}, {Dotter}, {Conroy}, {Cantiello},
  {Paxton}, \& {Johnson}}]{choi2016}
{Choi}, J., {Dotter}, A., {Conroy}, C., {et~al.} 2016, \apj, 823, 102,
  \dodoi{10.3847/0004-637X/823/2/102}

\bibitem[{{Clayton}(2007)}]{clayton2007}
{Clayton}, D. 2007, {Handbook of Isotopes in the Cosmos}

\bibitem[{{de los Reyes} {et~al.}(2020){de los Reyes}, {Kirby}, {Seitenzahl},
  \& {Shen}}]{delosreyes20}
{de los Reyes}, M. A.~C., {Kirby}, E.~N., {Seitenzahl}, I.~R., \& {Shen}, K.~J.
  2020, \apj, 891, 85, \dodoi{10.3847/1538-4357/ab736f}

\bibitem[{{Dotter}(2016)}]{dotter2016}
{Dotter}, A. 2016, \apjs, 222, 8, \dodoi{10.3847/0067-0049/222/1/8}

\bibitem[{{Erkal} {et~al.}(2019){Erkal}, {Boubert}, {Gualandris}, {Evans}, \&
  {Antonini}}]{erkal2019}
{Erkal}, D., {Boubert}, D., {Gualandris}, A., {Evans}, N.~W., \& {Antonini}, F.
  2019, \mnras, 483, 2007, \dodoi{10.1093/mnras/sty2674}

\bibitem[{{Evans} {et~al.}(2018){Evans}, {Riello}, {De Angeli}, {Carrasco},
  {Montegriffo}, {Fabricius}, {Jordi}, {Palaversa}, {Diener}, {Busso},
  {Cacciari}, {van Leeuwen}, {Burgess}, {Davidson}, {Harrison}, {Hodgkin},
  {Pancino}, {Richards}, {Altavilla}, {Balaguer-N{\'u}{\~n}ez}, {Barstow},
  {Bellazzini}, {Brown}, {Castellani}, {Cocozza}, {De Luise}, {Delgado},
  {Ducourant}, {Galleti}, {Gilmore}, {Giuffrida}, {Holl}, {Kewley}, {Koposov},
  {Marinoni}, {Marrese}, {Osborne}, {Piersimoni}, {Portell}, {Pulone},
  {Ragaini}, {Sanna}, {Terrett}, {Walton}, {Wevers}, \&
  {Wyrzykowski}}]{evans2018}
{Evans}, D.~W., {Riello}, M., {De Angeli}, F., {et~al.} 2018, \aap, 616, A4,
  \dodoi{10.1051/0004-6361/201832756}

\bibitem[{{Evans} {et~al.}(2021){Evans}, {Marchetti}, {Rossi}, {Baggen}, \&
  {Bloot}}]{evans2021}
{Evans}, F.~A., {Marchetti}, T., {Rossi}, E.~M., {Baggen}, J. F.~W., \&
  {Bloot}, S. 2021, arXiv e-prints, arXiv:2108.01100.
\newblock \doarXiv{2108.01100}

\bibitem[{{Feroz} \& {Hobson}(2008)}]{feroz2008}
{Feroz}, F., \& {Hobson}, M.~P. 2008, \mnras, 384, 449,
  \dodoi{10.1111/j.1365-2966.2007.12353.x}

\bibitem[{{Feroz} {et~al.}(2009){Feroz}, {Hobson}, \& {Bridges}}]{feroz2009}
{Feroz}, F., {Hobson}, M.~P., \& {Bridges}, M. 2009, \mnras, 398, 1601,
  \dodoi{10.1111/j.1365-2966.2009.14548.x}

\bibitem[{{Feroz} {et~al.}(2019){Feroz}, {Hobson}, {Cameron}, \&
  {Pettitt}}]{feroz2019}
{Feroz}, F., {Hobson}, M.~P., {Cameron}, E., \& {Pettitt}, A.~N. 2019, The Open
  Journal of Astrophysics, 2, 10, \dodoi{10.21105/astro.1306.2144}

\bibitem[{{Frebel} {et~al.}(2007){Frebel}, {Christlieb}, {Norris}, {Thom},
  {Beers}, \& {Rhee}}]{frebel07}
{Frebel}, A., {Christlieb}, N., {Norris}, J.~E., {et~al.} 2007, \apjl, 660,
  L117, \dodoi{10.1086/518122}

\bibitem[{{Gaia Collaboration} {et~al.}(2020){Gaia Collaboration}, {Brown},
  {Vallenari}, {Prusti}, {de Bruijne}, {Babusiaux}, \& {Biermann}}]{gaia2020}
{Gaia Collaboration}, {Brown}, A.~G.~A., {Vallenari}, A., {et~al.} 2020, arXiv
  e-prints, arXiv:2012.01533.
\newblock \doarXiv{2012.01533}

\bibitem[{{Gaia Collaboration} {et~al.}(2016){Gaia Collaboration}, {Prusti},
  {de Bruijne}, {Brown}, {Vallenari}, {Babusiaux}, {Bailer-Jones}, {Bastian},
  {Biermann}, {Evans}, {Eyer}, {Jansen}, {Jordi}, {Klioner}, {Lammers},
  {Lindegren}, {Luri}, {Mignard}, {Milligan}, {Panem}, {Poinsignon},
  {Pourbaix}, {Randich}, {Sarri}, {Sartoretti}, {Siddiqui}, {Soubiran},
  {Valette}, {van Leeuwen}, {Walton}, {Aerts}, {Arenou}, {Cropper}, {Drimmel},
  {H{\o}g}, {Katz}, {Lattanzi}, {O'Mullane}, {Grebel}, {Holland}, {Huc},
  {Passot}, {Bramante}, {Cacciari}, {Casta{\~n}eda}, {Chaoul}, {Cheek}, {De
  Angeli}, {Fabricius}, {Guerra}, {Hern{\'a}ndez}, {Jean-Antoine-Piccolo},
  {Masana}, {Messineo}, {Mowlavi}, {Nienartowicz}, {Ord{\'o}{\~n}ez-Blanco},
  {Panuzzo}, {Portell}, {Richards}, {Riello}, {Seabroke}, {Tanga},
  {Th{\'e}venin}, {Torra}, {Els}, {Gracia-Abril}, {Comoretto},
  {Garcia-Reinaldos}, {Lock}, {Mercier}, {Altmann}, {Andrae}, {Astraatmadja},
  {Bellas-Velidis}, {Benson}, {Berthier}, {Blomme}, {Busso}, {Carry},
  {Cellino}, {Clementini}, {Cowell}, {Creevey}, {Cuypers}, {Davidson}, {De
  Ridder}, {de Torres}, {Delchambre}, {Dell'Oro}, {Ducourant}, {Fr{\'e}mat},
  {Garc{\'\i}a-Torres}, {Gosset}, {Halbwachs}, {Hambly}, {Harrison}, {Hauser},
  {Hestroffer}, {Hodgkin}, {Huckle}, {Hutton}, {Jasniewicz}, {Jordan},
  {Kontizas}, {Korn}, {Lanzafame}, {Manteiga}, {Moitinho}, {Muinonen},
  {Osinde}, {Pancino}, {Pauwels}, {Petit}, {Recio-Blanco}, {Robin}, {Sarro},
  {Siopis}, {Smith}, {Smith}, {Sozzetti}, {Thuillot}, {van Reeven}, {Viala},
  {Abbas}, {Abreu Aramburu}, {Accart}, {Aguado}, {Allan}, {Allasia},
  {Altavilla}, {{\'A}lvarez}, {Alves}, {Anderson}, {Andrei}, {Anglada Varela},
  {Antiche}, {Antoja}, {Ant{\'o}n}, {Arcay}, {Atzei}, {Ayache}, {Bach},
  {Baker}, {Balaguer-N{\'u}{\~n}ez}, {Barache}, {Barata}, {Barbier}, {Barblan},
  {Baroni}, {Barrado y Navascu{\'e}s}, {Barros}, {Barstow}, {Becciani},
  {Bellazzini}, {Bellei}, {Bello Garc{\'\i}a}, {Belokurov}, {Bendjoya},
  {Berihuete}, {Bianchi}, {Bienaym{\'e}}, {Billebaud}, {Blagorodnova},
  {Blanco-Cuaresma}, {Boch}, {Bombrun}, {Borrachero}, {Bouquillon}, {Bourda},
  {Bouy}, {Bragaglia}, {Breddels}, {Brouillet}, {Br{\"u}semeister},
  {Bucciarelli}, {Budnik}, {Burgess}, {Burgon}, {Burlacu}, {Busonero}, {Buzzi},
  {Caffau}, {Cambras}, {Campbell}, {Cancelliere}, {Cantat-Gaudin}, {Carlucci},
  {Carrasco}, {Castellani}, {Charlot}, {Charnas}, {Charvet}, {Chassat},
  {Chiavassa}, {Clotet}, {Cocozza}, {Collins}, {Collins}, {Costigan}, {Crifo},
  {Cross}, {Crosta}, {Crowley}, {Dafonte}, {Damerdji}, {Dapergolas}, {David},
  {David}, {De Cat}, {de Felice}, {de Laverny}, {De Luise}, {De March}, {de
  Martino}, {de Souza}, {Debosscher}, {del Pozo}, {Delbo}, {Delgado},
  {Delgado}, {di Marco}, {Di Matteo}, {Diakite}, {Distefano}, {Dolding}, {Dos
  Anjos}, {Drazinos}, {Dur{\'a}n}, {Dzigan}, {Ecale}, {Edvardsson}, {Enke},
  {Erdmann}, {Escolar}, {Espina}, {Evans}, {Eynard Bontemps}, {Fabre},
  {Fabrizio}, {Faigler}, {Falc{\~a}o}, {Farr{\`a}s Casas}, {Faye}, {Federici},
  {Fedorets}, {Fern{\'a}ndez-Hern{\'a}ndez}, {Fernique}, {Fienga}, {Figueras},
  {Filippi}, {Findeisen}, {Fonti}, {Fouesneau}, {Fraile}, {Fraser}, {Fuchs},
  {Furnell}, {Gai}, {Galleti}, {Galluccio}, {Garabato}, {Garc{\'\i}a-Sedano},
  {Gar{\'e}}, {Garofalo}, {Garralda}, {Gavras}, {Gerssen}, {Geyer}, {Gilmore},
  {Girona}, {Giuffrida}, {Gomes}, {Gonz{\'a}lez-Marcos},
  {Gonz{\'a}lez-N{\'u}{\~n}ez}, {Gonz{\'a}lez-Vidal}, {Granvik}, {Guerrier},
  {Guillout}, {Guiraud}, {G{\'u}rpide}, {Guti{\'e}rrez-S{\'a}nchez}, {Guy},
  {Haigron}, {Hatzidimitriou}, {Haywood}, {Heiter}, {Helmi}, {Hobbs},
  {Hofmann}, {Holl}, {Holland}, {Hunt}, {Hypki}, {Icardi}, {Irwin}, {Jevardat
  de Fombelle}, {Jofr{\'e}}, {Jonker}, {Jorissen}, {Julbe}, {Karampelas},
  {Kochoska}, {Kohley}, {Kolenberg}, {Kontizas}, {Koposov}, {Kordopatis},
  {Koubsky}, {Kowalczyk}, {Krone-Martins}, {Kudryashova}, {Kull}, {Bachchan},
  {Lacoste-Seris}, {Lanza}, {Lavigne}, {Le Poncin-Lafitte}, {Lebreton},
  {Lebzelter}, {Leccia}, {Leclerc}, {Lecoeur-Taibi}, {Lemaitre}, {Lenhardt},
  {Leroux}, {Liao}, {Licata}, {Lindstr{\o}m}, {Lister}, {Livanou}, {Lobel},
  {L{\"o}ffler}, {L{\'o}pez}, {Lopez-Lozano}, {Lorenz}, {Loureiro},
  {MacDonald}, {Magalh{\~a}es Fernandes}, {Managau}, {Mann}, {Mantelet},
  {Marchal}, {Marchant}, {Marconi}, {Marie}, {Marinoni}, {Marrese},
  {Marschalk{\'o}}, {Marshall}, {Mart{\'\i}n-Fleitas}, {Martino}, {Mary},
  {Matijevi{\v{c}}}, {Mazeh}, {McMillan}, {Messina}, {Mestre}, {Michalik},
  {Millar}, {Miranda}, {Molina}, {Molinaro}, {Molinaro}, {Moln{\'a}r},
  {Moniez}, {Montegriffo}, {Monteiro}, {Mor}, {Mora}, {Morbidelli}, {Morel},
  {Morgenthaler}, {Morley}, {Morris}, {Mulone}, {Muraveva}, {Musella},
  {Narbonne}, {Nelemans}, {Nicastro}, {Noval}, {Ord{\'e}novic},
  {Ordieres-Mer{\'e}}, {Osborne}, {Pagani}, {Pagano}, {Pailler}, {Palacin},
  {Palaversa}, {Parsons}, {Paulsen}, {Pecoraro}, {Pedrosa}, {Pentik{\"a}inen},
  {Pereira}, {Pichon}, {Piersimoni}, {Pineau}, {Plachy}, {Plum}, {Poujoulet},
  {Pr{\v{s}}a}, {Pulone}, {Ragaini}, {Rago}, {Rambaux}, {Ramos-Lerate},
  {Ranalli}, {Rauw}, {Read}, {Regibo}, {Renk}, {Reyl{\'e}}, {Ribeiro},
  {Rimoldini}, {Ripepi}, {Riva}, {Rixon}, {Roelens}, {Romero-G{\'o}mez},
  {Rowell}, {Royer}, {Rudolph}, {Ruiz-Dern}, {Sadowski}, {Sagrist{\`a}
  Sell{\'e}s}, {Sahlmann}, {Salgado}, {Salguero}, {Sarasso}, {Savietto},
  {Schnorhk}, {Schultheis}, {Sciacca}, {Segol}, {Segovia}, {Segransan},
  {Serpell}, {Shih}, {Smareglia}, {Smart}, {Smith}, {Solano}, {Solitro},
  {Sordo}, {Soria Nieto}, {Souchay}, {Spagna}, {Spoto}, {Stampa}, {Steele},
  {Steidelm{\"u}ller}, {Stephenson}, {Stoev}, {Suess}, {S{\"u}veges}, {Surdej},
  {Szabados}, {Szegedi-Elek}, {Tapiador}, {Taris}, {Tauran}, {Taylor},
  {Teixeira}, {Terrett}, {Tingley}, {Trager}, {Turon}, {Ulla}, {Utrilla},
  {Valentini}, {van Elteren}, {Van Hemelryck}, {van Leeuwen}, {Varadi},
  {Vecchiato}, {Veljanoski}, {Via}, {Vicente}, {Vogt}, {Voss}, {Votruba},
  {Voutsinas}, {Walmsley}, {Weiler}, {Weingrill}, {Werner}, {Wevers},
  {Whitehead}, {Wyrzykowski}, {Yoldas}, {{\v{Z}}erjal}, {Zucker}, {Zurbach},
  {Zwitter}, {Alecu}, {Allen}, {Allende Prieto}, {Amorim},
  {Anglada-Escud{\'e}}, {Arsenijevic}, {Azaz}, {Balm}, {Beck}, {Bernstein},
  {Bigot}, {Bijaoui}, {Blasco}, {Bonfigli}, {Bono}, {Boudreault}, {Bressan},
  {Brown}, {Brunet}, {Bunclark}, {Buonanno}, {Butkevich}, {Carret}, {Carrion},
  {Chemin}, {Ch{\'e}reau}, {Corcione}, {Darmigny}, {de Boer}, {de Teodoro}, {de
  Zeeuw}, {Delle Luche}, {Domingues}, {Dubath}, {Fodor}, {Fr{\'e}zouls},
  {Fries}, {Fustes}, {Fyfe}, {Gallardo}, {Gallegos}, {Gardiol}, {Gebran},
  {Gomboc}, {G{\'o}mez}, {Grux}, {Gueguen}, {Heyrovsky}, {Hoar}, {Iannicola},
  {Isasi Parache}, {Janotto}, {Joliet}, {Jonckheere}, {Keil}, {Kim},
  {Klagyivik}, {Klar}, {Knude}, {Kochukhov}, {Kolka}, {Kos}, {Kutka}, {Lainey},
  {LeBouquin}, {Liu}, {Loreggia}, {Makarov}, {Marseille}, {Martayan},
  {Martinez-Rubi}, {Massart}, {Meynadier}, {Mignot}, {Munari}, {Nguyen},
  {Nordlander}, {Ocvirk}, {O'Flaherty}, {Olias Sanz}, {Ortiz}, {Osorio},
  {Oszkiewicz}, {Ouzounis}, {Palmer}, {Park}, {Pasquato}, {Peltzer}, {Peralta},
  {P{\'e}turaud}, {Pieniluoma}, {Pigozzi}, {Poels}, {Prat}, {Prod'homme},
  {Raison}, {Rebordao}, {Risquez}, {Rocca-Volmerange}, {Rosen}, {Ruiz-Fuertes},
  {Russo}, {Sembay}, {Serraller Vizcaino}, {Short}, {Siebert}, {Silva},
  {Sinachopoulos}, {Slezak}, {Soffel}, {Sosnowska}, {Strai{\v{z}}ys}, {ter
  Linden}, {Terrell}, {Theil}, {Tiede}, {Troisi}, {Tsalmantza}, {Tur},
  {Vaccari}, {Vachier}, {Valles}, {Van Hamme}, {Veltz}, {Virtanen}, {Wallut},
  {Wichmann}, {Wilkinson}, {Ziaeepour}, \& {Zschocke}}]{gaia2016}
{Gaia Collaboration}, {Prusti}, T., {de Bruijne}, J.~H.~J., {et~al.} 2016,
  \aap, 595, A1, \dodoi{10.1051/0004-6361/201629272}

\bibitem[{{Gaia Collaboration} {et~al.}(2018){Gaia Collaboration}, {Brown},
  {Vallenari}, {Prusti}, {de Bruijne}, {Babusiaux}, {Bailer-Jones}, {Biermann},
  {Evans}, {Eyer}, {Jansen}, {Jordi}, {Klioner}, {Lammers}, {Lindegren},
  {Luri}, {Mignard}, {Panem}, {Pourbaix}, {Randich}, {Sartoretti}, {Siddiqui},
  {Soubiran}, {van Leeuwen}, {Walton}, {Arenou}, {Bastian}, {Cropper},
  {Drimmel}, {Katz}, {Lattanzi}, {Bakker}, {Cacciari}, {Casta{\~n}eda},
  {Chaoul}, {Cheek}, {De Angeli}, {Fabricius}, {Guerra}, {Holl}, {Masana},
  {Messineo}, {Mowlavi}, {Nienartowicz}, {Panuzzo}, {Portell}, {Riello},
  {Seabroke}, {Tanga}, {Th{\'e}venin}, {Gracia-Abril}, {Comoretto},
  {Garcia-Reinaldos}, {Teyssier}, {Altmann}, {Andrae}, {Audard},
  {Bellas-Velidis}, {Benson}, {Berthier}, {Blomme}, {Burgess}, {Busso},
  {Carry}, {Cellino}, {Clementini}, {Clotet}, {Creevey}, {Davidson}, {De
  Ridder}, {Delchambre}, {Dell'Oro}, {Ducourant},
  {Fern{\'a}ndez-Hern{\'a}ndez}, {Fouesneau}, {Fr{\'e}mat}, {Galluccio},
  {Garc{\'\i}a-Torres}, {Gonz{\'a}lez-N{\'u}{\~n}ez}, {Gonz{\'a}lez-Vidal},
  {Gosset}, {Guy}, {Halbwachs}, {Hambly}, {Harrison}, {Hern{\'a}ndez},
  {Hestroffer}, {Hodgkin}, {Hutton}, {Jasniewicz}, {Jean-Antoine-Piccolo},
  {Jordan}, {Korn}, {Krone-Martins}, {Lanzafame}, {Lebzelter}, {L{\"o}ffler},
  {Manteiga}, {Marrese}, {Mart{\'\i}n-Fleitas}, {Moitinho}, {Mora}, {Muinonen},
  {Osinde}, {Pancino}, {Pauwels}, {Petit}, {Recio-Blanco}, {Richards},
  {Rimoldini}, {Robin}, {Sarro}, {Siopis}, {Smith}, {Sozzetti}, {S{\"u}veges},
  {Torra}, {van Reeven}, {Abbas}, {Abreu Aramburu}, {Accart}, {Aerts},
  {Altavilla}, {{\'A}lvarez}, {Alvarez}, {Alves}, {Anderson}, {Andrei},
  {Anglada Varela}, {Antiche}, {Antoja}, {Arcay}, {Astraatmadja}, {Bach},
  {Baker}, {Balaguer-N{\'u}{\~n}ez}, {Balm}, {Barache}, {Barata}, {Barbato},
  {Barblan}, {Barklem}, {Barrado}, {Barros}, {Barstow}, {Bartholom{\'e}
  Mu{\~n}oz}, {Bassilana}, {Becciani}, {Bellazzini}, {Berihuete}, {Bertone},
  {Bianchi}, {Bienaym{\'e}}, {Blanco-Cuaresma}, {Boch}, {Boeche}, {Bombrun},
  {Borrachero}, {Bossini}, {Bouquillon}, {Bourda}, {Bragaglia}, {Bramante},
  {Breddels}, {Bressan}, {Brouillet}, {Br{\"u}semeister}, {Brugaletta},
  {Bucciarelli}, {Burlacu}, {Busonero}, {Butkevich}, {Buzzi}, {Caffau},
  {Cancelliere}, {Cannizzaro}, {Cantat-Gaudin}, {Carballo}, {Carlucci},
  {Carrasco}, {Casamiquela}, {Castellani}, {Castro-Ginard}, {Charlot},
  {Chemin}, {Chiavassa}, {Cocozza}, {Costigan}, {Cowell}, {Crifo}, {Crosta},
  {Crowley}, {Cuypers}, {Dafonte}, {Damerdji}, {Dapergolas}, {David}, {David},
  {de Laverny}, {De Luise}, {De March}, {de Martino}, {de Souza}, {de Torres},
  {Debosscher}, {del Pozo}, {Delbo}, {Delgado}, {Delgado}, {Di Matteo},
  {Diakite}, {Diener}, {Distefano}, {Dolding}, {Drazinos}, {Dur{\'a}n},
  {Edvardsson}, {Enke}, {Eriksson}, {Esquej}, {Eynard Bontemps}, {Fabre},
  {Fabrizio}, {Faigler}, {Falc{\~a}o}, {Farr{\`a}s Casas}, {Federici},
  {Fedorets}, {Fernique}, {Figueras}, {Filippi}, {Findeisen}, {Fonti},
  {Fraile}, {Fraser}, {Fr{\'e}zouls}, {Gai}, {Galleti}, {Garabato},
  {Garc{\'\i}a-Sedano}, {Garofalo}, {Garralda}, {Gavel}, {Gavras}, {Gerssen},
  {Geyer}, {Giacobbe}, {Gilmore}, {Girona}, {Giuffrida}, {Glass}, {Gomes},
  {Granvik}, {Gueguen}, {Guerrier}, {Guiraud}, {Guti{\'e}rrez-S{\'a}nchez},
  {Haigron}, {Hatzidimitriou}, {Hauser}, {Haywood}, {Heiter}, {Helmi}, {Heu},
  {Hilger}, {Hobbs}, {Hofmann}, {Holland}, {Huckle}, {Hypki}, {Icardi},
  {Jan{\ss}en}, {Jevardat de Fombelle}, {Jonker}, {Juh{\'a}sz}, {Julbe},
  {Karampelas}, {Kewley}, {Klar}, {Kochoska}, {Kohley}, {Kolenberg},
  {Kontizas}, {Kontizas}, {Koposov}, {Kordopatis}, {Kostrzewa-Rutkowska},
  {Koubsky}, {Lambert}, {Lanza}, {Lasne}, {Lavigne}, {Le Fustec}, {Le
  Poncin-Lafitte}, {Lebreton}, {Leccia}, {Leclerc}, {Lecoeur-Taibi},
  {Lenhardt}, {Leroux}, {Liao}, {Licata}, {Lindstr{\o}m}, {Lister}, {Livanou},
  {Lobel}, {L{\'o}pez}, {Managau}, {Mann}, {Mantelet}, {Marchal}, {Marchant},
  {Marconi}, {Marinoni}, {Marschalk{\'o}}, {Marshall}, {Martino}, {Marton},
  {Mary}, {Massari}, {Matijevi{\v{c}}}, {Mazeh}, {McMillan}, {Messina},
  {Michalik}, {Millar}, {Molina}, {Molinaro}, {Moln{\'a}r}, {Montegriffo},
  {Mor}, {Morbidelli}, {Morel}, {Morris}, {Mulone}, {Muraveva}, {Musella},
  {Nelemans}, {Nicastro}, {Noval}, {O'Mullane}, {Ord{\'e}novic},
  {Ord{\'o}{\~n}ez-Blanco}, {Osborne}, {Pagani}, {Pagano}, {Pailler},
  {Palacin}, {Palaversa}, {Panahi}, {Pawlak}, {Piersimoni}, {Pineau}, {Plachy},
  {Plum}, {Poggio}, {Poujoulet}, {Pr{\v{s}}a}, {Pulone}, {Racero}, {Ragaini},
  {Rambaux}, {Ramos-Lerate}, {Regibo}, {Reyl{\'e}}, {Riclet}, {Ripepi}, {Riva},
  {Rivard}, {Rixon}, {Roegiers}, {Roelens}, {Romero-G{\'o}mez}, {Rowell},
  {Royer}, {Ruiz-Dern}, {Sadowski}, {Sagrist{\`a} Sell{\'e}s}, {Sahlmann},
  {Salgado}, {Salguero}, {Sanna}, {Santana-Ros}, {Sarasso}, {Savietto},
  {Schultheis}, {Sciacca}, {Segol}, {Segovia}, {S{\'e}gransan}, {Shih},
  {Siltala}, {Silva}, {Smart}, {Smith}, {Solano}, {Solitro}, {Sordo}, {Soria
  Nieto}, {Souchay}, {Spagna}, {Spoto}, {Stampa}, {Steele},
  {Steidelm{\"u}ller}, {Stephenson}, {Stoev}, {Suess}, {Surdej}, {Szabados},
  {Szegedi-Elek}, {Tapiador}, {Taris}, {Tauran}, {Taylor}, {Teixeira},
  {Terrett}, {Teyssand ier}, {Thuillot}, {Titarenko}, {Torra Clotet}, {Turon},
  {Ulla}, {Utrilla}, {Uzzi}, {Vaillant}, {Valentini}, {Valette}, {van Elteren},
  {Van Hemelryck}, {van Leeuwen}, {Vaschetto}, {Vecchiato}, {Veljanoski},
  {Viala}, {Vicente}, {Vogt}, {von Essen}, {Voss}, {Votruba}, {Voutsinas},
  {Walmsley}, {Weiler}, {Wertz}, {Wevers}, {Wyrzykowski}, {Yoldas},
  {{\v{Z}}erjal}, {Ziaeepour}, {Zorec}, {Zschocke}, {Zucker}, {Zurbach}, \&
  {Zwitter}}]{gaia2018}
{Gaia Collaboration}, {Brown}, A.~G.~A., {Vallenari}, A., {et~al.} 2018, \aap,
  616, A1, \dodoi{10.1051/0004-6361/201833051}

\bibitem[{{Geisler} {et~al.}(2005){Geisler}, {Smith}, {Wallerstein},
  {Gonzalez}, \& {Charbonnel}}]{geisler2005}
{Geisler}, D., {Smith}, V.~V., {Wallerstein}, G., {Gonzalez}, G., \&
  {Charbonnel}, C. 2005, \aj, 129, 1428, \dodoi{10.1086/427540}

\bibitem[{{Gravity Collaboration} {et~al.}(2018){Gravity Collaboration},
  {Abuter}, {Amorim}, {Anugu}, {Baub{\"o}ck}, {Benisty}, {Berger}, {Blind},
  {Bonnet}, {Brandner}, {Buron}, {Collin}, {Chapron}, {Cl{\'e}net}, {Coud{\'e}
  Du Foresto}, {de Zeeuw}, {Deen}, {Delplancke-Str{\"o}bele}, {Dembet},
  {Dexter}, {Duvert}, {Eckart}, {Eisenhauer}, {Finger}, {F{\"o}rster
  Schreiber}, {F{\'e}dou}, {Garcia}, {Garcia Lopez}, {Gao}, {Gendron},
  {Genzel}, {Gillessen}, {Gordo}, {Habibi}, {Haubois}, {Haug}, {Hau{\ss}mann},
  {Henning}, {Hippler}, {Horrobin}, {Hubert}, {Hubin}, {Jimenez Rosales},
  {Jochum}, {Jocou}, {Kaufer}, {Kellner}, {Kendrew}, {Kervella}, {Kok},
  {Kulas}, {Lacour}, {Lapeyr{\`e}re}, {Lazareff}, {Le Bouquin}, {L{\'e}na},
  {Lippa}, {Lenzen}, {M{\'e}rand}, {M{\"u}ler}, {Neumann}, {Ott}, {Palanca},
  {Paumard}, {Pasquini}, {Perraut}, {Perrin}, {Pfuhl}, {Plewa}, {Rabien},
  {Ram{\'\i}rez}, {Ramos}, {Rau}, {Rodr{\'\i}guez-Coira}, {Rohloff}, {Rousset},
  {Sanchez-Bermudez}, {Scheithauer}, {Sch{\"o}ller}, {Schuler}, {Spyromilio},
  {Straub}, {Straubmeier}, {Sturm}, {Tacconi}, {Tristram}, {Vincent}, {von
  Fellenberg}, {Wank}, {Waisberg}, {Widmann}, {Wieprecht}, {Wiest},
  {Wiezorrek}, {Woillez}, {Yazici}, {Ziegler}, \& {Zins}}]{gravity2018}
{Gravity Collaboration}, {Abuter}, R., {Amorim}, A., {et~al.} 2018, \aap, 615,
  L15, \dodoi{10.1051/0004-6361/201833718}

\bibitem[{{Green} {et~al.}(2019){Green}, {Schlafly}, {Zucker}, {Speagle}, \&
  {Finkbeiner}}]{green2019}
{Green}, G.~M., {Schlafly}, E., {Zucker}, C., {Speagle}, J.~S., \&
  {Finkbeiner}, D. 2019, \apj, 887, 93, \dodoi{10.3847/1538-4357/ab5362}

\bibitem[{{Hansen} {et~al.}(2013){Hansen}, {Bergemann}, {Cescutti},
  {Fran{\c{c}}ois}, {Arcones}, {Karakas}, {Lind}, \& {Chiappini}}]{hansen2013}
{Hansen}, C.~J., {Bergemann}, M., {Cescutti}, G., {et~al.} 2013, \aap, 551,
  A57, \dodoi{10.1051/0004-6361/201220584}

\bibitem[{Harris {et~al.}(2020)Harris, Millman, van~der Walt, Gommers,
  Virtanen, Cournapeau, Wieser, Taylor, Berg, Smith, Kern, Picus, Hoyer, van
  Kerkwijk, Brett, Haldane, del R{\'{i}}o, Wiebe, Peterson,
  G{\'{e}}rard-Marchant, Sheppard, Reddy, Weckesser, Abbasi, Gohlke, \&
  Oliphant}]{harris2020}
Harris, C.~R., Millman, K.~J., van~der Walt, S.~J., {et~al.} 2020, Nature, 585,
  357, \dodoi{10.1038/s41586-020-2649-2}

\bibitem[{{Hattori} {et~al.}(2018{\natexlab{a}}){Hattori}, {Valluri}, {Bell},
  \& {Roederer}}]{hattori2018a}
{Hattori}, K., {Valluri}, M., {Bell}, E.~F., \& {Roederer}, I.~U.
  2018{\natexlab{a}}, \apj, 866, 121, \dodoi{10.3847/1538-4357/aadee5}

\bibitem[{{Hattori} {et~al.}(2018{\natexlab{b}}){Hattori}, {Valluri}, \&
  {Castro}}]{hattori2018b}
{Hattori}, K., {Valluri}, M., \& {Castro}, N. 2018{\natexlab{b}}, \apj, 869,
  33, \dodoi{10.3847/1538-4357/aaed22}

\bibitem[{{Hawkins} \& {Wyse}(2018)}]{hawkins2018}
{Hawkins}, K., \& {Wyse}, R. F.~G. 2018, \mnras, 481, 1028,
  \dodoi{10.1093/mnras/sty2282}

\bibitem[{{Hayes} {et~al.}(2018){Hayes}, {Majewski}, {Shetrone},
  {Fern{\'a}ndez-Alvar}, {Allende Prieto}, {Schuster}, {Carigi}, {Cunha},
  {Smith}, {Sobeck}, {Almeida}, {Beers}, {Carrera}, {Fern{\'a}ndez-Trincado},
  {Garc{\'\i}a-Hern{\'a}ndez}, {Geisler}, {Lane}, {Lucatello}, {Matthews},
  {Minniti}, {Nitschelm}, {Tang}, {Tissera}, \& {Zamora}}]{hayes2018}
{Hayes}, C.~R., {Majewski}, S.~R., {Shetrone}, M., {et~al.} 2018, \apj, 852,
  49, \dodoi{10.3847/1538-4357/aa9cec}

\bibitem[{{Herzog-Arbeitman} {et~al.}(2018){Herzog-Arbeitman}, {Lisanti}, \&
  {Necib}}]{herzog2018}
{Herzog-Arbeitman}, J., {Lisanti}, M., \& {Necib}, L. 2018, \jcap, 2018, 052,
  \dodoi{10.1088/1475-7516/2018/04/052}

\bibitem[{{Hills}(1988)}]{hills1988}
{Hills}, J.~G. 1988, \nat, 331, 687, \dodoi{10.1038/331687a0}

\bibitem[{{Hills}(1991)}]{hills1991}
---. 1991, \aj, 102, 704, \dodoi{10.1086/115905}

\bibitem[{{Hills}(1992)}]{hills1992}
---. 1992, \aj, 103, 1955, \dodoi{10.1086/116204}

\bibitem[{Hunter(2007)}]{hunter2007}
Hunter, J.~D. 2007, Computing in Science \& Engineering, 9, 90,
  \dodoi{10.1109/MCSE.2007.55}

\bibitem[{{Irrgang} {et~al.}(2021){Irrgang}, {Dimpel}, {Heber}, \&
  {Raddi}}]{irrgang2021}
{Irrgang}, A., {Dimpel}, M., {Heber}, U., \& {Raddi}, R. 2021, \aap, 646, L4,
  \dodoi{10.1051/0004-6361/202040178}

\bibitem[{{Jacobson} {et~al.}(2015){Jacobson}, {Keller}, {Frebel}, {Casey},
  {Asplund}, {Bessell}, {Da Costa}, {Lind}, {Marino}, {Norris}, {Pe{\~n}a},
  {Schmidt}, {Tisserand}, {Walsh}, {Yong}, \& {Yu}}]{jacobson15}
{Jacobson}, H.~R., {Keller}, S., {Frebel}, A., {et~al.} 2015, \apj, 807, 171,
  \dodoi{10.1088/0004-637X/807/2/171}

\bibitem[{{Ji} {et~al.}(2016){Ji}, {Frebel}, {Chiti}, \& {Simon}}]{ji2016}
{Ji}, A.~P., {Frebel}, A., {Chiti}, A., \& {Simon}, J.~D. 2016, \nat, 531, 610,
  \dodoi{10.1038/nature17425}

\bibitem[{{Ji} {et~al.}(2020){Ji}, {Li}, {Hansen}, {Casey}, {Koposov}, {Pace},
  {Mackey}, {Lewis}, {Simpson}, {Bland-Hawthorn}, {Cullinane}, {Da Costa},
  {Hattori}, {Martell}, {Kuehn}, {Erkal}, {Shipp}, {Wan}, \& {Zucker}}]{ji2020}
{Ji}, A.~P., {Li}, T.~S., {Hansen}, T.~T., {et~al.} 2020, \aj, 160, 181,
  \dodoi{10.3847/1538-3881/abacb6}

\bibitem[{{Juri{\'c}} {et~al.}(2008){Juri{\'c}}, {Ivezi{\'c}}, {Brooks},
  {Lupton}, {Schlegel}, {Finkbeiner}, {Padmanabhan}, {Bond}, {Sesar},
  {Rockosi}, {Knapp}, {Gunn}, {Sumi}, {Schneider}, {Barentine}, {Brewington},
  {Brinkmann}, {Fukugita}, {Harvanek}, {Kleinman}, {Krzesinski}, {Long},
  {Neilsen}, {Nitta}, {Snedden}, \& {York}}]{juric2008}
{Juri{\'c}}, M., {Ivezi{\'c}}, {\v{Z}}., {Brooks}, A., {et~al.} 2008, \apj,
  673, 864, \dodoi{10.1086/523619}

\bibitem[{{Kelson}(2003)}]{kelson2003}
{Kelson}, D.~D. 2003, \pasp, 115, 688, \dodoi{10.1086/375502}

\bibitem[{{Kelson} {et~al.}(2000){Kelson}, {Illingworth}, {van Dokkum}, \&
  {Franx}}]{kelson2000}
{Kelson}, D.~D., {Illingworth}, G.~D., {van Dokkum}, P.~G., \& {Franx}, M.
  2000, \apj, 531, 159, \dodoi{10.1086/308445}

\bibitem[{{Kelson} {et~al.}(2014){Kelson}, {Williams}, {Dressler}, {McCarthy},
  {Shectman}, {Mulchaey}, {Villanueva}, {Crane}, \& {Quadri}}]{kelson2014}
{Kelson}, D.~D., {Williams}, R.~J., {Dressler}, A., {et~al.} 2014, \apj, 783,
  110, \dodoi{10.1088/0004-637X/783/2/110}

\bibitem[{{Kirby} {et~al.}(2009){Kirby}, {Guhathakurta}, {Bolte}, {Sneden}, \&
  {Geha}}]{kirby2009}
{Kirby}, E.~N., {Guhathakurta}, P., {Bolte}, M., {Sneden}, C., \& {Geha}, M.~C.
  2009, \apj, 705, 328, \dodoi{10.1088/0004-637X/705/1/328}

\bibitem[{{Klose} {et~al.}(2002){Klose}, {Fuhr}, \& {Wiese}}]{klose2002}
{Klose}, J.~Z., {Fuhr}, J.~R., \& {Wiese}, W.~L. 2002, Journal of Physical and
  Chemical Reference Data, 31, 217, \dodoi{10.1063/1.1448482}

\bibitem[{{Kobayashi} {et~al.}(2020{\natexlab{a}}){Kobayashi}, {Karakas}, \&
  {Lugaro}}]{kobayashi2020}
{Kobayashi}, C., {Karakas}, A.~I., \& {Lugaro}, M. 2020{\natexlab{a}}, \apj,
  900, 179, \dodoi{10.3847/1538-4357/abae65}

\bibitem[{{Kobayashi} {et~al.}(2020{\natexlab{b}}){Kobayashi}, {Leung}, \&
  {Nomoto}}]{kobayashi2020b}
{Kobayashi}, C., {Leung}, S.-C., \& {Nomoto}, K. 2020{\natexlab{b}}, \apj, 895,
  138, \dodoi{10.3847/1538-4357/ab8e44}

\bibitem[{{Koposov} {et~al.}(2020){Koposov}, {Boubert}, {Li}, {Erkal}, {Da
  Costa}, {Zucker}, {Ji}, {Kuehn}, {Lewis}, {Mackey}, {Simpson}, {Shipp},
  {Wan}, {Belokurov}, {Bland-Hawthorn}, {Martell}, {Nordlander}, {Pace}, {De
  Silva}, {Wang}, \& {S5 Collaboration}}]{koposov2020}
{Koposov}, S.~E., {Boubert}, D., {Li}, T.~S., {et~al.} 2020, \mnras, 491, 2465,
  \dodoi{10.1093/mnras/stz3081}

\bibitem[{{Kunder} {et~al.}(2017){Kunder}, {Kordopatis}, {Steinmetz},
  {Zwitter}, {McMillan}, {Casagrande}, {Enke}, {Wojno}, {Valentini},
  {Chiappini}, {Matijevi{\v{c}}}, {Siviero}, {de Laverny}, {Recio-Blanco},
  {Bijaoui}, {Wyse}, {Binney}, {Grebel}, {Helmi}, {Jofre}, {Antoja}, {Gilmore},
  {Siebert}, {Famaey}, {Bienaym{\'e}}, {Gibson}, {Freeman}, {Navarro},
  {Munari}, {Seabroke}, {Anguiano}, {{\v{Z}}erjal}, {Minchev}, {Reid},
  {Bland-Hawthorn}, {Kos}, {Sharma}, {Watson}, {Parker}, {Scholz}, {Burton},
  {Cass}, {Hartley}, {Fiegert}, {Stupar}, {Ritter}, {Hawkins}, {Gerhard},
  {Chaplin}, {Davies}, {Elsworth}, {Lund}, {Miglio}, \& {Mosser}}]{kunder17}
{Kunder}, A., {Kordopatis}, G., {Steinmetz}, M., {et~al.} 2017, \aj, 153, 75,
  \dodoi{10.3847/1538-3881/153/2/75}

\bibitem[{{Leonard}(1991)}]{leonard1991}
{Leonard}, P. J.~T. 1991, \aj, 101, 562, \dodoi{10.1086/115704}

\bibitem[{{Letarte} {et~al.}(2010){Letarte}, {Hill}, {Tolstoy}, {Jablonka},
  {Shetrone}, {Venn}, {Spite}, {Irwin}, {Battaglia}, {Helmi}, {Primas},
  {Fran{\c{c}}ois}, {Kaufer}, {Szeifert}, {Arimoto}, \&
  {Sadakane}}]{letarte2010}
{Letarte}, B., {Hill}, V., {Tolstoy}, E., {et~al.} 2010, \aap, 523, A17,
  \dodoi{10.1051/0004-6361/200913413}

\bibitem[{{Lind} {et~al.}(2011){Lind}, {Asplund}, {Barklem}, \&
  {Belyaev}}]{lind2011}
{Lind}, K., {Asplund}, M., {Barklem}, P.~S., \& {Belyaev}, A.~K. 2011, \aap,
  528, A103, \dodoi{10.1051/0004-6361/201016095}

\bibitem[{{Lindegren} {et~al.}(2020){Lindegren}, {Bastian}, {Biermann},
  {Bombrun}, {de Torres}, {Gerlach}, {Geyer}, {Hern{\'a}ndez}, {Hilger},
  {Hobbs}, {Klioner}, {Lammers}, {McMillan}, {Ramos-Lerate},
  {Steidelm{\"u}ller}, {Stephenson}, \& {van Leeuwen}}]{lindegren2020}
{Lindegren}, L., {Bastian}, U., {Biermann}, M., {et~al.} 2020, arXiv e-prints,
  arXiv:2012.01742.
\newblock \doarXiv{2012.01742}

\bibitem[{{Mainzer} {et~al.}(2011){Mainzer}, {Grav}, {Bauer}, {Masiero},
  {McMillan}, {Cutri}, {Walker}, {Wright}, {Eisenhardt}, {Tholen}, {Spahr},
  {Jedicke}, {Denneau}, {DeBaun}, {Elsbury}, {Gautier}, {Gomillion}, {Hand},
  {Mo}, {Watkins}, {Wilkins}, {Bryngelson}, {Del Pino Molina}, {Desai},
  {G{\'o}mez Camus}, {Hidalgo}, {Konstantopoulos}, {Larsen}, {Maleszewski},
  {Malkan}, {Mauduit}, {Mullan}, {Olszewski}, {Pforr}, {Saro}, {Scotti}, \&
  {Wasserman}}]{mainzer2011}
{Mainzer}, A., {Grav}, T., {Bauer}, J., {et~al.} 2011, \apj, 743, 156,
  \dodoi{10.1088/0004-637X/743/2/156}

\bibitem[{{Marchetti} {et~al.}(2018{\natexlab{a}}){Marchetti}, {Contigiani},
  {Rossi}, {Albert}, {Brown}, \& {Sesana}}]{marchetti2018b}
{Marchetti}, T., {Contigiani}, O., {Rossi}, E.~M., {et~al.} 2018{\natexlab{a}},
  \mnras, 476, 4697, \dodoi{10.1093/mnras/sty579}

\bibitem[{{Marchetti} {et~al.}(2018{\natexlab{b}}){Marchetti}, {Rossi},
  {Kordopatis}, {Brown}, {Rimoldi}, {Starkenburg}, {Youakim}, \&
  {Ashley}}]{marchetti2018a}
{Marchetti}, T., {Rossi}, E.~M., {Kordopatis}, G., {et~al.} 2018{\natexlab{b}},
  in Astrometry and Astrophysics in the Gaia Sky, ed. A.~{Recio-Blanco}, P.~{de
  Laverny}, A.~G.~A. {Brown}, \& T.~{Prusti}, Vol. 330, 181--184,
  \dodoi{10.1017/S1743921317005671}

\bibitem[{{Matteucci} \& {Brocato}(1990)}]{matteucci1990}
{Matteucci}, F., \& {Brocato}, E. 1990, \apj, 365, 539, \dodoi{10.1086/169508}

\bibitem[{{McKinney}(2010)}]{mckinney2010}
{McKinney}, W. 2010, in {P}roceedings of the 9th {P}ython in {S}cience
  {C}onference, ed. {S}t\'efan van~der {W}alt \& {J}arrod {M}illman, 56 -- 61,
  \dodoi{10.25080/Majora-92bf1922-00a}

\bibitem[{{McWilliam}(1998)}]{mcwilliam1998}
{McWilliam}, A. 1998, \aj, 115, 1640, \dodoi{10.1086/300289}

\bibitem[{{Miyamoto} \& {Nagai}(1975)}]{miyamoto1975}
{Miyamoto}, M., \& {Nagai}, R. 1975, \pasj, 27, 533

\bibitem[{{Monaco} {et~al.}(2005){Monaco}, {Bellazzini}, {Bonifacio},
  {Ferraro}, {Marconi}, {Pancino}, {Sbordone}, \& {Zaggia}}]{monaco2005}
{Monaco}, L., {Bellazzini}, M., {Bonifacio}, P., {et~al.} 2005, \aap, 441, 141,
  \dodoi{10.1051/0004-6361:20053333}

\bibitem[{{Morton}(2015)}]{morton2015}
{Morton}, T.~D. 2015, {isochrones: Stellar model grid package}.
\newblock \doeprint{1503.010}

\bibitem[{{Naidu} {et~al.}(2020){Naidu}, {Conroy}, {Bonaca}, {Johnson}, {Ting},
  {Caldwell}, {Zaritsky}, \& {Cargile}}]{naidu2020}
{Naidu}, R.~P., {Conroy}, C., {Bonaca}, A., {et~al.} 2020, \apj, 901, 48,
  \dodoi{10.3847/1538-4357/abaef4}

\bibitem[{{Naidu} {et~al.}(2021){Naidu}, {Ji}, {Conroy}, {Bonaca}, {Ting},
  {Zaritsky}, {van Son}, {Broekgaarden}, {Tacchella}, {Chandra}, {Caldwell},
  {Cargile}, \& {Speagle}}]{naidu2021}
{Naidu}, R.~P., {Ji}, A.~P., {Conroy}, C., {et~al.} 2021, arXiv e-prints,
  arXiv:2110.14652.
\newblock \doarXiv{2110.14652}

\bibitem[{{Navarro} {et~al.}(1996){Navarro}, {Frenk}, \& {White}}]{navarro1996}
{Navarro}, J.~F., {Frenk}, C.~S., \& {White}, S. D.~M. 1996, \apj, 462, 563,
  \dodoi{10.1086/177173}

\bibitem[{{Necib} {et~al.}(2020){Necib}, {Ostdiek}, {Lisanti}, {Cohen},
  {Freytsis}, \& {Garrison-Kimmel}}]{necib2020}
{Necib}, L., {Ostdiek}, B., {Lisanti}, M., {et~al.} 2020, \apj, 903, 25,
  \dodoi{10.3847/1538-4357/abb814}

\bibitem[{{Ness} {et~al.}(2015){Ness}, {Hogg}, {Rix}, {Ho}, \&
  {Zasowski}}]{ness15}
{Ness}, M., {Hogg}, D.~W., {Rix}, H.~W., {Ho}, A. Y.~Q., \& {Zasowski}, G.
  2015, \apj, 808, 16, \dodoi{10.1088/0004-637X/808/1/16}

\bibitem[{{Ness} {et~al.}(2016){Ness}, {Hogg}, {Rix}, {Martig}, {Pinsonneault},
  \& {Ho}}]{ness16}
{Ness}, M., {Hogg}, D.~W., {Rix}, H.~W., {et~al.} 2016, \apj, 823, 114,
  \dodoi{10.3847/0004-637X/823/2/114}

\bibitem[{{Nidever} {et~al.}(2020){Nidever}, {Hasselquist}, {Hayes}, {Hawkins},
  {Povick}, {Majewski}, {Smith}, {Anguiano}, {Stringfellow}, {Sobeck}, {Cunha},
  {Beers}, {Bestenlehner}, {Cohen}, {Garcia-Hernandez}, {J{\"o}nsson},
  {Nitschelm}, {Shetrone}, {Lacerna}, {Allende Prieto}, {Beaton}, {Dell'Agli},
  {Fern{\'a}ndez-Trincado}, {Feuillet}, {Gallart}, {Hearty}, {Holtzman},
  {Manchado}, {Mu{\~n}oz}, {O'Connell}, \& {Rosado}}]{nidever2020}
{Nidever}, D.~L., {Hasselquist}, S., {Hayes}, C.~R., {et~al.} 2020, \apj, 895,
  88, \dodoi{10.3847/1538-4357/ab7305}

\bibitem[{{Nissen} \& {Schuster}(2010)}]{nissen2010}
{Nissen}, P.~E., \& {Schuster}, W.~J. 2010, \aap, 511, L10,
  \dodoi{10.1051/0004-6361/200913877}

\bibitem[{{Nordlander} \& {Lind}(2017)}]{nordlander2017}
{Nordlander}, T., \& {Lind}, K. 2017, \aap, 607, A75,
  \dodoi{10.1051/0004-6361/201730427}

\bibitem[{{pandas Development Team}(2020)}]{reback2020}
{pandas Development Team}. 2020, pandas-dev/pandas: Pandas, latest,  Zenodo,
  \dodoi{10.5281/zenodo.3509134}

\bibitem[{{Paxton} {et~al.}(2011){Paxton}, {Bildsten}, {Dotter}, {Herwig},
  {Lesaffre}, \& {Timmes}}]{paxton2011}
{Paxton}, B., {Bildsten}, L., {Dotter}, A., {et~al.} 2011, \apjs, 192, 3,
  \dodoi{10.1088/0067-0049/192/1/3}

\bibitem[{{Paxton} {et~al.}(2013){Paxton}, {Cantiello}, {Arras}, {Bildsten},
  {Brown}, {Dotter}, {Mankovich}, {Montgomery}, {Stello}, {Timmes}, \&
  {Townsend}}]{paxton2013}
{Paxton}, B., {Cantiello}, M., {Arras}, P., {et~al.} 2013, \apjs, 208, 4,
  \dodoi{10.1088/0067-0049/208/1/4}

\bibitem[{{Paxton} {et~al.}(2015){Paxton}, {Marchant}, {Schwab}, {Bauer},
  {Bildsten}, {Cantiello}, {Dessart}, {Farmer}, {Hu}, {Langer}, {Townsend},
  {Townsley}, \& {Timmes}}]{paxton2015}
{Paxton}, B., {Marchant}, P., {Schwab}, J., {et~al.} 2015, \apjs, 220, 15,
  \dodoi{10.1088/0067-0049/220/1/15}

\bibitem[{{Placco} {et~al.}(2021){Placco}, {Sneden}, {Roederer}, {Lawler}, {Den
  Hartog}, {Hejazi}, {Maas}, \& {Bernath}}]{placco2021}
{Placco}, V.~M., {Sneden}, C., {Roederer}, I.~U., {et~al.} 2021, Research Notes
  of the American Astronomical Society, 5, 92, \dodoi{10.3847/2515-5172/abf651}

\bibitem[{{Poveda} {et~al.}(1967){Poveda}, {Ruiz}, \& {Allen}}]{poveda1967}
{Poveda}, A., {Ruiz}, J., \& {Allen}, C. 1967, Boletin de los Observatorios
  Tonantzintla y Tacubaya, 4, 86

\bibitem[{{Prantzos} {et~al.}(2018){Prantzos}, {Abia}, {Limongi}, {Chieffi}, \&
  {Cristallo}}]{prantzos2018}
{Prantzos}, N., {Abia}, C., {Limongi}, M., {Chieffi}, A., \& {Cristallo}, S.
  2018, \mnras, 476, 3432, \dodoi{10.1093/mnras/sty316}

\bibitem[{{Ram{\'\i}rez} {et~al.}(2014){Ram{\'\i}rez}, {Mel{\'e}ndez}, {Bean},
  {Asplund}, {Bedell}, {Monroe}, {Casagrande}, {Schirbel}, {Dreizler}, {Teske},
  {Tucci Maia}, {Alves-Brito}, \& {Baumann}}]{ramirez2014}
{Ram{\'\i}rez}, I., {Mel{\'e}ndez}, J., {Bean}, J., {et~al.} 2014, \aap, 572,
  A48, \dodoi{10.1051/0004-6361/201424244}

\bibitem[{{Reggiani} \& {Mel{\'e}ndez}(2018)}]{reggiani2018}
{Reggiani}, H., \& {Mel{\'e}ndez}, J. 2018, \mnras, 475, 3502,
  \dodoi{10.1093/mnras/sty104}

\bibitem[{{Reggiani} {et~al.}(2017){Reggiani}, {Mel{\'e}ndez}, {Kobayashi},
  {Karakas}, \& {Placco}}]{reggiani2017}
{Reggiani}, H., {Mel{\'e}ndez}, J., {Kobayashi}, C., {Karakas}, A., \&
  {Placco}, V. 2017, \aap, 608, A46, \dodoi{10.1051/0004-6361/201730750}

\bibitem[{{Reggiani} {et~al.}(2020){Reggiani}, {Schlaufman}, {Casey}, \&
  {Ji}}]{reggiani2020}
{Reggiani}, H., {Schlaufman}, K.~C., {Casey}, A.~R., \& {Ji}, A.~P. 2020, \aj,
  160, 173, \dodoi{10.3847/1538-3881/aba948}

\bibitem[{{Reggiani} {et~al.}(2021){Reggiani}, {Schlaufman}, {Casey}, {Simon},
  \& {Ji}}]{reggiani2021}
{Reggiani}, H., {Schlaufman}, K.~C., {Casey}, A.~R., {Simon}, J.~D., \& {Ji},
  A.~P. 2021, arXiv e-prints, arXiv:2108.10880.
\newblock \doarXiv{2108.10880}

\bibitem[{{Reggiani} {et~al.}(2022){Reggiani}, {Schlaufman}, {Healy},
  {Lothringer}, \& {Sing}}]{reggiani2022}
{Reggiani}, H., {Schlaufman}, K.~C., {Healy}, B.~F., {Lothringer}, J.~D., \&
  {Sing}, D.~K. 2022, arXiv e-prints, arXiv:2201.08508.
\newblock \doarXiv{2201.08508}

\bibitem[{{Reggiani} {et~al.}(2019){Reggiani}, {Amarsi}, {Lind}, {Barklem},
  {Zatsarinny}, {Bartschat}, {Fursa}, {Bray}, {Spina}, \&
  {Mel{\'e}ndez}}]{reggiani2019}
{Reggiani}, H., {Amarsi}, A.~M., {Lind}, K., {et~al.} 2019, \aap, 627, A177,
  \dodoi{10.1051/0004-6361/201935156}

\bibitem[{{Sakari} {et~al.}(2018){Sakari}, {Placco}, {Farrell}, {Roederer},
  {Wallerstein}, {Beers}, {Ezzeddine}, {Frebel}, {Hansen}, {Holmbeck},
  {Sneden}, {Cowan}, {Venn}, {Davis}, {Matijevi{\v{c}}}, {Wyse},
  {Bland-Hawthorn}, {Chiappini}, {Freeman}, {Gibson}, {Grebel}, {Helmi},
  {Kordopatis}, {Kunder}, {Navarro}, {Reid}, {Seabroke}, {Steinmetz}, \&
  {Watson}}]{sakari2018}
{Sakari}, C.~M., {Placco}, V.~M., {Farrell}, E.~M., {et~al.} 2018, \apj, 868,
  110, \dodoi{10.3847/1538-4357/aae9df}

\bibitem[{{Sanders} {et~al.}(2021){Sanders}, {Belokurov}, \& {Man}}]{sanders21}
{Sanders}, J.~L., {Belokurov}, V., \& {Man}, K. T.~F. 2021, \mnras, 506, 4321,
  \dodoi{10.1093/mnras/stab1951}

\bibitem[{{Shectman} \& {Johns}(2003)}]{shectman2003}
{Shectman}, S.~A., \& {Johns}, M. 2003, in Society of Photo-Optical
  Instrumentation Engineers (SPIE) Conference Series, Vol. 4837, \procspie, ed.
  J.~M. {Oschmann} \& L.~M. {Stepp}, 910--918, \dodoi{10.1117/12.457909}

\bibitem[{{Shen} {et~al.}(2018){Shen}, {Boubert}, {G{\"a}nsicke}, {Jha},
  {Andrews}, {Chomiuk}, {Foley}, {Fraser}, {Gromadzki}, {Guillochon}, {Kotze},
  {Maguire}, {Siebert}, {Smith}, {Strader}, {Badenes}, {Kerzendorf}, {Koester},
  {Kromer}, {Miles}, {Pakmor}, {Schwab}, {Toloza}, {Toonen}, {Townsley}, \&
  {Williams}}]{shen2018}
{Shen}, K.~J., {Boubert}, D., {G{\"a}nsicke}, B.~T., {et~al.} 2018, \apj, 865,
  15, \dodoi{10.3847/1538-4357/aad55b}

\bibitem[{{Shetrone} {et~al.}(2003){Shetrone}, {Venn}, {Tolstoy}, {Primas},
  {Hill}, \& {Kaufer}}]{shetrone2003}
{Shetrone}, M., {Venn}, K.~A., {Tolstoy}, E., {et~al.} 2003, \aj, 125, 684,
  \dodoi{10.1086/345966}

\bibitem[{{Skrutskie} {et~al.}(2006){Skrutskie}, {Cutri}, {Stiening},
  {Weinberg}, {Schneider}, {Carpenter}, {Beichman}, {Capps}, {Chester},
  {Elias}, {Huchra}, {Liebert}, {Lonsdale}, {Monet}, {Price}, {Seitzer},
  {Jarrett}, {Kirkpatrick}, {Gizis}, {Howard}, {Evans}, {Fowler}, {Fullmer},
  {Hurt}, {Light}, {Kopan}, {Marsh}, {McCallon}, {Tam}, {Van Dyk}, \&
  {Wheelock}}]{skrutskie2006}
{Skrutskie}, M.~F., {Cutri}, R.~M., {Stiening}, R., {et~al.} 2006, \aj, 131,
  1163, \dodoi{10.1086/498708}

\bibitem[{{Skrutskie, M. F.; Cutri, R. M.; Stiening, R.; Weinberg, M. D.;
  Schneider, S.; Carpenter, J. M.; Beichman, C.; Capps, R.; Chester, T.; Elias,
  J.; Huchra, J.; Liebert, J.; Lonsdale, C.; Monet, D. G.; Price, S.; Seitzer,
  P.; Jarrett, T.; Kirkpatrick, J. D.; Gizis, J. E.; Howard, E.; Evans, T.;
  Fowler, J.; Fullmer, L.; Hurt, R.; Light, R.; Kopan, E. L.; Marsh, K. A.;
  McCallon, H. L.; Tam, R.; Van Dyk, S.; Wheelock, S.}(2019)}]{ipac_2mass}
{Skrutskie, M. F.; Cutri, R. M.; Stiening, R.; Weinberg, M. D.; Schneider, S.;
  Carpenter, J. M.; Beichman, C.; Capps, R.; Chester, T.; Elias, J.; Huchra,
  J.; Liebert, J.; Lonsdale, C.; Monet, D. G.; Price, S.; Seitzer, P.; Jarrett,
  T.; Kirkpatrick, J. D.; Gizis, J. E.; Howard, E.; Evans, T.; Fowler, J.;
  Fullmer, L.; Hurt, R.; Light, R.; Kopan, E. L.; Marsh, K. A.; McCallon, H.
  L.; Tam, R.; Van Dyk, S.; Wheelock, S.} 2019, 2MASS All-Sky Point Source
  Catalog,  IPAC, \dodoi{10.26131/IRSA2}

\bibitem[{{Sneden} {et~al.}(2009){Sneden}, {Lawler}, {Cowan}, {Ivans}, \& {Den
  Hartog}}]{sneden2009}
{Sneden}, C., {Lawler}, J.~E., {Cowan}, J.~J., {Ivans}, I.~I., \& {Den Hartog},
  E.~A. 2009, \apjs, 182, 80, \dodoi{10.1088/0067-0049/182/1/80}

\bibitem[{{Sneden} {et~al.}(2016){Sneden}, {Lawler}, {den Hartog}, \&
  {Wood}}]{sneden2016}
{Sneden}, C., {Lawler}, J.~E., {den Hartog}, E.~A., \& {Wood}, M.~E. 2016, IAU
  Focus Meeting, 29A, 287, \dodoi{10.1017/S1743921316003069}

\bibitem[{{Sneden}(1973)}]{sneden1973}
{Sneden}, C.~A. 1973, PhD thesis, THE UNIVERSITY OF TEXAS AT AUSTIN.

\bibitem[{{Suda} {et~al.}(2017){Suda}, {Hidaka}, {Aoki}, {Katsuta}, {Yamada},
  {Fujimoto}, {Ohtani}, {Masuyama}, {Noda}, \& {Wada}}]{suda2017}
{Suda}, T., {Hidaka}, J., {Aoki}, W., {et~al.} 2017, \pasj, 69, 76,
  \dodoi{10.1093/pasj/psx059}

\bibitem[{{Tinsley}(1979)}]{tinsley1979}
{Tinsley}, B.~M. 1979, \apj, 229, 1046, \dodoi{10.1086/157039}

\bibitem[{{Tody}(1986)}]{iraf1986}
{Tody}, D. 1986, Society of Photo-Optical Instrumentation Engineers (SPIE)
  Conference Series, Vol. 627, {The IRAF Data Reduction and Analysis System},
  ed. D.~L. {Crawford}, 733, \dodoi{10.1117/12.968154}

\bibitem[{{Tody}(1993)}]{iraf1993}
---. 1993, Astronomical Society of the Pacific Conference Series, Vol.~52,
  {IRAF in the Nineties}, ed. R.~J. {Hanisch}, R.~J.~V. {Brissenden}, \&
  J.~{Barnes}, 173

\bibitem[{{Virtanen} {et~al.}(2020){Virtanen}, {Gommers}, {Oliphant},
  {Haberland}, {Reddy}, {Cournapeau}, {Burovski}, {Peterson}, {Weckesser},
  {Bright}, {van der Walt}, {Brett}, {Wilson}, {Millman}, {Mayorov}, {Nelson},
  {Jones}, {Kern}, {Larson}, {Carey}, {Polat}, {Feng}, {Moore}, {VanderPlas},
  {Laxalde}, {Perktold}, {Cimrman}, {Henriksen}, {Quintero}, {Harris},
  {Archibald}, {Ribeiro}, {Pedregosa}, {van Mulbregt}, \& {SciPy 1. 0
  Contributors}}]{virtanen2020}
{Virtanen}, P., {Gommers}, R., {Oliphant}, T.~E., {et~al.} 2020, Nature
  Methods, 17, 261, \dodoi{10.1038/s41592-019-0686-2}

\bibitem[{{Wolf} {et~al.}(2018){Wolf}, {Onken}, {Luvaul}, {Schmidt}, {Bessell},
  {Chang}, {Da Costa}, {Mackey}, {Martin-Jones}, {Murphy}, {Preston}, {Scalzo},
  {Shao}, {Smillie}, {Tisserand}, {White}, \& {Yuan}}]{wolf2018}
{Wolf}, C., {Onken}, C.~A., {Luvaul}, L.~C., {et~al.} 2018, \pasa, 35, e010,
  \dodoi{10.1017/pasa.2018.5}

\bibitem[{{Wright} {et~al.}(2010){Wright}, {Eisenhardt}, {Mainzer}, {Ressler},
  {Cutri}, {Jarrett}, {Kirkpatrick}, {Padgett}, {McMillan}, {Skrutskie},
  {Stanford}, {Cohen}, {Walker}, {Mather}, {Leisawitz}, {Gautier}, {McLean},
  {Benford}, {Lonsdale}, {Blain}, {Mendez}, {Irace}, {Duval}, {Liu}, {Royer},
  {Heinrichsen}, {Howard}, {Shannon}, {Kendall}, {Walsh}, {Larsen}, {Cardon},
  {Schick}, {Schwalm}, {Abid}, {Fabinsky}, {Naes}, \& {Tsai}}]{wright2010}
{Wright}, E.~L., {Eisenhardt}, P. R.~M., {Mainzer}, A.~K., {et~al.} 2010, \aj,
  140, 1868, \dodoi{10.1088/0004-6256/140/6/1868}

\bibitem[{{Wright, Edward L.; Eisenhardt, Peter R. M.; Mainzer, Amy K.;
  Ressler, Michael E.; Cutri, Roc M.; Jarrett, Thomas; Kirkpatrick, J. Davy;
  Padgett, Deborah; McMillan, Robert S.; Skrutskie, Michael; Stanford, S. A.;
  Cohen, Martin; Walker, Russell G.; Mather, John C.; Leisawitz, David;
  Gautier, Thomas N., III; McLean, Ian; Benford, Dominic; Lonsdale, Carol J.;
  Blain, Andrew; Mendez, Bryan; Irace, William R.; Duval, Valerie; Liu,
  Fengchuan; Royer, Don; Heinrichsen, Ingolf; Howard, Joan; et
  al.}(2019)}]{ipac_allwise}
{Wright, Edward L.; Eisenhardt, Peter R. M.; Mainzer, Amy K.; Ressler, Michael
  E.; Cutri, Roc M.; Jarrett, Thomas; Kirkpatrick, J. Davy; Padgett, Deborah;
  McMillan, Robert S.; Skrutskie, Michael; Stanford, S. A.; Cohen, Martin;
  Walker, Russell G.; Mather, John C.; Leisawitz, David; Gautier, Thomas N.,
  III; McLean, Ian; Benford, Dominic; Lonsdale, Carol J.; Blain, Andrew;
  Mendez, Bryan; Irace, William R.; Duval, Valerie; Liu, Fengchuan; Royer, Don;
  Heinrichsen, Ingolf; Howard, Joan; et al.} 2019, AllWISE Source Catalog,
  IPAC, \dodoi{10.26131/IRSA1}

\bibitem[{{Yu} \& {Tremaine}(2003)}]{yu2003}
{Yu}, Q., \& {Tremaine}, S. 2003, \apj, 599, 1129, \dodoi{10.1086/379546}

\bibitem[{{Zhao} {et~al.}(2016){Zhao}, {Mashonkina}, {Yan}, {Alexeeva},
  {Kobayashi}, {Pakhomov}, {Shi}, {Sitnova}, {Tan}, {Zhang}, {Zhang}, {Zhou},
  {Bolte}, {Chen}, {Li}, {Liu}, \& {Zhai}}]{zhao2016}
{Zhao}, G., {Mashonkina}, L., {Yan}, H.~L., {et~al.} 2016, \apj, 833, 225,
  \dodoi{10.3847/1538-4357/833/2/225}

\end{thebibliography}

\begin{deluxetable*}{llccccc}
\tablecaption{Atomic data, Equivalent Widths and line Abundances. Full version online.\label{measured_ews}}
\tablewidth{0pt}
\tablehead{
\colhead{Gaia EDR3 Source ID} & \colhead{Wavelength} & \colhead{Species} &
\colhead{Excitation Potential} & \colhead{log($gf$)} &
\colhead{EW} & \colhead{$\log_\epsilon(\rm{X})$} \\ 
 & \colhead{(\AA)} &  & \colhead{(eV)} & & (m\AA) & }
\startdata
3252546886080448384 & $5889.951$ & \ion{Na}{1} & $0.000$ & $0.108$ & $177.00$ & $4.497$\\ 
3252546886080448384 & $5895.924$ & \ion{Na}{1} & $0.000$ & $-0.194$ & $131.30$ & $4.315$\\ 
2629296824480015744 & $5682.633$ & \ion{Na}{1} & $2.102$ & $-0.706$ & $36.70$ & $4.971$\\ 
2629296824480015744 & $5688.203$ & \ion{Na}{1} & $2.104$ & $-0.406$ & $47.90$ & $4.852$\\ 
6505889848642319872 & $5682.633$ & \ion{Na}{1} & $2.102$ & $-0.706$ & $18.80$ & $4.388$\\ 
6505889848642319872 & $5688.203$ & \ion{Na}{1} & $2.104$ & $-0.406$ & $32.70$ & $4.397$\\ 
3252546886080448384 & $3986.753$ & \ion{Mg}{1} & $4.346$ & $-1.060$ & $38.40$ & $5.770$\\ 
3252546886080448384 & $4057.505$ & \ion{Mg}{1} & $4.346$ & $-0.900$ & $22.70$ & $5.322$\\ 
3252546886080448384 & $4167.271$ & \ion{Mg}{1} & $4.346$ & $-0.745$ & $60.30$ & $5.796$\\ 
3252546886080448384 & $4702.991$ & \ion{Mg}{1} & $4.346$ & $-0.440$ & $80.10$ & $5.718$\\ 
3252546886080448384 & $4283.011$ & \ion{Ca}{1} & $1.886$ & $-0.224$ & $61.40$ & $4.616$
\enddata
\tablecomments{This table is published in its entirety in the machine-readable format.  A portion is shown here for guidance regarding its form and content.} 
\end{deluxetable*}

\begin{longrotatetable} 
\begin{deluxetable*}{lcccccc} 
\tablecaption{Stellar Properties and Adopted Parameters\label{stellar_params}} 
\tablewidth{0pt} 
\tabletypesize{\tiny} 
\tablehead{ 
\colhead{} & \colhead{} & \colhead{} & \colhead{} & \colhead{} & \colhead{} & \colhead{}} 
\startdata
\hline 
\hline 
Property & 2853089398265954432 & 330414789019026944 & 2260163008363761664 & 2233912206910720000 & 1765600930139450752 &  Units \\ 
\hline 
\textbf{Photometric Properties} & & & & & &\\ 
 External ID & TYC 1732-2222-1$^{\rm{c}}$ & TYC 2319-713-1$^{\rm{c}}$ & ES26$^{\rm{c}}$ & ES10$^{\rm{c}}$ & TYC 1126-382-1 & \\ 
Gaia EDR3 parallax & $0.830\pm0.053$ & $0.496\pm0.034$ & $0.336\pm0.012$ & $0.354\pm0.014$ & $0.613\pm0.022$ & $mas$ \\ 
Gaia DR2 G & $11.012\pm0.002$ & $11.664\pm0.002$ & $11.821\pm0.002$ & $12.959\pm0.002$ & $11.865\pm0.002$ & Vega mag \\ 
Gaia DR2 BP & $11.569\pm0.002$ & $12.150\pm0.002$ & $12.318\pm0.002$ & $13.391\pm0.002$ & $12.341\pm0.002$ & Vega mag \\ 
Gaia DR2 RP & $10.347\pm0.002$ & $11.021\pm0.002$ & $11.178\pm0.002$ & $12.351\pm0.002$ & $11.233\pm0.002$ & Vega mag \\ 
SkyMapper $u$ & $\cdots$ & $\cdots$ & $\cdots$ & $\cdots$ & $\cdots$ & AB mag \\ 
SkyMapper $v$ & $\cdots$ & $\cdots$ & $\cdots$ & $\cdots$ & $\cdots$ & AB mag \\ 
SkyMapper $g$ & $\cdots$ & $\cdots$ & $\cdots$ & $\cdots$ & $\cdots$ & AB mag \\ 
SkyMapper $r$ & $\cdots$ & $\cdots$ & $\cdots$ & $\cdots$ & $\cdots$ & AB mag \\ 
SkyMapper $i$ & $\cdots$ & $\cdots$ & $\cdots$ & $\cdots$ & $\cdots$ & AB mag \\ 
SkyMapper $z$ & $\cdots$ & $\cdots$ & $\cdots$ & $\cdots$ & $\cdots$ & AB mag \\ 
2MASS $J$ & $9.499\pm0.022$ & $10.158\pm0.020$ & $10.332\pm0.021$ & $11.589\pm0.024$ & $10.419\pm0.022$ & Vega mag \\ 
2MASS $H$ & $9.069\pm0.022$ & $9.708\pm0.023$ & $9.818\pm0.017$ & $11.164\pm0.027$ & $9.935\pm0.023$ & Vega mag \\ 
2MASS $K$ & $8.923\pm0.022$ & $9.602\pm0.021$ & $9.740\pm0.016$ & $11.085\pm0.021$ & $9.836\pm0.021$ & Vega mag \\ 
WISE W1 & $8.837\pm0.022$ & $9.510\pm0.022$ & $9.667\pm0.023$ & $11.019\pm0.023$ & $9.769\pm0.023$ & Vega mag \\ 
WISE W2 & $8.896\pm0.020$ & $9.537\pm0.020$ & $9.692\pm0.020$ & $11.035\pm0.020$ & $9.782\pm0.020$ & Vega mag \\ 
\hline
\textbf{Stellar Properties} & & & & & & \\ 
Distance $d_{\text{iso}}$ & $1.2^{+0.1}_{-0.1}$ & $2.0^{+0.1}_{-0.1}$ & $2.8^{+0.1}_{-0.1}$ & $2.8^{+0.1}_{-0.1}$ & $1.6^{+0.1}_{-0.1}$ & kpc \\ 
Mass $M_{\odot}$ & $0.90^{+0.04}_{-0.04}$ & $0.95^{+0.06}_{-0.07}$ & $0.78^{+0.01}_{-0.01}$ & $0.78^{+0.01}_{-0.01}$ & $0.78^{+0.01}_{-0.01}$ & M$_{\odot}$ \\ 
Extinction A$_{\text{V}}$ & $0.312^{+0.006}_{-0.012}$ & $0.448^{+0.009}_{-0.016}$ & $0.178^{+0.002}_{-0.003}$ & $0.353^{+0.028}_{-0.039}$ & $0.319^{+0.001}_{-0.002}$ & mag \\ 
Effective temperature $T_{\text{eff}}$ & $4887\pm100$ & $5204\pm100$ & $4927\pm100$ & $5378\pm100$ & $5146\pm100$ & K \\ 
Surface gravity $\log{g}$ & $2.33\pm0.15$ & $2.23\pm0.15$ & $1.90\pm0.15$ & $2.48\pm0.15$ & $2.45\pm0.15$ & cm s$^{-2}$ \\ 
Metallicity $[\text{Fe/H}]$ & $-1.13\pm0.27$ & $-2.54\pm0.27$ & $-2.04\pm0.37$ & $-0.95\pm0.23$ & $-1.97\pm0.28$ &  \\ 
Microturbulence $\xi$ & $1.59\pm0.10$ & $1.62\pm0.10$ & $1.69\pm0.10$ & $1.56\pm0.10$ & $1.57\pm0.10$ & km s$^{-1}$ \\ 
\hline
\textbf{Orbital Properties} & & & & & & \\ 
Radial Velocity RV & $-303.5\pm2.0$ & $-120.6\pm1.5$ & $14.6\pm1.0$ & $-343.9\pm1.1$ & $-271.8\pm0.5$ & km s$^{-1}$ \\ 
Total Galactic velocity $v$ & $359.8^{+12.6}_{-12.2}$ & $519.8^{+5.8}_{-9.4}$ & $385.0^{+2.1}_{-4.2}$ & $436.7^{+2.5}_{-4.3}$ & $484.4^{+3.0}_{-5.5}$ & km s$^{-1}$ \\ 
Pericenter $\mathrm{R_{peri}}$ & $7.88^{+0.63}_{-2.86}$ & $9.58^{+0.02}_{-0.03}$ & $8.41^{+0.60}_{-3.31}$ & $8.55^{+0.01}_{-0.01}$ & $7.75^{+0.01}_{-0.01}$ & kpc \\ 
Apocenter $\mathrm{R_{apo}}$ & $31.14^{+4.26}_{-3.40}$ & $257.91^{+10.35}_{-17.23}$ & $44.40^{+1.46}_{-2.34}$ & $74.57^{+3.09}_{-4.87}$ & $140.62^{+7.20}_{-13.91}$ & kpc \\ 
Eccentricity $e$ & $0.61^{+0.10}_{-0.05}$ & $0.93^{+0.01}_{-0.01}$ & $0.68^{+0.11}_{-0.01}$ & $0.79^{+0.01}_{-0.01}$ & $0.90^{+0.01}_{-0.01}$ & \\ 
Maximum distance from Galactic plane $\mathrm{z_{max}}$ & $6.38^{+1.27}_{-0.92}$ & $180.59^{+10.37}_{-17.16}$ & $39.32^{+1.22}_{-2.19}$ & $71.49^{+3.09}_{-3.97}$ & $32.36^{+2.23}_{-3.57}$ & kpc \\ 
Total orbital energy E$\mathrm{_{tot}}$ & $-9893^{+4720}_{-4405}$ & $67044^{+3111}_{-4976}$ & $3367^{+879}_{-1718}$ & $21039^{+1130}_{-1914}$ & $37869^{+1433}_{-2642}$ & km$ ^2$ s$^{-2}$ \\ 
Angular momentum L$_Z$ & $2021^{+78}_{-76}$ & $514^{+22}_{-37}$ & $144^{+17}_{-35}$ & $227^{+23}_{-23}$ & $-616^{+10}_{-7}$ & kpc km s$^{-1}$ \\ 
\hline 
\hline 
Property & 3252546886080448384 & 5212110596595560192 & 5212817273334550016 & 3599974177996428032 & 6317828550897175936 &  Units \\ 
\hline 
\textbf{Photometric Properties} & & & & & & \\ 
 External ID & ES1 & ES8 & ES5$^{\rm{c}}$ & TYC 4934-700-1 & HE 1523$-$0901$^{\rm{c}}$ & \\ 
Gaia EDR3 parallax & $1.321\pm0.043$ & $0.444\pm0.009$ & $0.357\pm0.010$ & $0.362\pm0.017$ & $0.328\pm0.020$ & $mas$ \\ 
Gaia DR2 G & $12.073\pm0.002$ & $10.713\pm0.002$ & $10.895\pm0.002$ & $10.771\pm0.002$ & $10.744\pm0.002$ & Vega mag \\ 
Gaia DR2 BP & $12.391\pm0.002$ & $11.328\pm0.002$ & $11.655\pm0.002$ & $11.325\pm0.002$ & $11.391\pm0.002$ & Vega mag \\ 
Gaia DR2 RP & $11.584\pm0.002$ & $9.982\pm0.002$ & $10.069\pm0.002$ & $10.092\pm0.002$ & $9.992\pm0.002$ & Vega mag \\ 
SkyMapper $u$ & $13.509\pm0.012$ & $13.494\pm0.020$ & $14.421\pm0.012$ & $13.409\pm0.002$ & $13.722\pm0.007$ & AB mag \\ 
SkyMapper $v$ & $13.132\pm0.011$ & $12.931\pm0.006$ & $13.924\pm0.007$ & $12.726\pm0.006$ & $13.058\pm0.005$ & AB mag \\ 
SkyMapper $g$ & $\cdots$ & $11.308\pm0.006$ & $11.678\pm0.005$ & $11.301\pm0.007$ & $11.394\pm0.005$ & AB mag \\ 
SkyMapper $r$ & $\cdots$ & $10.750\pm0.004$ & $10.964\pm0.006$ & $10.810\pm0.008$ & $10.797\pm0.005$ & AB mag \\ 
SkyMapper $i$ & $\cdots$ & $10.323\pm0.004$ & $10.420\pm0.007$ & $\cdots$ & $10.336\pm0.007$ & AB mag \\ 
SkyMapper $z$ & $\cdots$ & $10.137\pm0.004$ & $10.178\pm0.007$ & $\cdots$ & $10.149\pm0.011$ & AB mag \\ 
2MASS $J$ & $10.957\pm0.024$ & $9.000\pm0.023$ & $8.881\pm0.024$ & $9.238\pm0.026$ & $9.027\pm0.028$ & Vega mag \\ 
2MASS $H$ & $10.635\pm0.024$ & $8.410\pm0.031$ & $8.199\pm0.023$ & $8.699\pm0.055$ & $8.421\pm0.031$ & Vega mag \\ 
2MASS $K$ & $10.523\pm0.021$ & $8.289\pm0.033$ & $8.022\pm0.017$ & $8.551\pm0.021$ & $8.352\pm0.027$ & Vega mag \\ 
WISE W1 & $10.521\pm0.023$ & $8.157\pm0.023$ & $7.930\pm0.026$ & $8.466\pm0.022$ & $8.200\pm0.024$ & Vega mag \\ 
WISE W2 & $10.514\pm0.020$ & $8.191\pm0.020$ & $7.988\pm0.020$ & $8.497\pm0.020$ & $8.199\pm0.021$ & Vega mag \\ 
\hline
\textbf{Stellar Properties} & & & & & & \\ 
Distance $d_{\text{iso}}$ & $0.8^{+0.1}_{-0.1}$ & $2.2^{+0.1}_{-0.1}$ & $2.7^{+0.1}_{-0.1}$ & $2.6^{+0.1}_{-0.1}$ & $2.9^{+0.1}_{-0.1}$ & kpc \\ 
Mass $M_{\odot}$ & $0.79^{+0.02}_{-0.01}$ & $0.98^{+0.07}_{-0.05}$ & $1.95^{+0.03}_{-0.02}$ & $0.78^{+0.01}_{-0.01}$ & $0.80^{+0.01}_{-0.01}$ & M$_{\odot}$ \\ 
Extinction A$_{\text{V}}$ & $0.029^{+0.028}_{-0.018}$ & $0.374^{+0.022}_{-0.017}$ & $0.659^{+0.001}_{-0.001}$ & $0.100^{+0.001}_{-0.001}$ & $0.370^{+0.001}_{-0.001}$ & mag \\ 
Effective temperature $T_{\text{eff}}$ & $5620\pm100$ & $4688\pm100$ & $4481\pm100$ & $4783\pm100$ & $4742\pm100$ & K \\ 
Surface gravity $\log{g}$ & $3.45\pm0.15$ & $1.60\pm0.15$ & $1.61\pm0.15$ & $1.51\pm0.15$ & $1.29\pm0.15$ & cm s$^{-2}$ \\ 
Metallicity $[\text{Fe/H}]$ & $-1.90\pm0.10$ & $-1.56\pm0.12$ & $-0.84\pm0.19$ & $-2.33\pm0.10$ & $-2.65\pm0.22$ &  \\ 
Microturbulence $\xi$ & $1.34\pm0.10$ & $1.76\pm0.10$ & $1.76\pm0.10$ & $1.78\pm0.10$ & $1.83\pm0.10$ & km s$^{-1}$ \\ 
\hline
\textbf{Orbital Properties} & & & & & \\ 
Radial Velocity RV & $1.7\pm1.2$ & $298.2\pm1.5$ & $159.9\pm0.5$ & $291.5\pm0.9$ & $-162.4\pm1.5$ & km s$^{-1}$ \\ 
Total Galactic velocity $v$ & $326.2^{+5.9}_{-8.6}$ & $460.2^{+3.2}_{-4.2}$ & $367.5^{+0.3}_{-0.2}$ & $327.2^{+0.5}_{-0.5}$ & $287.0^{+0.6}_{-0.6}$ & km s$^{-1}$ \\ 
Pericenter $\mathrm{R_{peri}}$ & $4.98^{+1.53}_{-0.44}$ & $7.75^{+0.01}_{-0.01}$ & $7.82^{+0.01}_{-0.01}$ & $8.45^{+0.01}_{-0.01}$ & $5.34^{+0.08}_{-0.02}$ & kpc \\ 
Apocenter $\mathrm{R_{apo}}$ & $23.30^{+0.58}_{-1.30}$ & $93.90^{+4.70}_{-6.51}$ & $31.18^{+0.06}_{-0.05}$ & $21.56^{+0.04}_{-0.05}$ & $10.43^{+0.05}_{-0.04}$ & kpc \\ 
Eccentricity $e$ & $0.65^{+0.03}_{-0.09}$ & $0.85^{+0.01}_{-0.01}$ & $0.60^{+0.01}_{-0.01}$ & $0.44^{+0.01}_{-0.01}$ & $0.32^{+0.01}_{-0.01}$ & \\ 
Maximum distance from Galactic plane $\mathrm{z_{max}}$ & $12.85^{+0.24}_{-1.37}$ & $29.34^{+1.44}_{-1.96}$ & $30.52^{+0.08}_{-0.08}$ & $6.21^{+0.03}_{-0.03}$ & $5.21^{+0.07}_{-0.05}$ & kpc \\ 
Total orbital energy E$\mathrm{_{tot}}$ & $-20298^{+2015}_{-2849}$ & $26766^{+1466}_{-1959}$ & $-10979^{+99}_{-84}$ & $-19605^{+168}_{-161}$ & $-48894^{+162}_{-155}$ & km$ ^2$ s$^{-2}$ \\ 
Angular momentum L$_Z$ & $1331^{+49}_{-69}$ & $336^{+11}_{-12}$ & $123^{+7}_{-7}$ & $2561^{+5}_{-4}$ & $1356^{+1}_{-1}$ & kpc km s$^{-1}$ \\ 
\hline 
\hline 
Property & 6479574961975897856 & 6558932694746826240 & 2629296824480015744 & 6505889848642319872 & 6556192329517108352 &  Units \\ 
\hline 
\textbf{Photometric Properties} & & & & & & \\ 
External ID & TYC 8429-663-1 & TYC 8443-1432-1$^{\rm{c}}$ & ES3$^{\rm{c}}$ & ES4 & TYC 7509-604-1 & \\ 
Gaia EDR3 parallax & $0.304\pm0.013$ & $0.341\pm0.018$ & $1.522\pm0.042$ & $0.379\pm0.019$ & $0.253\pm0.018$ & $mas$ \\ 
Gaia DR2 G & $10.583\pm0.002$ & $11.458\pm0.002$ & $11.365\pm0.002$ & $10.604\pm0.002$ & $11.499\pm0.002$ & Vega mag \\ 
Gaia DR2 BP & $11.174\pm0.002$ & $11.985\pm0.002$ & $11.843\pm0.002$ & $11.170\pm0.002$ & $12.050\pm0.002$ & Vega mag \\ 
Gaia DR2 RP & $9.878\pm0.002$ & $10.796\pm0.002$ & $10.742\pm0.002$ & $9.917\pm0.002$ & $10.814\pm0.002$ & Vega mag \\ 
SkyMapper $u$ & $13.440\pm0.011$ & $14.007\pm0.007$ & $13.524\pm0.011$ & $13.376\pm0.007$ & $14.235\pm0.005$ & AB mag \\ 
SkyMapper $v$ & $12.753\pm0.007$ & $13.420\pm0.018$ & $13.123\pm0.011$ & $12.790\pm0.009$ & $13.598\pm0.005$ & AB mag \\ 
SkyMapper $g$ & $11.168\pm0.018$ & $11.953\pm0.008$ & $11.792\pm0.007$ & $11.146\pm0.011$ & $12.011\pm0.006$ & AB mag \\ 
SkyMapper $r$ & $10.614\pm0.007$ & $11.472\pm0.007$ & $11.366\pm0.007$ & $10.627\pm0.011$ & $11.509\pm0.006$ & AB mag \\ 
SkyMapper $i$ & $10.227\pm0.011$ & $11.123\pm0.007$ & $11.074\pm0.005$ & $10.241\pm0.011$ & $11.137\pm0.006$ & AB mag \\ 
SkyMapper $z$ & $\cdots$ & $10.995\pm0.007$ & $10.963\pm0.006$ & $10.105\pm0.011$ & $10.991\pm0.006$ & AB mag \\ 
2MASS $J$ & $8.969\pm0.020$ & $9.936\pm0.022$ & $9.934\pm0.023$ & $9.047\pm0.026$ & $9.945\pm0.024$ & Vega mag \\ 
2MASS $H$ & $8.366\pm0.026$ & $9.423\pm0.029$ & $9.510\pm0.023$ & $8.509\pm0.047$ & $9.409\pm0.026$ & Vega mag \\ 
2MASS $K$ & $8.314\pm0.024$ & $9.305\pm0.019$ & $9.384\pm0.023$ & $8.324\pm0.033$ & $9.269\pm0.024$ & Vega mag \\ 
WISE W1 & $8.200\pm0.022$ & $9.228\pm0.023$ & $9.313\pm0.022$ & $8.261\pm0.023$ & $9.191\pm0.023$ & Vega mag \\ 
WISE W2 & $8.222\pm0.020$ & $9.268\pm0.021$ & $9.350\pm0.021$ & $8.303\pm0.020$ & $9.226\pm0.019$ & Vega mag \\ 
\hline
\textbf{Stellar Properties} & & & & & & \\ 
Distance $d_{\text{iso}}$ & $3.1^{+0.1}_{-0.1}$ & $2.9^{+0.1}_{-0.1}$ & $0.6^{+0.1}_{-0.1}$ & $2.6^{+0.1}_{-0.1}$ & $3.8^{+0.1}_{-0.1}$ & kpc \\ 
Mass $M_{\odot}$ & $0.78^{+0.01}_{-0.01}$ & $0.79^{+0.02}_{-0.01}$ & $0.86^{+0.02}_{-0.02}$ & $0.85^{+0.03}_{-0.04}$ & $1.42^{+0.04}_{-0.07}$ & M$_{\odot}$ \\ 
Extinction A$_{\text{V}}$ & $0.095^{+0.005}_{-0.004}$ & $0.013^{+0.011}_{-0.008}$ & $0.131^{+0.019}_{-0.017}$ & $0.006^{+0.008}_{-0.005}$ & $0.148^{+0.008}_{-0.011}$ & mag \\ 
Effective temperature $T_{\text{eff}}$ & $4599\pm100$ & $4686\pm100$ & $4958\pm100$ & $4574\pm100$ & $4708\pm100$ & K \\ 
Surface gravity $\log{g}$ & $1.18\pm0.15$ & $1.69\pm0.15$ & $3.07\pm0.15$ & $1.42\pm0.15$ & $1.69\pm0.15$ & cm s$^{-2}$ \\ 
Metallicity $[\text{Fe/H}]$ & $-2.17\pm0.13$ & $-1.60\pm0.15$ & $-1.13\pm0.28$ & $-1.67\pm0.14$ & $-1.29\pm0.19$ &  \\ 
Microturbulence $\xi$ & $1.86\pm0.10$ & $1.74\pm0.10$ & $1.43\pm0.10$ & $1.80\pm0.10$ & $1.74\pm0.10$ & km s$^{-1}$ \\ 
\hline
\textbf{Orbital Properties} & & & & & & \\ 
Radial Velocity RV & $194.2\pm0.6$ & $-33.5\pm0.7$ & $-219.7\pm0.6$ & $-38.2\pm0.4$ & $11.2\pm0.7$ & km s$^{-1}$ \\ 
Total Galactic velocity $v$ & $296.8^{+0.7}_{-1.0}$ & $372.5^{+4.3}_{-7.6}$ & $431.5^{+1.4}_{-2.9}$ & $387.4^{+7.1}_{-11.7}$ & $274.2^{+1.5}_{-0.9}$ & km s$^{-1}$ \\ 
Pericenter $\mathrm{R_{peri}}$ & $5.07^{+0.05}_{-0.02}$ & $6.69^{+0.01}_{-0.01}$ & $7.93^{+0.01}_{-1.92}$ & $7.05^{+0.08}_{-5.13}$ & $5.40^{+0.81}_{-1.40}$ & kpc \\ 
Apocenter $\mathrm{R_{apo}}$ & $11.81^{+0.18}_{-0.29}$ & $25.66^{+1.18}_{-3.20}$ & $62.55^{+1.39}_{-2.15}$ & $35.21^{+2.89}_{-3.41}$ & $15.20^{+0.05}_{-0.06}$ & kpc \\ 
Eccentricity $e$ & $0.40^{+0.01}_{-0.02}$ & $0.59^{+0.01}_{-0.05}$ & $0.78^{+0.04}_{-0.01}$ & $0.69^{+0.20}_{-0.04}$ & $0.48^{+0.11}_{-0.06}$ & \\ 
Maximum distance from Galactic plane $\mathrm{z_{max}}$ & $10.01^{+0.14}_{-0.23}$ & $10.96^{+0.54}_{-0.53}$ & $18.16^{+0.50}_{-0.97}$ & $10.51^{+1.10}_{-0.89}$ & $5.13^{+0.27}_{-0.51}$ & kpc \\ 
Total orbital energy E$\mathrm{_{tot}}$ & $-44625^{+215}_{-268}$ & $-15410^{+1576}_{-2767}$ & $14360^{+608}_{-1244}$ & $-6840^{+2769}_{-4433}$ & $-40268^{+419}_{-242}$ & km$ ^2$ s$^{-2}$ \\ 
Angular momentum L$_Z$ & $787^{+2}_{-4}$ & $2126^{+21}_{-40}$ & $125^{+7}_{-9}$ & $70^{+23}_{-38}$ & $1005^{+8}_{-4}$ & kpc km s$^{-1}$
\enddata 
\tablenotetext{a}{We report random uncertainties derived under the 
unlikely assumption that the MIST isochrone grid perfectly reproduces 
all stellar properties.  There are almost certainly larger systematic 
uncertainties that we have not investigated, though the excellent 
agreement between our analysis and previous results of star HE1523-0901 
seems to indicate that any systematic uncertainties in our analysis cannot be 
too large.}
\tablenotetext{b}{External IDs ``ESX'' are stars identified in \cite{hattori2018a}.}
\tablenotetext{c}{Stars more likely to be of extragalactic origin.}
\end{deluxetable*} 
\end{longrotatetable}

\begin{longrotatetable}
\begin{deluxetable*}{lcccccccccccccccccccc}
\tablecaption{Chemical Abundances\label{chem_abundances_tbl}}
\tablecolumns{21}
\tabletypesize{\tiny}
\tablewidth{0pt}
\tablehead{
\colhead{Species} & 
\colhead{n} & \colhead{log($\epsilon_X$)} &
\colhead{[X/Fe]} & \colhead{$\sigma_{\mathrm{[X/Fe]}}$} &
\colhead{n} & \colhead{log($\epsilon_X$)} &
\colhead{[X/Fe]} & \colhead{$\sigma_{\mathrm{[X/Fe]}}$} &
\colhead{n} & \colhead{log($\epsilon_X$)} &
\colhead{[X/Fe]} & \colhead{$\sigma_{\mathrm{[X/Fe]}}$} &
\colhead{n} & \colhead{log($\epsilon_X$)} &
\colhead{[X/Fe]} & \colhead{$\sigma_{\mathrm{[X/Fe]}}$} &
\colhead{n} & \colhead{log($\epsilon_X$)} &
\colhead{[X/Fe]} & \colhead{$\sigma_{\mathrm{[X/Fe]}}$} }
\startdata
\hline 
 & \multicolumn{4}{c}{Gaia EDR3 2853089398265954432} & \multicolumn{4}{c}{Gaia EDR3 330414789019026944} & \multicolumn{4}{c}{Gaia EDR3 2260163008363761664} & \multicolumn{4}{c}{Gaia EDR3 2233912206910720000} & \multicolumn{4}{c}{Gaia EDR3 1765600930139450752} \\ 
\ion{Na}{1} & $2$ & $4.684$ & $-0.179$ & $0.150$ & $2$ & $2.991$ & $-0.864$ & $0.185$ & $\cdots$ & $\cdots$ & $\cdots$ & $\cdots$ & $2$ & $4.922$ & $-0.365$ & $0.135$ & $2$ & $4.082$ & $-0.165$ & $0.095$  \\ 
\ion{Na}{1}$_{\rm{NLTE}}$ & $2$ & $4.727$ & $-0.136$ & $\cdots$ & $2$ & $\cdots$ & $\cdots$ & $\cdots$ & $\cdots$ & $\cdots$ & $\cdots$ & $\cdots$ & $2$ & $4.805$ & $-0.482$ & $\cdots$ & $2$ & $3.430$ & $-0.817$ & $\cdots$  \\ 
\ion{Mg}{1} & $6$ & $6.252$ & $0.059$ & $0.074$ & $5$ & $5.100$ & $-0.085$ & $0.246$ & $2$ & $5.864$ & $-0.007$ & $0.030$ & $7$ & $6.582$ & $-0.035$ & $0.089$ & $4$ & $5.642$ & $0.065$ & $0.227$  \\ 
\ion{Al}{1} & $1$ & $3.674$ & $-1.399$ & $0.068$ & $\cdots$ & $\cdots$ & $\cdots$ & $\cdots$ & $\cdots$ & $\cdots$ & $\cdots$ & $\cdots$ & $\cdots$ & $\cdots$ & $\cdots$ & $\cdots$ & $\cdots$ & $\cdots$ & $\cdots$ & $\cdots$  \\ 
\ion{Al}{1}$_{\rm{NLTE}}$ & $1$ & $3.530$ & $-1.543$ & $\cdots$ & $\cdots$ & $\cdots$ & $\cdots$ & $\cdots$ & $\cdots$ & $\cdots$ & $\cdots$ & $\cdots$ & $\cdots$ & $\cdots$ & $\cdots$ & $\cdots$ & $\cdots$ & $\cdots$ & $\cdots$ & $\cdots$  \\ 
\ion{Si}{1} & $4$ & $6.816$ & $0.663$ & $0.245$ & $\cdots$ & $\cdots$ & $\cdots$ & $\cdots$ & $1$ & $5.813$ & $-0.018$ & $0.002$ & $6$ & $6.585$ & $0.008$ & $0.104$ & $1$ & $5.235$ & $-0.302$ & $0.004$  \\ 
\ion{K}{1}$_{\rm{NLTE}}$ & $1$ & $3.538$ & $-0.175$ & $\cdots$ & $1$ & $3.401$ & $0.696$ & $\cdots$ & $1$ & $\cdots$ & $\cdots$ & $\cdots$ & $1$ & $5.200$ & $1.063$ & $\cdots$ & $1$ & $3.007$ & $-0.090$ & $\cdots$  \\ 
\ion{Ca}{1} & $16$ & $5.410$ & $0.467$ & $0.156$ & $16$ & $4.270$ & $0.335$ & $0.068$ & $12$ & $4.843$ & $0.222$ & $0.157$ & $23$ & $5.519$ & $0.152$ & $0.067$ & $21$ & $4.554$ & $0.227$ & $0.079$  \\ 
\ion{Sc}{2} & $4$ & $2.126$ & $0.343$ & $0.288$ & $1$ & $0.589$ & $-0.186$ & $0.014$ & $\cdots$ & $\cdots$ & $\cdots$ & $\cdots$ & $4$ & $2.038$ & $-0.169$ & $0.125$ & $2$ & $0.727$ & $-0.440$ & $0.238$  \\ 
\ion{Ti}{1} & $25$ & $3.826$ & $0.213$ & $0.136$ & $8$ & $2.920$ & $0.315$ & $0.129$ & $8$ & $3.362$ & $0.071$ & $0.130$ & $34$ & $4.042$ & $0.005$ & $0.061$ & $12$ & $3.184$ & $0.187$ & $0.085$  \\ 
\ion{Ti}{2} & $22$ & $3.956$ & $0.343$ & $0.201$ & $22$ & $2.699$ & $0.094$ & $0.107$ & $27$ & $3.032$ & $-0.259$ & $0.112$ & $48$ & $4.246$ & $0.209$ & $0.095$ & $24$ & $3.234$ & $0.237$ & $0.069$  \\ 
\ion{Cr}{1} & $11$ & $4.425$ & $0.162$ & $0.225$ & $6$ & $3.094$ & $-0.161$ & $0.201$ & $4$ & $3.948$ & $0.007$ & $0.163$ & $14$ & $4.399$ & $-0.288$ & $0.092$ & $5$ & $3.193$ & $-0.454$ & $0.187$  \\ 
\ion{Cr}{2} & $3$ & $4.495$ & $0.232$ & $0.277$ & $2$ & $3.183$ & $-0.072$ & $0.027$ & $3$ & $3.357$ & $-0.584$ & $0.164$ & $4$ & $4.850$ & $0.163$ & $0.056$ & $3$ & $3.740$ & $0.093$ & $0.204$  \\ 
\ion{Mn}{1} & $1$ & $3.517$ & $-0.546$ & $0.002$ & $\cdots$ & $\cdots$ & $\cdots$ & $\cdots$ & $\cdots$ & $\cdots$ & $\cdots$ & $\cdots$ & $5$ & $3.844$ & $-0.643$ & $0.093$ & $\cdots$ & $\cdots$ & $\cdots$ & $\cdots$  \\ 
\ion{Fe}{1} & $51$ & $6.026$ & $\cdots$ & $\cdots$  & $52$ & $5.121$ & $\cdots$ & $\cdots$  & $26$ & $5.865$ & $\cdots$ & $\cdots$  & $70$ & $6.534$ & $\cdots$ & $\cdots$  & $53$ & $5.486$ & $\cdots$ & $\cdots$   \\ 
\ion{Fe}{2} & $17$ & $6.331$ & $\cdots$ & $\cdots$ & $8$ & $4.923$ & $\cdots$ & $\cdots$ & $6$ & $5.415$ & $\cdots$ & $\cdots$ & $26$ & $6.507$ & $\cdots$ & $\cdots$ & $13$ & $5.488$ & $\cdots$ & $\cdots$  \\ 
\ion{Co}{1} & $2$ & $4.058$ & $0.475$ & $0.133$ & $\cdots$ & $\cdots$ & $\cdots$ & $\cdots$ & $1$ & $3.791$ & $0.530$ & $0.010$ & $3$ & $4.077$ & $0.070$ & $0.096$ & $\cdots$ & $\cdots$ & $\cdots$ & $\cdots$  \\ 
\ion{Ni}{1} & $11$ & $4.737$ & $-0.106$ & $0.131$ & $11$ & $3.961$ & $0.126$ & $0.118$ & $3$ & $4.176$ & $-0.345$ & $0.077$ & $24$ & $4.932$ & $-0.335$ & $0.045$ & $10$ & $4.176$ & $-0.051$ & $0.099$  \\ 
\ion{Cu}{1} & $\cdots$ & $\cdots$ & $\cdots$ & $\cdots$ & $\cdots$ & $\cdots$ & $\cdots$ & $\cdots$ & $\cdots$ & $\cdots$ & $\cdots$ & $\cdots$ & $\cdots$ & $\cdots$ & $\cdots$ & $\cdots$ & $\cdots$ & $\cdots$ & $\cdots$ & $\cdots$  \\ 
\ion{Zn}{1} & $1$ & $3.043$ & $-0.160$ & $0.114$ & $\cdots$ & $\cdots$ & $\cdots$ & $\cdots$ & $\cdots$ & $\cdots$ & $\cdots$ & $\cdots$ & $2$ & $3.405$ & $-0.222$ & $0.118$ & $\cdots$ & $\cdots$ & $\cdots$ & $\cdots$  \\ 
\ion{Sr}{1} & $\cdots$ & $\cdots$ & $\cdots$ & $\cdots$ & $\cdots$ & $\cdots$ & $\cdots$ & $\cdots$ & $\cdots$ & $\cdots$ & $\cdots$ & $\cdots$ & $1$ & $2.047$ & $0.150$ & $0.018$ & $\cdots$ & $\cdots$ & $\cdots$ & $\cdots$  \\ 
\ion{Sr}{2} & $\cdots$ & $\cdots$ & $\cdots$ & $\cdots$ & $2$ & $-0.018$ & $-0.483$ & $0.314$ & $\cdots$ & $\cdots$ & $\cdots$ & $\cdots$ & $1$ & $1.887$ & $-0.010$ & $0.077$ & $1$ & $0.831$ & $-0.026$ & $0.065$  \\ 
\ion{Y}{2} & $5$ & $2.201$ & $1.348$ & $0.200$ & $1$ & $-0.413$ & $-0.258$ & $0.011$ & $1$ & $0.708$ & $0.177$ & $0.072$ & $5$ & $1.059$ & $-0.218$ & $0.194$ & $\cdots$ & $\cdots$ & $\cdots$ & $\cdots$  \\ 
\ion{Ba}{2} & $4$ & $2.196$ & $1.283$ & $0.161$ & $3$ & $-0.822$ & $-0.727$ & $0.049$ & $2$ & $0.558$ & $-0.033$ & $0.136$ & $4$ & $1.667$ & $0.330$ & $0.136$ & $4$ & $0.089$ & $-0.208$ & $0.100$  \\ 
\ion{La}{2} & $\cdots$ & $\cdots$ & $\cdots$ & $\cdots$ & $\cdots$ & $\cdots$ & $\cdots$ & $\cdots$ & $\cdots$ & $\cdots$ & $\cdots$ & $\cdots$ & $\cdots$ & $\cdots$ & $\cdots$ & $\cdots$ & $\cdots$ & $\cdots$ & $\cdots$ & $\cdots$  \\ 
\ion{Eu}{2} & $1$ & $-0.918$ & $-0.081$ & $0.300$ & $2$ & $-1.012$ & $0.833$ & $0.300$ & $3$ & $-0.504$ & $0.655$ & $0.300$ & $5$ & $-0.213$ & $0.200$ & $0.300$ & $2$ & $-1.303$ & $0.150$ & $0.300$  \\ 
\hline 
 & \multicolumn{4}{c}{Gaia EDR3 3252546886080448384} & \multicolumn{4}{c}{Gaia EDR3 5212110596595560192} & \multicolumn{4}{c}{Gaia EDR3 5212817273334550016} & \multicolumn{4}{c}{Gaia EDR3 3599974177996428032} & \multicolumn{4}{c}{Gaia EDR3 6317828550897175936} \\ 
\ion{Na}{1} & $2$ & $4.406$ & $0.229$ & $0.092$ & $2$ & $4.245$ & $-0.275$ & $0.015$ & $2$ & $4.801$ & $-0.356$ & $0.075$ & $2$ & $3.978$ & $-0.020$ & $0.254$ & $2$ & $3.742$ & $-0.035$ & $0.232$  \\ 
\ion{Na}{1}$_{\rm{NLTE}}$ & $2$ & $3.810$ & $-0.367$ & $\cdots$ & $2$ & $4.158$ & $-0.362$ & $\cdots$ & $2$ & $4.701$ & $-0.456$ & $\cdots$ & $2$ & $3.885$ & $-0.113$ & $\cdots$ & $2$ & $3.342$ & $-0.435$ & $\cdots$  \\ 
\ion{Mg}{1} & $6$ & $5.699$ & $0.192$ & $0.077$ & $6$ & $6.195$ & $0.345$ & $0.044$ & $4$ & $6.586$ & $0.099$ & $0.140$ & $6$ & $5.541$ & $0.213$ & $0.071$ & $4$ & $5.263$ & $0.156$ & $0.101$  \\ 
\ion{Al}{1} & $2$ & $3.327$ & $-1.060$ & $0.448$ & $\cdots$ & $\cdots$ & $\cdots$ & $\cdots$ & $\cdots$ & $\cdots$ & $\cdots$ & $\cdots$ & $1$ & $3.365$ & $-0.843$ & $0.034$ & $1$ & $2.832$ & $-1.155$ & $0.043$  \\ 
\ion{Al}{1}$_{\rm{NLTE}}$ & $2$ & $3.888$ & $-0.499$ & $\cdots$ & $\cdots$ & $\cdots$ & $\cdots$ & $\cdots$ & $\cdots$ & $\cdots$ & $\cdots$ & $\cdots$ & $1$ & $3.260$ & $-0.948$ & $\cdots$ & $1$ & $2.980$ & $-1.007$ & $\cdots$  \\ 
\ion{Si}{1} & $1$ & $5.392$ & $-0.075$ & $0.018$ & $6$ & $6.137$ & $0.327$ & $0.073$ & $8$ & $6.679$ & $0.232$ & $0.106$ & $3$ & $5.573$ & $0.285$ & $0.121$ & $3$ & $5.099$ & $0.032$ & $0.201$  \\ 
\ion{K}{1}$_{\rm{NLTE}}$ & $1$ & $3.230$ & $0.203$ & $\cdots$ & $1$ & $3.510$ & $0.140$ & $\cdots$ & $1$ & $3.969$ & $-0.038$ & $\cdots$ & $1$ & $3.095$ & $0.247$ & $\cdots$ & $1$ & $\cdots$ & $\cdots$ & $\cdots$  \\ 
\ion{Ca}{1} & $22$ & $4.668$ & $0.411$ & $0.018$ & $22$ & $4.863$ & $0.263$ & $0.074$ & $22$ & $5.294$ & $0.057$ & $0.053$ & $22$ & $4.209$ & $0.131$ & $0.034$ & $15$ & $4.094$ & $0.237$ & $0.175$  \\ 
\ion{Sc}{2} & $2$ & $1.121$ & $0.024$ & $0.036$ & $3$ & $1.687$ & $0.247$ & $0.084$ & $3$ & $2.251$ & $0.174$ & $0.106$ & $5$ & $0.776$ & $-0.142$ & $0.059$ & $5$ & $0.454$ & $-0.243$ & $0.065$  \\ 
\ion{Ti}{1} & $16$ & $3.056$ & $0.129$ & $0.032$ & $36$ & $3.324$ & $0.054$ & $0.060$ & $47$ & $3.873$ & $-0.034$ & $0.065$ & $43$ & $2.737$ & $-0.011$ & $0.037$ & $28$ & $2.841$ & $0.314$ & $0.131$  \\ 
\ion{Ti}{2} & $34$ & $3.213$ & $0.286$ & $0.030$ & $49$ & $3.641$ & $0.371$ & $0.080$ & $50$ & $4.418$ & $0.511$ & $0.113$ & $58$ & $2.927$ & $0.179$ & $0.036$ & $46$ & $2.800$ & $0.273$ & $0.068$  \\ 
\ion{Cr}{1} & $8$ & $3.348$ & $-0.229$ & $0.029$ & $15$ & $3.646$ & $-0.274$ & $0.055$ & $15$ & $4.523$ & $-0.034$ & $0.123$ & $16$ & $3.092$ & $-0.306$ & $0.034$ & $15$ & $2.844$ & $-0.333$ & $0.063$  \\ 
\ion{Cr}{2} & $2$ & $3.779$ & $0.202$ & $0.015$ & $3$ & $4.167$ & $0.247$ & $0.214$ & $3$ & $4.543$ & $-0.014$ & $0.110$ & $4$ & $3.193$ & $-0.205$ & $0.033$ & $3$ & $2.882$ & $-0.295$ & $0.092$  \\ 
\ion{Mn}{1} & $1$ & $2.874$ & $-0.503$ & $0.007$ & $4$ & $3.073$ & $-0.647$ & $0.115$ & $5$ & $3.565$ & $-0.792$ & $0.072$ & $2$ & $2.827$ & $-0.371$ & $0.293$ & $\cdots$ & $\cdots$ & $\cdots$ & $\cdots$  \\ 
\ion{Fe}{1} & $42$ & $5.361$ & $\cdots$ & $\cdots$  & $59$ & $5.720$ & $\cdots$ & $\cdots$  & $49$ & $6.325$ & $\cdots$ & $\cdots$  & $74$ & $5.272$ & $\cdots$ & $\cdots$  & $61$ & $5.083$ & $\cdots$ & $\cdots$   \\ 
\ion{Fe}{2} & $16$ & $5.560$ & $\cdots$ & $\cdots$ & $17$ & $5.902$ & $\cdots$ & $\cdots$ & $16$ & $6.612$ & $\cdots$ & $\cdots$ & $24$ & $5.134$ & $\cdots$ & $\cdots$ & $20$ & $4.814$ & $\cdots$ & $\cdots$  \\ 
\ion{Co}{1} & $\cdots$ & $\cdots$ & $\cdots$ & $\cdots$ & $2$ & $2.962$ & $-0.278$ & $0.107$ & $5$ & $3.816$ & $-0.061$ & $0.059$ & $1$ & $2.541$ & $-0.177$ & $0.010$ & $\cdots$ & $\cdots$ & $\cdots$ & $\cdots$  \\ 
\ion{Ni}{1} & $9$ & $4.014$ & $-0.143$ & $0.036$ & $22$ & $4.334$ & $-0.166$ & $0.043$ & $26$ & $4.966$ & $-0.171$ & $0.109$ & $26$ & $3.811$ & $-0.167$ & $0.037$ & $11$ & $3.982$ & $0.225$ & $0.275$  \\ 
\ion{Cu}{1} & $\cdots$ & $\cdots$ & $\cdots$ & $\cdots$ & $1$ & $1.856$ & $-0.624$ & $0.010$ & $1$ & $3.001$ & $-0.116$ & $0.034$ & $\cdots$ & $\cdots$ & $\cdots$ & $\cdots$ & $\cdots$ & $\cdots$ & $\cdots$ & $\cdots$  \\ 
\ion{Zn}{1} & $1$ & $2.224$ & $-0.293$ & $0.008$ & $1$ & $2.938$ & $0.078$ & $0.037$ & $1$ & $3.504$ & $0.007$ & $0.068$ & $2$ & $2.241$ & $-0.097$ & $0.008$ & $\cdots$ & $\cdots$ & $\cdots$ & $\cdots$  \\ 
\ion{Sr}{1} & $\cdots$ & $\cdots$ & $\cdots$ & $\cdots$ & $1$ & $0.944$ & $-0.186$ & $0.013$ & $1$ & $1.133$ & $-0.634$ & $0.032$ & $1$ & $0.227$ & $-0.381$ & $0.006$ & $\cdots$ & $\cdots$ & $\cdots$ & $\cdots$  \\ 
\ion{Sr}{2} & $2$ & $0.819$ & $0.032$ & $0.152$ & $2$ & $1.240$ & $0.110$ & $0.071$ & $2$ & $1.764$ & $-0.003$ & $0.113$ & $2$ & $0.442$ & $-0.166$ & $0.098$ & $2$ & $0.782$ & $0.395$ & $0.075$  \\ 
\ion{Y}{2} & $\cdots$ & $\cdots$ & $\cdots$ & $\cdots$ & $2$ & $0.557$ & $0.047$ & $0.104$ & $3$ & $1.182$ & $0.035$ & $0.324$ & $3$ & $-0.399$ & $-0.387$ & $0.027$ & $4$ & $0.102$ & $0.335$ & $0.036$  \\ 
\ion{Ba}{2} & $1$ & $0.342$ & $0.115$ & $0.038$ & $2$ & $0.781$ & $0.211$ & $0.113$ & $2$ & $1.519$ & $0.312$ & $0.145$ & $4$ & $-0.089$ & $-0.137$ & $0.100$ & $4$ & $0.796$ & $0.969$ & $0.083$  \\ 
\ion{La}{2} & $\cdots$ & $\cdots$ & $\cdots$ & $\cdots$ & $1$ & $-0.212$ & $0.378$ & $0.033$ & $1$ & $0.350$ & $0.303$ & $0.071$ & $\cdots$ & $\cdots$ & $\cdots$ & $\cdots$ & $1$ & $-0.133$ & $1.200$ & $0.013$  \\ 
\ion{Eu}{2} & $3$ & $-1.262$ & $0.261$ & $0.200$ & $4$ & $-0.904$ & $0.276$ & $0.200$ & $5$ & $-0.411$ & $0.132$ & $0.200$ & $4$ & $-1.210$ & $0.492$ & $0.300$ & $4$ & $-0.024$ & $1.900$ & $0.300$  \\ 
\hline 
 & \multicolumn{4}{c}{Gaia EDR3 6479574961975897856} & \multicolumn{4}{c}{Gaia EDR3 6558932694746826240} & \multicolumn{4}{c}{Gaia EDR3 2629296824480015744} & \multicolumn{4}{c}{Gaia EDR3 6505889848642319872} & \multicolumn{4}{c}{Gaia EDR3 6556192329517108352} \\ 
\ion{Na}{1} & $\cdots$ & $\cdots$ & $\cdots$ & $\cdots$ & $2$ & $4.011$ & $-0.480$ & $0.130$ & $2$ & $4.912$ & $0.066$ & $0.060$ & $2$ & $4.393$ & $-0.073$ & $0.020$ & $2$ & $4.383$ & $-0.322$ & $0.099$  \\ 
\ion{Na}{1}$_{\rm{NLTE}}$ & $\cdots$ & $\cdots$ & $\cdots$ & $\cdots$ & $2$ & $3.938$ & $-0.553$ & $\cdots$ & $2$ & $4.798$ & $-0.048$ & $\cdots$ & $2$ & $4.298$ & $-0.168$ & $\cdots$ & $2$ & $4.307$ & $-0.398$ & $\cdots$  \\ 
\ion{Mg}{1} & $7$ & $5.751$ & $0.499$ & $0.078$ & $8$ & $5.929$ & $0.108$ & $0.042$ & $5$ & $6.322$ & $0.146$ & $0.079$ & $5$ & $6.184$ & $0.388$ & $0.070$ & $7$ & $6.249$ & $0.214$ & $0.056$  \\ 
\ion{Al}{1} & $1$ & $3.387$ & $-0.745$ & $0.115$ & $1$ & $3.583$ & $-1.118$ & $0.002$ & $\cdots$ & $\cdots$ & $\cdots$ & $\cdots$ & $\cdots$ & $\cdots$ & $\cdots$ & $\cdots$ & $1$ & $3.933$ & $-0.982$ & $0.024$  \\ 
\ion{Al}{1}$_{\rm{NLTE}}$ & $1$ & $3.231$ & $-0.901$ & $\cdots$ & $1$ & $3.405$ & $-1.296$ & $\cdots$ & $\cdots$ & $\cdots$ & $\cdots$ & $\cdots$ & $\cdots$ & $\cdots$ & $\cdots$ & $\cdots$ & $1$ & $3.737$ & $-1.178$ & $\cdots$  \\ 
\ion{Si}{1} & $7$ & $5.774$ & $0.562$ & $0.049$ & $6$ & $6.028$ & $0.247$ & $0.041$ & $8$ & $6.573$ & $0.437$ & $0.090$ & $7$ & $6.071$ & $0.315$ & $0.087$ & $7$ & $6.201$ & $0.206$ & $0.055$  \\ 
\ion{K}{1}$_{\rm{NLTE}}$ & $1$ & $\cdots$ & $\cdots$ & $\cdots$ & $1$ & $3.536$ & $0.195$ & $\cdots$ & $1$ & $3.697$ & $0.001$ & $\cdots$ & $1$ & $3.648$ & $0.332$ & $\cdots$ & $1$ & $3.646$ & $0.091$ & $\cdots$  \\ 
\ion{Ca}{1} & $21$ & $4.346$ & $0.344$ & $0.096$ & $23$ & $4.773$ & $0.202$ & $0.061$ & $22$ & $5.193$ & $0.267$ & $0.039$ & $22$ & $4.922$ & $0.376$ & $0.073$ & $23$ & $5.057$ & $0.272$ & $0.069$  \\ 
\ion{Sc}{2} & $5$ & $0.989$ & $0.147$ & $0.092$ & $5$ & $1.443$ & $0.032$ & $0.076$ & $5$ & $2.104$ & $0.338$ & $0.121$ & $3$ & $1.547$ & $0.161$ & $0.089$ & $5$ & $1.922$ & $0.297$ & $0.101$  \\ 
\ion{Ti}{1} & $54$ & $2.879$ & $0.207$ & $0.091$ & $58$ & $3.232$ & $-0.009$ & $0.047$ & $46$ & $3.655$ & $0.059$ & $0.031$ & $51$ & $3.416$ & $0.200$ & $0.069$ & $65$ & $3.445$ & $-0.010$ & $0.047$  \\ 
\ion{Ti}{2} & $59$ & $3.140$ & $0.468$ & $0.054$ & $61$ & $3.575$ & $0.334$ & $0.057$ & $46$ & $3.977$ & $0.381$ & $0.104$ & $52$ & $3.601$ & $0.385$ & $0.086$ & $57$ & $3.956$ & $0.501$ & $0.075$  \\ 
\ion{Cr}{1} & $15$ & $3.013$ & $-0.309$ & $0.125$ & $17$ & $3.569$ & $-0.322$ & $0.040$ & $14$ & $4.084$ & $-0.162$ & $0.040$ & $15$ & $3.713$ & $-0.153$ & $0.077$ & $17$ & $3.788$ & $-0.317$ & $0.059$  \\ 
\ion{Cr}{2} & $4$ & $3.460$ & $0.138$ & $0.064$ & $5$ & $4.057$ & $0.166$ & $0.088$ & $4$ & $4.474$ & $0.228$ & $0.114$ & $3$ & $3.968$ & $0.102$ & $0.052$ & $4$ & $4.196$ & $0.091$ & $0.084$  \\ 
\ion{Mn}{1} & $6$ & $2.768$ & $-0.354$ & $0.077$ & $6$ & $3.078$ & $-0.613$ & $0.054$ & $5$ & $3.685$ & $-0.361$ & $0.033$ & $6$ & $3.176$ & $-0.490$ & $0.037$ & $6$ & $3.175$ & $-0.730$ & $0.064$  \\ 
\ion{Fe}{1} & $48$ & $5.109$ & $\cdots$ & $\cdots$  & $72$ & $5.686$ & $\cdots$ & $\cdots$  & $63$ & $6.000$ & $\cdots$ & $\cdots$  & $65$ & $5.675$ & $\cdots$ & $\cdots$  & $76$ & $5.865$ & $\cdots$ & $\cdots$   \\ 
\ion{Fe}{2} & $20$ & $5.293$ & $\cdots$ & $\cdots$ & $25$ & $5.860$ & $\cdots$ & $\cdots$ & $22$ & $6.325$ & $\cdots$ & $\cdots$ & $23$ & $5.793$ & $\cdots$ & $\cdots$ & $28$ & $6.162$ & $\cdots$ & $\cdots$  \\ 
\ion{Co}{1} & $3$ & $2.746$ & $0.104$ & $0.112$ & $2$ & $3.178$ & $-0.033$ & $0.146$ & $1$ & $3.622$ & $0.056$ & $0.026$ & $2$ & $2.987$ & $-0.199$ & $0.160$ & $3$ & $3.260$ & $-0.165$ & $0.022$  \\ 
\ion{Ni}{1} & $26$ & $3.843$ & $-0.059$ & $0.073$ & $31$ & $4.191$ & $-0.280$ & $0.069$ & $27$ & $4.738$ & $-0.088$ & $0.056$ & $28$ & $4.318$ & $-0.128$ & $0.033$ & $30$ & $4.530$ & $-0.155$ & $0.035$  \\ 
\ion{Cu}{1} & $\cdots$ & $\cdots$ & $\cdots$ & $\cdots$ & $1$ & $1.594$ & $-0.857$ & $0.005$ & $1$ & $2.472$ & $-0.334$ & $0.030$ & $1$ & $1.921$ & $-0.505$ & $0.014$ & $1$ & $1.520$ & $-1.145$ & $0.008$  \\ 
\ion{Zn}{1} & $2$ & $2.459$ & $0.197$ & $0.039$ & $2$ & $2.832$ & $0.001$ & $0.043$ & $2$ & $3.370$ & $0.184$ & $0.081$ & $1$ & $2.794$ & $-0.012$ & $0.044$ & $2$ & $3.129$ & $0.084$ & $0.060$  \\ 
\ion{Sr}{1} & $1$ & $0.429$ & $-0.103$ & $0.013$ & $1$ & $0.685$ & $-0.416$ & $0.011$ & $1$ & $1.215$ & $-0.241$ & $0.011$ & $1$ & $0.862$ & $-0.214$ & $0.014$ & $1$ & $0.667$ & $-0.648$ & $0.017$  \\ 
\ion{Sr}{2} & $2$ & $0.683$ & $0.151$ & $0.092$ & $2$ & $0.999$ & $-0.102$ & $0.108$ & $2$ & $1.520$ & $0.064$ & $0.180$ & $2$ & $1.158$ & $0.082$ & $0.102$ & $2$ & $1.358$ & $0.043$ & $0.112$  \\ 
\ion{Y}{2} & $4$ & $-0.253$ & $-0.165$ & $0.103$ & $5$ & $0.261$ & $-0.220$ & $0.053$ & $4$ & $0.958$ & $0.122$ & $0.136$ & $5$ & $0.438$ & $-0.018$ & $0.076$ & $5$ & $0.592$ & $-0.103$ & $0.072$  \\ 
\ion{Ba}{2} & $3$ & $0.095$ & $0.123$ & $0.098$ & $4$ & $0.585$ & $0.044$ & $0.110$ & $2$ & $1.280$ & $0.384$ & $0.142$ & $2$ & $0.565$ & $0.049$ & $0.144$ & $4$ & $0.993$ & $0.238$ & $0.108$  \\ 
\ion{La}{2} & $2$ & $-0.830$ & $0.358$ & $0.029$ & $3$ & $-0.402$ & $0.217$ & $0.072$ & $1$ & $0.366$ & $0.630$ & $0.097$ & $2$ & $-0.327$ & $0.317$ & $0.040$ & $3$ & $-0.203$ & $0.202$ & $0.199$  \\ 
\ion{Eu}{2} & $4$ & $-1.472$ & $0.305$ & $0.300$ & $4$ & $-1.028$ & $0.180$ & $0.300$ & $3$ & $-0.588$ & $0.266$ & $0.300$ & $5$ & $-0.900$ & $0.334$ & $0.200$ & $4$ & $-0.895$ & $0.100$ & $0.300$
\enddata
\end{deluxetable*}
\end{longrotatetable}

\end{document}